\documentclass[12pt]{iopart}

\makeatletter
\expandafter\let\csname equation*\endcsname\relax
\expandafter\let\csname endequation*\endcsname\relax

\makeatother

\usepackage{amsmath}
\usepackage{mathtools}
\usepackage{iopams}

\newcommand{\beq}{\begin{equation}}
\newcommand{\eeq}{\end{equation}}

\usepackage{tabularx}
\usepackage{amsthm}
\usepackage{amscd}       
\usepackage{mathrsfs}
\usepackage{bm}
\usepackage{boxedminipage}
\usepackage{graphicx}
\usepackage{appendix}
\usepackage{physics}
\usepackage{multirow}
\usepackage{epstopdf}
\usepackage{float}

\usepackage[ruled,vlined]{algorithm2e}  
\usepackage{subcaption}
\usepackage{comment}
\usepackage{csquotes}
\MakeOuterQuote{"}

\usepackage[final]{hyperref} 
\hypersetup{
  colorlinks=true, linkcolor=blue, citecolor=blue, filecolor=magenta, urlcolor=blue
}

\theoremstyle{plain} 

\theoremstyle{definition}

\newcommand{\R}{\mathbb{R}}
\newcommand{\N}{\mathcal{N}}
\newcommand{\T}{\mathcal{T}}
\newcommand{\CL}{\mathcal{L}}

\newcommand{\bu}{\bm{u}}
\newcommand{\bfX}{\mathbf{X}}
\newcommand{\bfx}{\mathbf{x}}
\newcommand{\bfy}{\mathbf{y}}

\newcommand{\bmtheta}{\bm{\theta}}
\newcommand{\bp}{\mathbf{p}}
\newcommand{\bq}{\mathbf{q}}
\newcommand{\ii}{\mathrm{i}}
\DeclareMathOperator{\arccosh}{arccosh}
\DeclareMathOperator{\arcsinh}{arcsinh}

\newtheorem{theorem}{Theorem}
\newtheorem*{thm*}{Theorem}
\newtheorem*{lemma*}{Lemma}

\usepackage{array}

\begin{document}

\title[Nonlinear Waves Meet Machine Learning]{Machine learning of nonlinear waves: data-driven methods for computer-assisted discovery of equations, symmetries, conservation laws, and integrability}
\author{Jimmie Adriazola}
\address{School of Mathematical and Statistical Sciences, Arizona State University}
\ead{jimmie.adriazola@asu.edu}

\author{Panayotis G. Kevrekidis$^{*}$}
\address{Department of Mathematics and Statistics, University of Massachusetts, Amherst}
\ead{kevrekid@umass.edu}

\author{Vassilis Koukouloyannis}
\address{Department of Mathematics, University of the Aegean, Karlovassi, Greece}
\ead{vkouk@aegean.gr}

\author{Wei Zhu}
\address{School of Mathematics, Georgia Institute of Technology}
\ead{weizhu@gatech.edu}

\begin{abstract}
The purpose of this article is to provide a perspective---admittedly, a rather subjective one---of recent developments at the interface of machine learning (ML)/data-driven methods and nonlinear wave studies. We review some recent pillars of the rapidly evolving landscape of scientific ML, including deep learning, data-driven equation discovery, Koopman-based methods, and operator learning, among others. We then showcase these methods in applications ranging from learning lattice dynamical models and reduced order modeling of effective dynamics to discovery of conservation laws and potential identification of integrability of ordinary differential equations (DEs) and partial DE models. Our intention is to make clear that these ML methods are complementary to the preexisting powerful tools of the nonlinear waves community, and should be integrated into this toolkit to augment and enable mathematical discoveries and computational capabilities in the age of data.
\end{abstract}



\begin{footnotesize}
$^{*}$Corresponding author.
\end{footnotesize}

\section{Introduction}

Nonlinear wave systems lie at the heart of many of the most profound and important problems in physics. From shallow water waves and tsunamis, to optical solitons in fibres, to matter waves in Bose--Einstein condensates (BECs), these systems exhibit rich and intricate dynamics including solitons, dispersive shocks, wave turbulence, and modulational instabilities~\cite{ablowitz2011nonlinear, el2016dispersive, sulem1999nonlinear,CarreteroGonzalez2024}. These phenomena typically emerge from the delicate interplay between nonlinearity and dispersion (and often, in practice, from the balance of non-conservative contributions therein such as damping and forcing). For decades, the nonlinear waves community has led the way towards developing powerful analytical tools and numerical methods to understand these features, often leveraging deep insights provided by integrability, conservation laws, and spectral decompositions~\cite{whitham1999linear, ablowitz2011nonlinear}.

And yet, we are now in the age of data; this is an era in which massive simulations, high-fidelity experiments, and large-scale sensor networks are producing data at unprecedented scales~\cite{brunton2019data}. The traditional tools of nonlinear wave theory, powerful as they may be, were not designed with this data-rich regime in mind. At the same time, the rise of machine learning (ML) offers a complementary paradigm of learning patterns, dynamics, and even governing equations directly from data~\cite{brunton2016discovering,rudy2017data, raissi2019physics, karniadakis2021physics}. We emphasize that ML is not a replacement for the theoretical insights of nonlinear wave physics. Indeed, an aim of this review is to advocate that it can enable us to extend what we know, accelerate what we can simulate, and reveal what we have yet to understand. In that sense, it constitutes an invitation to the Nonlinear Waves community to (further) explore and integrate this new toolkit, and to the scientific ML (SciML) community to (further) adapt and purpose these methods towards the remarkable challenges of the field of Nonlinear Waves.

Recent years have seen an explosion of interest in leveraging ML methods to address fundamental challenges in wave physics~\cite{malkiel2017deeplearningdesignretrieval}. Supervised learning approaches including recurrent neural networks (RNNs), long short-term memory (LSTM) networks, and attention-based architectures have demonstrated success in forecasting complex wave dynamics; see, e.g.,~\cite{li2023transformer}. For example, data-driven models trained on time series have been used to predict rogue waves, phase modulations, or dispersive spreading with accuracy that rivals traditional solvers, particularly on short time horizons. Moreover, unsupervised learning techniques have shown value in uncovering low-dimensional structures within high-dimensional wave fields~\cite{taira2017modal}. Principal component analysis, dynamic mode decomposition (DMD), and nonlinear autoencoders can identify dominant modes, suitable reduced order models, and enable the study of coherent structures and attractors in wave turbulence and modulationally unstable media~\cite{taira2017modal}. Similarly in spirit to traditional phase space methods that defined an entire era of nonlinear science, these data-driven methods offer an intuitive geometric picture in which to view dynamics that may defy analytical simplification~\cite{strogatz2018nonlinear}.

Meanwhile, physics-informed approaches such as physics-informed neural networks (PINNs) offer a compelling hybrid framework~\cite{raissi2019physics,wang2021eigenvector}. These approaches are relevant not only to forward problems---especially in high-dimensional settings---but are perhaps even more impactful in solving inverse problems. In the latter, the data inform the identification of parameters in the physically inspired model terms. By encoding partial differential equation (PDEs) directly into the architecture or loss function of a neural network, they bridge the gap between data and physics. This allows for the solution of forward and inverse problems, even in regimes with limited or corrupted data~\cite{wang2021eigenvector}, on equations such as the nonlinear Schr\"odinger (NLS), Korteweg--de Vries (KdV), and Sine--Gordon (SG) models~\cite{raissi2019physics,raissi2017physicsinformeddeeplearningi,raissi2017physicsinformeddeeplearningii}. Moreover, the true power of PINNs is evident in their use as surrogate models in inverse design problems where forward solvers are extremely costly to compute when evaluating a quantity of interest~\cite{goswami2022physics}.

When boundary or initial input data are not fixed during training, operator learning techniques such as deep operator networks (DeepONets) and Fourier neural operators (FNOs) replace the use of PINNs~\cite{lu2021learning}. PINNs are trained to learn functions that simultaneously satisfy the PDE operator as well as specified boundary/initial data. On the other hand, operator learning is the numerical construction of a mapping between a class of boundary/initial data and solutions in infinite-dimensional spaces~\cite{Kovachki2021Survey}. Neural operators have recently demonstrated impressive capabilities in simulating nonlinear wave phenomena. For example, MITONet, an autoregressive neural operator, emulates two-dimensional shallow-water dynamics under varying boundary and bathymetry conditions, allowing for rapid and accurate forecasting of tide-driven coastal flows~\cite{rivera2025mitonet}. Other studies apply neural operators for long-time integration of canonical wave equations, for example, Lei et al use a recurrent integration scheme with neural operators to accurately simulate the KdV, SG and Klein--Gordon equations over extended time horizons~\cite{lei2024longtime}. The latter task naturally poses significant challenges in connection to the preservation of the different conserved quantities and the fundamental (e.g. symplectic) structural features of such integrable, as well as near-integrable systems.

In experimental settings, neural operators have also been used to recover and interpret the emergent behaviour of KdV--Burgers in nonlinear droplet systems, effectively extracting underlying Green's functions directly from the data~\cite{fruchart2024kdv_burgers}. Neural operators have also found novel applications in soliton identification. In particular, Zhang, et al develop a method that learns a mapping between initial and final states of quasi-one-dimensional BEC systems, recovering ground states and solitonic dynamics from data while respecting the Gross--Pitaevskii equation~\cite{zhang2023cno_bec}.

Other particularly exciting directions include the automated discovery of governing equations from data~\cite{brunton2016discovering, Champion2019}. Techniques such as sparse regression and symbolic neural models are being used to learn the form of evolution equations from observed dynamics~\cite{brunton2016discovering, cranmer2020lagrangian}. These tools offer an attractive possibility, namely that we may reverse the traditional modeling pipeline, where we infer laws from observations rather than the other way around~\cite{Champion2019}. Importantly, as will be discussed in detail below, such techniques and tailored variants thereof have also been used in order to discover conservation laws~\cite{teg1,liu2024interpretable,zhu2023machine} and eventually the potential full integrability~\cite{krippendorf2021integrability,de2024data,adriazola2025computer} of the models at hand.

This presents, in our view, not a ``passing trend'', but rather a real opportunity to address fundamental open problems and to pave new avenues of exploration. ML methods can act as surrogate models for rapid exploration, as interpretable embeddings for discovering new structures, and as inverse tools for designing experiments or reconstructing models or parameters thereof. 
They may enable insights into the solution of difficult problems such as the discovery of conservation laws, of Lax pairs~\cite{adriazola2025computer}, of suitable low-dimensional reductions~\cite{brunton2019data,bramburger2024data} or, e.g. of modulation equations when these are unknown or seemingly too complex to extract (e.g. in higher dimensions)~\cite{el2016dispersive}. Of course, challenges remain. Generalization to unseen data, interpretability, extrapolation, and physical consistency, among them, are precisely the kinds of foundational problems that have emerged~\cite{baker2019workshop} and which will continue to challenge the nonlinear wave and ML communities in years to come.

We believe that the future of nonlinear wave science will rest not only on analytical rigor and numerical precision, but also on insights derived from data-driven methods, hybrid models, and intelligent inference. We anticipate that the next generation of nonlinear wave scientists may benefit from integrating a deeper understanding of this diverse toolbox and its potential. Toward this vision, we aspire to summarize (in, admittedly, a way bearing a strong imprint of our personal taste) some of the recent advances in ML methods and the substantial promise they hold for applications in nonlinear wave science.

We begin, in section~\ref{sect:Pillars}, by reviewing the recent pillars of SciML. Section~\ref{sec3} discusses some of the recent advances based on PINNs, while section~\ref{sec4} focuses on reduced-order modeling, spearheaded by sparse identification of nonlinear dynamic (SINDy)~\cite{brunton2016discovering} and related approaches. Section 5 explains autoencoder-based model reduction. Section~\ref{sec5} addresses the learning of structural properties of models from data (e.g. conservation laws, Hamiltonian structure). Section 7 discusses applications of Koopman theory, and section~\ref{sec6} is dedicated to the discovery of integrability and its associated features. Finally, in section~\ref{sec7}, we summarize our findings and offer an outlook on potential future studies.

\section{Recent pillars of SciML}\label{sect:Pillars}
\subsection{Deep learning}
\subsubsection{Neural network architectures}

Modern deep learning is grounded in a diverse family of neural network architectures, each designed to capture particular structures in data. We aim to briefly summarize some of these structures, and trust that the interested reader can learn about them in detail from the listed citations. The most basic network architecture is perhaps the \emph{fully connected feedforward network} (also called multilayer perceptron, MLP), which consists of layers of linear transformations followed by nonlinear activation functions. These models are universal approximators~\cite{cybenko1989approximation,barron_approx,yarotsky2018optimal} and remain widely used for tabular data and as baseline models in SciML \cite{goodfellow2016deep,Bishop2006}. Their main strengths are flexibility and simplicity, but they often require large amounts of data and struggle with high-dimensional structured inputs.  

\emph{Convolutional neural networks} (CNNs) introduce weight sharing and local connectivity to exploit spatial structure. Originally developed for image recognition \cite{LeCun1998}, CNNs are now widely applied to problems with translation invariance and local correlation structure, including fluid dynamics and PDE surrogate modeling \cite{Guo2016,Thuerey2018}. Their strengths lie in parameter efficiency and strong inductive bias for spatial data, though they may fail on data without clear grid-like structure~\cite{Bronstein2017,Cohen2016,Zhou2020}.

RNNs are designed for sequential data, processing inputs recursively while maintaining a hidden state. They have been widely applied in time-series analysis and natural language processing \cite{Rumelhart1986}. However, standard RNNs often suffer from vanishing or exploding gradients during training \cite{Pascanu2013}. To overcome this, architectures such as the LSTM \cite{Hochreiter1997} and gated recurrent unit (GRU) \cite{Cho2014} were introduced, enabling networks to capture long-range dependencies more effectively. While LSTMs and GRUs remain influential, their sequential nature limits parallelization, and future work in this direction is warranted, should they find further application in large scale nonlinear wave modeling \cite{Shi2015}.

In recent years, \emph{transformer architectures} \cite{Vaswani2017} have displaced recurrent networks in many domains. By relying on self-attention mechanisms \cite{Bahdanau2015} rather than recurrence, transformers scale more efficiently and excel at modeling long-range dependencies. Their strengths include parallelizable training and superior performance across language, vision, and multimodal learning, though their large parameter counts and training costs are potential drawbacks.

Beyond these canonical classes, specialized architectures continue to expand the landscape. Graph neural networks (GNNs) generalize convolutions to non-Euclidean data \cite{Bronstein2017}, making them powerful tools for modeling physical systems defined on meshes or networks. Autoencoders and variational autoencoders (VAEs) \cite{Kingma2014} provide unsupervised representation learning, while generative adversarial networks (GANs) \cite{Goodfellow2014} offer powerful generative models. Each architecture introduces inductive biases and trade-offs, and the choice depends strongly on the problem domain.  

We emphasize the strengths and weaknesses of the major families of neural network architectures. MLPs serve as general-purpose approximators, CNNs excel at spatially structured data, RNNs (and their variants LSTMs and GRUs) are well-suited for sequential data, transformers capture long-range dependencies, and GNNs effectively model relational structures. Together, these architectures form the foundation of modern SciML applications, some of which we highlight in this review. Naturally, it is impossible to cover every domain where these architectures arise, and we refer the reader to the broader literature for further exploration.

\subsubsection{Learning functions through PINNs}
\label{sec:dl_of_functions}
In recent years, a novel class of NN-based techniques, the so-called PINNs, has emerged, designed to solve (forward problem) and discover (inverse problem) differential equations (DE). They were introduced in \cite{raissi2019physics}, although they had been announced earlier in two arXiv preprints~\cite{raissi2017physicsinformeddeeplearningi,raissi2017physicsinformeddeeplearningii}. The key idea is to embed physical laws, expressed through PDEs, directly into the loss function used to train the networks. This approach enables the learning of complex physical phenomena from sparse or noisy data, providing a compelling alternative to traditional numerical solvers and facilitating the systematic discovery of physical models from data.

Traditional methods for solving PDEs, such as finite difference, finite volume, and finite element methods, can be computationally expensive, especially in high dimensions, and require extensive domain-specific knowledge. 
They also heavily rely on mathematically 
well-posed formulations.
Conversely, PINNs offer mesh-free solutions that are generalizable and (potentially) scalable, opening new avenues in engineering, physics, and biomedical sciences.
Importantly, also, we note in passing that 
they heavily rely on 
optimization methods that may provide answers
(potentially pertaining to local extrema etc.)
in the case of incomplete/non-well-posed data.
It does not elude us that this feature formulates some novel deep mathematical analysis
questions in its own right.

More concretely, in PINNs, given a PDE of the form:
$$
\mathcal{N}[u(x,t)] = 0, \quad x \in \Omega, \ t \in [0,T],
$$
where $\mathcal{N}$ is a (usually nonlinear) differential operator and $u(x,t)$ is the unknown solution, the neural network is trained to minimize a composite loss function consisting of the sum of two parts: a self-supervised one and a supervised one. 
$$\mathcal{L}=\mathcal{L}_{ss}+\mathcal{L}_{s}$$
For the self-supervised part of the loss function, which contains the physics originating information, we consider a set of $N_{ss}$ points $(x^{ss}_i, t^{ss}_i)$ in the $\Omega\times[0, T]$ domain and can be for example of the form 
$$
\mathcal{L}_{ss}=\frac{1}{N_{ss}}\sum_{i=1}^{N_{ss}} \mathcal{N}[u(x_i^{ss},t_i^{ss})]^2,
$$

while for the supervised part we consider $N_s$ points $(x^s_j, t^s_j)$ in the boundary of the $\Omega\times[0,T]$ domain containing the information taken by the known initial and boundary conditions as well as in a possible number of points where the solution could be known inside the $\Omega\times[0,T]$ domain. For instance, for Dirichlet boundary conditions or, in general, when the solution is known at the collocation points $(x^s_j, t^s_j)$,
this part may take the form
$$
\mathcal{L}_{s}=\frac{1}{N_s}\sum_{j=1}^{N_s} [u(x^s_j,t^s_j)-u_j]^2.
$$

Differentiations, when required, are typically carried out by automatic differentiation (AD) \cite{AD} which is a key enabler of PINNs. By computing derivatives analytically through the computational graph of the neural network, PINNs avoid potential numerical issues associated with finite difference approximations. Libraries such as TensorFlow \cite{tensorflow} and PyTorch \cite{pytorch} make AD straightforward and efficient.

For the implementation of the PINN, a fully connected feedforward neural network is typically used. The input layer receives spatial and temporal coordinates $(x,t)$, while the output layer predicts the physical quantity of interest $u(x,t)$. The network parameters are updated via backpropagation using stochastic gradient descent or variants like Adam or L-BFGS (see, e.g.,~\cite{goodfellow2016deep}). DeepXDE \cite{lu2021deepxde} is a powerful python package for PINNs which includes and automates many of the previously mentioned methods.  

One of the most powerful features of PINNs is their ability to solve inverse problems, such as identifying unknown parameters of a PDE under consideration or initial conditions from sparse observations \cite{raissi2019physics, lu2021deepxde}. In these works, we assume a prior knowledge or guess of the functional form of the PDE. On the other hand, in \cite{Raissi2018} a configuration of two NNs is used. The first is used for the ADs needed which are realized through TensorFlow while the second provides the approximation of the discovered PDE. This treatment has the advantage that it does not need any prior assumption on the form of the PDE but has the disadvantage that the NN approximation is a ``black box'' one and does not provide any closed form equation. 

The applicability of PINNs is extremely wide since they can be used in every system described by PDEs. 
A traditional class of examples involves
\textbf{Fluid Dynamics} where they have been used to solve Navier-Stokes equations in fluid flow problems \cite{raissi2019physics, raissi2020hidden}. For instance, the reconstruction of velocity and pressure fields from sparse measurements around an obstacle showcases their potential in real-time simulation and flow control. Another field of application is this of \textbf{Solid Mechanics}
and in particular in elasticity and plasticity modeling \cite{zhu2019physics}. For example, PINNs can model stress-strain relationships under dynamic loading without relying on dense meshes or predefined constitutive laws. 
In \textbf{heat conduction} problems, PINNs can predict temperature distributions over time with limited boundary condition data, offering faster alternatives to traditional solvers \cite{meng2020ppinn}, while they are also used in \textbf{Biomedical Applications} including 
blood flow modeling, cardiac electrophysiology, and tumour growth simulations \cite{kissas2020machine}. Their ability to integrate prior knowledge with sparse patient data makes them suitable for personalized medicine.

PINNs offer several benefits over conventional numerical solvers: First of all, there is no need for mesh-discretization of the integration domain, which simplifies implementation and improves flexibility. 
This can become especially valuable in suitably
high dimensional examples~\cite{raissi2019physics,shahab2025physicsinformedneuralnetworkshighdimensional}.
The method can also easily generalize from sparse or noisy datasets by leveraging physical laws. The ability of PINNs to estimate model parameters and unknown sources directly is also invaluable. Finally, they can easily be applied to a variety of linear and nonlinear PDEs across different scientific fields.

Despite the important advantages of PINNs there are also some limitations that have to be taken under consideration.
Training a PINN can be slow and occasionally unstable. The loss landscapes are often highly non-convex, 
thus rendering questionable the potential result
of convergence and occasionally needing numerous 
realizations to ensure a meaningful result.
Moreover, gradients from physics-based losses can vanish or explode, particularly in complex domains. PINNs often struggle with stiff PDEs and multi-scale problems due to the difficulty in capturing disparate scales with a single network. Adaptive sampling and domain decomposition techniques have been proposed to mitigate this \cite{jagtap2020adaptive}.
 The computational cost  can also be a serious caveat. Although PINNs avoid meshing, their computational cost can be high due to the need for evaluating derivatives via AD and performing global optimization over large parameter spaces. Finally, when the physics-based constraints are not sufficiently strong or the data is too sparse, PINNs may overfit or fail to generalize, especially in high-dimensional problems.

To address some of the above challenges, several variants of PINNs have been proposed, including but not restricted to the following: Adaptive PINNs which modify the sampling of collocation points to focus on areas with higher residuals~\cite{mcclenny2020selfadaptive}, XPINNs in which the domain is partitioned and the corresponding sub-networks are trained independently to improve scalability \cite{jagtap2020extended}. In the case of Fourier Features PINNs, Fourier embeddings are used to capture fine-scale features and improve convergence \cite{sallam2023ffpinn}, while Bayesian PINNs incorporate uncertainty quantification using Bayesian methods or dropout-based inference \cite{yang2021bpinns}.

\subsubsection{Deep operator learning}\label{section:OLearn}

In traditional supervised learning, neural networks approximate functions, mapping inputs to outputs drawn from finite-dimensional spaces (e.g., $\mathbb{R}^n \to \mathbb{R}^m$). In contrast, operator learning seeks to approximate mappings between infinite-dimensional function spaces, such as learning the solution operator that maps an initial condition or forcing term of a PDE to its solution. This distinction is crucial. 

While function learning captures pointwise relationships, operator learning captures entire transformations of functions. Unlike traditional neural networks, neural operators are, in principle, resolution-invariant, that is, once trained, they can accept new functions $a$ evaluated at the same sensor points, regardless of the original function's discretization. This makes operator learning especially suitable for SciML tasks where the underlying objects of interest are PDE solution operators rather than isolated function values. 

In this section, we briefly discuss two very successful approaches to operator learning. The first approach we discuss involves DeepONets, introduced by Lu, Jin, and Karniadakis \cite{lu2021learning}, which are neural architectures designed to learn nonlinear operators between Banach spaces. The key innovation of DeepONets is the separation of the input function and the evaluation location through two coordinated neural networks; one is called the \emph{branch network} and the other the \emph{trunk network}. 

To begin, let us state a more precise mathematical setting for the purposes of operator learning. Let $\mathcal{G}: \mathcal{A} \to \mathcal{B}$ be a nonlinear operator mapping between Banach spaces, where $\mathcal{A}$ is a function space that can be sampled from (e.g., $C(\Omega)$) and $\mathcal{B}$ is typically a space of scalar-valued functions on a domain $Y$. The goal is to learn the mapping $a \mapsto \mathcal{G}(a)(y)$ for any $a \in \mathcal{A}$ and any $y \in Y$. To construct a data-driven approximation of $\mathcal{G}$, we assume access to a finite collection of pairs $\{a_j, \mathcal{G}(a_j)(y_i)\}$, where $a_j$ are input functions sampled from $\mathcal{A}$ and $\{y_i\}$ are locations at which the output is evaluated.

The DeepONet is defined as a neural network that approximates the operator $\mathcal{G}$ via the ansatz
\begin{equation}
    \mathcal{G}_\theta(a)(y) := \sum_{k=1}^p b_k(a(x_1), \ldots, a(x_m)) \, t_k(y),
\end{equation}
where $a(x_1), \ldots, a(x_m)$ are the evaluations of the input function $a$ at fixed sensor locations $\{x_i\}_{i=1}^m \subset D$, $b_k: \mathbb{R}^m \to \mathbb{R}$ is the output of the \emph{branch network}, $t_k: Y \to \mathbb{R}$ is the output of the \emph{trunk network}, evaluated at $y$, and $p$ is the number of basis functions in the network, i.e., the dimension of the internal representation. The branch and trunk networks are typically standard feedforward neural networks, parameterized jointly by $\theta$. The architecture enforces a bilinear structure in the final layer.

A foundational result in \cite{lu2021learning} shows that DeepONets are universal approximators of continuous nonlinear operators:
\begin{theorem}[Universal Approximation of Operators]
Let $K \subset C(D)$ be a compact set, and let $\mathcal{G}: K \to \mathbb{R}$ be a continuous operator. Then, there exists a DeepONet $\mathcal{G}_\theta$ of the form above such that
$$
\sup_{a \in K} \left| \mathcal{G}_\theta(a)(y) - \mathcal{G}(a)(y) \right| < \varepsilon
$$
for any $\varepsilon > 0$ and fixed $y \in Y$.
\end{theorem}
For a more precisely stated version of this result, please see Theorem 5 in~\cite{lu2021learning}. Of course, this result relies on the classical universal approximation theorem for neural networks and relies on many of the assumptions therein. Nevertheless, the intention here is to communicate that the DeepONet ansatz is theoretically capable of representing arbitrary continuous operators.

Concerning training, the model is trained on a dataset $\{a^{(j)}, y^{(j)}, \mathcal{G}(a^{(j)})(y^{(j)})\}_{j=1}^N$ to minimize a loss function, typically the empirical $L^2$ loss:
\begin{equation}\label{eq:DeepONetLoss}
    \mathcal{L}(\theta) = \frac{1}{N} \sum_{j=1}^N \left| \mathcal{G}_\theta(a^{(j)})(y^{(j)}) - \mathcal{G}(a^{(j)})(y^{(j)}) \right|^2.
\end{equation}
Optimization is performed using standard techniques such as stochastic gradient descent. Key hyperparameters and choices in the DeepONet design include the number of sensor points $m$ used to discretize the input function, the number of basis functions $p$, determining the dimensionality of the internal representation, and the network architecture including the depth of branch/trunk subnetworks.

DeepONets have been applied to a wide variety of problems, including parametric PDE solvers \cite{lu2021learning}, surrogate modeling \cite{LuWang2024DeepONetRTE}, inverse problems \cite{Bhattacharya2021ModelReduction}, and control \cite{lin2023dynamicalresponse}. Their theoretical grounding and empirical success have positioned them as a foundational architecture for data-driven operator learning \cite{Kovachki2021Survey}.

FNOs, introduced by Li et al.~\cite{Li2021FNO}, are a class of neural architectures designed to learn mappings between function spaces by lifting functions into Fourier space, performing global convolutions, and mapping back to physical space. Just like DeepONets, FNOs are especially suited for learning solution operators of parametric PDEs, offering resolution-invariant and efficient surrogates for complex physical models.

In contrast to DeepONets, which approximate operators through a branch--trunk decomposition, the FNO realizes $\mathcal{G}$ as the composition of spectral integral operators. At layer $l$, the network represents the state as a vector-valued function 
$v^{(l)} : D \to \mathbb{R}^d,$
where for each spatial location $x \in D$, the vector $v^{(l)}(x)$ encodes $d$ latent features of the input. This representation can be interpreted as a graph signal on the discretization of $D$, with nodes given by spatial locations and features attached to each node. An FNO layer then performs message passing in the Fourier domain, updating the features according to
\begin{equation}\label{eq:FNOHyp}
    v^{(l+1)}(x) \;=\; \sigma\!\left( W v^{(l)}(x) \;+\; \mathcal{F}^{-1}\!\Big( R \cdot \mathcal{F}(v^{(l)}) \Big)(x) \right),
\end{equation}
where $W$ is a pointwise linear transformation shared across $x$, $\mathcal{F}$ and $\mathcal{F}^{-1}$ denote the Fourier and inverse Fourier transforms applied channel-wise, $R$ is a learnable Fourier-domain filter (typically diagonal in frequency), and $\sigma$ is a nonlinear activation. Stacking $L$ such layers defines the operator approximation $\mathcal{G}_\theta$. In this sense, the FNO may be viewed as a GNN with globally coupled message passing implemented efficiently through spectral convolution.

For concreteness, in one spatial dimension with input $v \in \mathbb{R}^{n_x \times d}$, a Fourier layer can be described in the following steps:
\begin{enumerate}
    \item Compute the Fourier transform $\widehat{v} = \mathcal{F}(v)$.
    \item Retain the lowest $K$ Fourier modes (spectral truncation).
    \item Apply learnable weights $\widehat{R}_k$ mode-wise: $\widehat{v}_k \mapsto \widehat{R}_k \widehat{v}_k$ for $|k| \le K$.
    \item Set higher modes to zero: $\widehat{v}_k = 0$ for $|k| > K$.
    \item Transform back to physical space with the inverse FFT.
\end{enumerate}
This spectral filtering amounts to a parameter-efficient, resolution-agnostic convolution with a global receptive field. 

To summarize, in this simplified setting, consider a one-dimensional input $a \in \mathbb{R}^{n_x \times d_a}$ defined on $n_x$ spatial grid points with $d_a$ input channels. The FNO architecture consists of three stages:  
\begin{enumerate}
    \item \textbf{Lifting layer:} apply a pointwise linear map $P: \mathbb{R}^{d_a} \to \mathbb{R}^d$ at each grid point. Concretely, for every $x_i$, the input vector $a(x_i) \in \mathbb{R}^{d_a}$ is mapped to a latent feature vector $v^{(0)}(x_i) = P\,a(x_i) \in \mathbb{R}^d$. This produces the initial feature representation $v^{(0)} \in \mathbb{R}^{n_x \times d}$.
    \item \textbf{Fourier layers:} propagate the latent features through $L$ stacked spectral layers of the form in Equation~\eqref{eq:FNOHyp}, which combine pointwise linear updates with spectral convolutions.
    \item \textbf{Projection layer:} apply a final pointwise linear map $Q: \mathbb{R}^d \to \mathbb{R}^{d_u}$ at each grid point to map the latent representation back to the physical output. For each $x_i$, $u(x_i) = Q\,v^{(L)}(x_i) \in \mathbb{R}^{d_u}$, yielding the network prediction $u \in \mathbb{R}^{n_x \times d_u}$.
\end{enumerate}
Just as with DeepONet, learning a FNO, that is, learning the lifting, Fourier, and projection layers, can be realized by solving  an empirical risk minimization problem that uses Equation~\eqref{eq:DeepONetLoss} as the loss function.

A notable property of FNOs is \emph{resolution invariance}: since the Fourier transform is naturally defined in function space, a model trained on one discretization can be evaluated on finer or coarser meshes without retraining. Computationally, FNOs scale as $O(n_x \log n_x)$ due to the FFT, with only $K \ll n_x$ Fourier modes retained. Compared to CNNs, FNOs achieve global receptive fields at every layer and demonstrate faster convergence on many PDE-based tasks.  

FNOs have been applied across a broad range of SciML tasks. They were first introduced as resolution-invariant solvers for parametric elliptic and time-dependent PDEs such as Burgers, Darcy flow, and the Navier--Stokes equations~\cite{Li2021FNO}, and have since been extended to modeling spatiotemporal turbulence~\cite{Atif2024FNOturbulence} and large-scale climate and weather dynamics, including urban microclimate simulation and ocean circulation prediction~\cite{Peng2023UrbanFNO,Choi2024OceanFNO}. Beyond forward modeling, FNOs have also been employed for faster-than-real-time PDE inference and control in fluid dynamics~\cite{Renn2023FNOControl}, as well as in PDE-constrained optimization tasks such as optimal boundary control for nonlinear optics~\cite{Margenberg2023FNOControlOptics}. Finally, stochastic adaptations of FNOs have been developed for surrogate modeling and uncertainty quantification in geophysical and stochastic systems~\cite{Choi2024OceanFNO}. Given their versatility in PDE solving, dynamical modeling, control, and stochastic inference, FNOs, and operator learning more broadly, are poised to become central tools in advancing the SciML of nonlinear waves.

\subsection{Interpretable learning with sparse and symbolic regression}
Modeling complex dynamical systems is a fundamental challenge across scientific and engineering disciplines. Traditional approaches typically rely on detailed first-principles models or large amounts of carefully curated empirical data. In many settings, however, neither is fully available; for example, the underlying mechanisms governing the observed dynamics may be only partially understood.

SINDy is a data-driven framework designed to infer governing equations directly from time-series measurements by leveraging the observation that many physical systems admit sparse representations in an appropriate function library.

While ML and black-box models such as neural networks have gained popularity in system identification, they often lack interpretability. The \textit{
SINDy
} framework, introduced 
in the seminal work of~\cite{brunton2016discovering}, offers an appealing alternative. It leverages sparse regression techniques to identify the fewest terms necessary to describe the system's dynamics, leading to interpretable and often physically meaningful models. SINDy is rooted in the assumption that most dynamical systems can be described by a small number of active terms out of a large possible/plausible function library. By combining time-series data and sparse regression, the method identifies these active terms, producing parsimonious models with 
potentially valuable physical interpretations.

The SINDy framework works by first considering a continuous-time dynamical system:
\begin{equation}
\frac{d\mathbf{x}}{dt} = \mathbf{f}(\mathbf{x}),
\label{ds}
\end{equation}
where $\mathbf{x}(t) \in \mathbb{R}^n$ is the state vector and $\mathbf{f}$ represents the unknown nonlinear dynamics. The key assumption is that $\mathbf{f}$ can be approximated as a sparse linear combination of functions from a predefined library, $\Theta(\mathbf{x}) = [\theta_1(\bfx), \ldots, \theta_p(\bfx)]$, where the basis functions $\theta_i(\bfx)$ may consist, e.g., of polynomials or trigonometric functions:
\begin{align}
\label{eq:f=thetaxxi}
\mathbf{f}(\mathbf{x})^T \approx \Theta(\mathbf{x}^T) \Xi,
\end{align}
where 
$\Theta(\mathbf{x}^T) \in \mathbb{R}^{1 \times p}$ is the library evaluated on the data, and $\Xi \in \mathbb{R}^{p \times n}$ contains the sparse coefficients. In order to calculate $\Xi$, we first construct the matrix $\mathbf{X} \in \mathbb{R}^{m \times n}$, by considering a set of $m$ snapshots $\mathbf{x}_i=\mathbf{x}(t_i)$ of the state vector,

$$
\mathbf{X} = 
\left[\begin{array}{cccc}
| & | & | & | \\
\mathbf{x}_1 & \mathbf{x}_2 & \ldots & \mathbf{x}_m\\
| & | & | & | 
\end{array}\right]^T
$$  
and the corresponding matrix $\mathbf{\dot{X}}$
$$
\mathbf{\dot{X}}=
\left[\begin{array}{cccc}
| & | & | & | \\
\mathbf{\dot{x}}_1 & \mathbf{\dot{x}}_2 & \ldots & \mathbf{\dot{x}}_m\\
| & | & | & | 
\end{array}\right]^T \approx \displaystyle\frac{d\mathbf{X}}{dt},
$$ 
which contains the corresponding time-derivatives. These can be computed using finite differences or smoothed estimates. 

Then, the key element of the method is to construct an appropriate library of candidate nonlinear functions evaluated on the data. A typical such matrix $\Theta(\mathbf{X})\in \mathbb{R}^{m \times p}$ contains basic functions like polynomials or trigonometric functions:
$$
\Theta(\mathbf{X}) = \left[ \mathbf{1}, \mathbf{X}, \mathbf{X}^2, \sin(\mathbf{X}), \cos(\mathbf{X}) \ldots \right].
$$

Here, higher polynomials are denoted as $\mathbf{X}^2$, $\mathbf{X}^3$, where
$\mathbf{X}^2$ denotes the quadratic nonlinearities in the state x:
$$
\mathbf{X}^2 =\left[\begin{array}{cccccc}
x_1^2(t_1)&x_1(t_1)x_2(t_1)&\cdots&x_2^2(t_1)&\cdots&x_n^2(t_1)\\
x_1^2(t_2)&x_1(t_2)x_2(t_2)&\cdots&x_2^2(t_2)&\cdots&x_n^2(t_2)\\
\vdots & \vdots & \ddots & \vdots & \ddots & \vdots\\
x_1^2(t_m)&x_1(t_m)x_2(t_m)&\cdots&x_2^2(t_m)&\cdots&x_n^2(t_m)
\end{array}\right]\\
$$
Note though, that the choice of the vocabulary elements is free and depends on the imagination of the researcher, or/and their
understanding of the physical/chemical/biological processes involved. The only objective is the best and most interpretable description of the actual system within the reach of the selected library.

Since we seek a matrix $\mathbf{X}$ for which it is $\dot{\mathbf{X}} \simeq \Theta(\mathbf{X}) \Xi,$
one could assume that it would be sufficient to solve a minimization problem of the form
\begin{equation}
\mathbf{\Xi}=\underset{\mathbf{\Xi}'}{\rm argmin} \| \dot{\mathbf{X}}-\mathbf{\Theta}(\mathbf{X})\mathbf{\Xi}' \|_F^2.  \label{eq:min1}
\end{equation}

But the heart of the SINDy method lies in assuming that this coefficient matrix should also be sparse. This way, the resulting dynamics will be, if not more physically relevant, then surely more interpretable. In order to perform this sparse regression one can consider the minimization problem:
\begin{equation}\mathbf{\Xi}=\underset{\mathbf{\Xi}'}{\rm argmin}\left( \| \dot{\mathbf{X}}-\mathbf{\Theta}(\mathbf{X})\mathbf{\Xi}'\|_F^2  +\lambda\|\mathbf{\Xi}'\|_1\right)\label{eq:min2}
\end{equation}
instead of \eqref{eq:min1}. The addition of the 1-norm penalizes the small entries in the $\mathbf{\Xi}$ matrix, while $\lambda$ is a sparsity parameter which controls the strength of the penalization, namely the larger the value of $\lambda$ the more coefficients become zero in $\mathbf{\Xi}$.  This problem is solved through the LASSO method \cite{hastie2015statistical}. Although LASSO is a very well known and popular method, for sparse results in minimization problems it can be computationally inefficient for SINDy since it can often produce very small but non zero coefficients \cite{SR3}. To circumvent this issue the Sequentially Thresholded Least Squares (STLSQ) method was introduced and it has also been used in original SINDy paper \cite{brunton2016discovering}.

In STLSQ, the original least squares problem  \eqref{eq:min1} for $\mathbf{\Xi}$ is solved as $\boldsymbol{\Xi} = \boldsymbol{\Theta}^\dagger \dot{\mathbf{X}}$. Then, all coefficients below the value of the sparsity coefficient (which now plays the role of a threshold) are zeroed out. After that we refit the model using only the remaining terms.
This process iterates until convergence, producing a sparse and interpretable dynamical model. Thus, STLSQ is straightforward, efficient and produces sparse results, yet still with some shortcomings.

One of the alternatives which has been proposed in order to to address some practical issues of LASSO and STLSQ is the Sparse Relaxed Regularized Regression (or SR3, for short) algorithm \cite{SR3,SINDySR3}, which is also used in PySINDy \cite{desilva2020pysindy,kaptanoglu2022pysindy}, a Python package for SINDy including many of the features mentioned here. There, by introducing an alternative function and an extra term in \eqref{eq:min2}, the algorithm manages to (i) better handle outliers and corrupt data within noisy sensor measurements, (ii) to consider the parametric dependencies in candidate library functions, and (iii) impose physical constraints of the problem.

A straightforward variant of SINDy involves its use to discover \textbf{discrete-time dynamical systems} (see, e.g.,~\cite{SINDYc}) of the form
\begin{equation}\mathbf{x}_{n+1}=\mathbf{F}(\mathbf{x}_n).\label{eq:system_discrete}
\end{equation}
These kinds of systems can either originate from time series exploring a naturally discrete phenomenon or the periodic measurement of a continuous procedure.
The corresponding minimization problem is
$$\mathbf{\Xi}=\underset{\mathbf{\Xi}'}{\rm argmin}\left( \| \mathbf{X}_{n+1}-\mathbf{\Theta}(\mathbf{X_n})\mathbf{\Xi}'\|_F^2\right)$$
where $\mathbf{X}_{n}$ is the matrix of the specific now (discrete-time) snapshots of the system and $\mathbf{X}_{n+1}$ its corresponding images through \eqref{eq:system_discrete}. For the sparse identification, all of the minimization techniques mentioned in the continuous case can be used.

Another obvious variant is to also consider the discovery of \textbf{PDEs}. This was done also in the original paper \cite{brunton2016discovering} as well in \cite{rudy2017data, SINDyPDE2}. In this case the methodology is the same but the size of the matrix of the candidate dictionary increases significantly. For example, in \cite{SINDyPDE2} where a PDE of the second order (in space) with second degree nonlinearities of the form $u_t=F(u,u_x,u_{xx})$ is considered, the library matrix should be of the form
$$\mathbf{\Theta}=\left[\mathbf{1}, \mathbf{U}, \mathbf{U}^2, \mathbf{U}_x, \mathbf{U}_x^2, \mathbf{U}\mathbf{U}_x, \mathbf{U}_{xx}, \mathbf{U}^2_{xx}, \mathbf{U}\mathbf{U}_{xx}, \mathbf{U}_x\mathbf{U}_{xx}\right],$$
where $\mathbf{U}$ is the spatially discretized version of $u$. One can easily understand that if one would consider higher derivatives or nonlinearities the library expands rapidly, and the identification of the sparse dynamics becomes quickly significantly harder.
One can also appreciate the relevant 
additional complications
further in the case where the  field $u$
is no longer scalar.
 
There are also several important extensions of the SINDy method.
First of all \textbf{SINDYc} which extends SINDy to systems with control inputs of the form
$$
\frac{d\mathbf{x}}{dt} = \mathbf{f}(\mathbf{x}, \mathbf{u}),
$$
where $\mathbf{u}(t)$ is the control input. The library is expanded to include terms involving both $\mathbf{x}$ as well as $\mathbf{u}$.

On the other hand, \textbf{Implicit SINDy} \cite{ImplicitSINDy} has been introduced to handle systems with algebraic constraints or implicit dynamics:
$$
\mathbf{g}(\mathbf{x}, \dot{\mathbf{x}}) = 0.
$$
The corresponding minimization problem is 
\begin{equation}
\mathbf{\Theta}(\mathbf{X}, \dot{\mathbf{X}})\mathbf{\Xi}=\mathbf{0}\label{eq:implicit}
\end{equation}
where the library matrix $\mathbf{\Theta}$ is generalized to include functions of $x$ and $\dot{x}$. However, this approach requires solving for a sparse matrix $\mathbf{\Xi}$ in the null space of $\mathbf{\Theta}(\mathbf{X}, \dot{\mathbf{X}})$ which leads to  highly ill-conditioned computations for noisy data. This causes the whole procedure to 
commonly relax  to the trivial solution. In order to stabilize the relevant procedure, the \textbf{SINDy-PI} method \cite{SINDY-PI} has been proposed which also includes a parallel realization. The method relies on the idea that if even
a single term of the dynamics is known which corresponds to a column $\theta_j(\mathbf{x}, \dot{\mathbf{x}})\in \mathbf{\Theta}(\mathbf{x}, \dot{\mathbf{x}})$, it is possible to rewrite \eqref{eq:implicit}
as
$$\theta_j(\mathbf{X}, \dot{\mathbf{X}})=\mathbf{\Theta}(\mathbf{X}, \dot{\mathbf{X}}|\theta_j(\mathbf{X}, \dot{\mathbf{X}}))\mathbf{\xi}_j,$$
where $\mathbf{\Theta}(\mathbf{X}, \dot{\mathbf{X}}|\theta_j(\mathbf{X}, \dot{\mathbf{X}}))$ is the library $\mathbf{\Theta}(\mathbf{X}, \dot{\mathbf{X}})\mathbf{\Xi}$ with the column $\theta_j$ removed. Now, the problem is not  implicit 
anymore and can be solved as a usual sparse minimization problem avoiding the relaxation to zero.
\textbf{Weak Form SINDy} \cite{WeakSINDy1,WeakSINDy2} has been proposed to deal with the problem of noise sensitivity of the method. Since numerical differentiation is very sensitive to noisy measurements, in these works the corresponding derivatives are expressed weakly and are more robust.

The method has already been applied to several disciplines including but not restricted to {Fluid Mechanics}, in modeling wake dynamics and vortex shedding \cite{brunton2016discovering}, {Epidemiology}, in deriving models for infectious disease spread \cite{Champion2019}, {Control Systems}, in learning dynamics for model predictive control (MPC) \cite{Kaiser2018}, {Biological Networks}, in reconstructing gene and neural systems from expression data \cite{SINDyGene}, and {Chemical Reactions} systems like the Belousov–Zhabotinsky reaction \cite{SINDY-PI}.

Despite its strengths, SINDy faces notable challenges.
\textit{Noise Sensitivity}: Derivative estimation is sensitive to noise. Approaches such as total variation regularization, Gaussian processes, and weak formulations help mitigate this issue. \textit{Library Selection}: Selecting an expressive yet efficient function library remains critical and often requires domain expertise. Symbolic regression and neural networks are emerging to automate this. \textit{Scalability and High Dimensions}: For high-dimensional systems such as PDEs, SINDy requires dimensionality reduction techniques like Proper Orthogonal Decomposition (POD) or manifold learning.

There are numerous promising future research directions. \textit{Neural-SINDy Hybrids}: this would involve integrating SINDy with neural networks for structure discovery while retaining interpretability. \textit{Bayesian SINDy}: this emerging
approach accounts for uncertainty in model structure and parameters. Finally, another challenge is that of real-time systems, where it can contribute to improving performance for real-time system identification and control.

In conclusion, SINDy provides a powerful framework for discovering interpretable models of nonlinear systems from data. By leveraging sparsity, it uncovers governing equations that offer both predictive accuracy and potentially physical insight. With continued development, particularly in noise robustness, scalability, and integration with deep learning, SINDy continues to bear significant potential as a pillar of data-driven scientific discovery.

{ \subsection{Koopman operator-based methods}\label{sec:koopman_nlw}

Nonlinear-wave PDE models are infinite-dimensional  and often exhibit a sharp separation between the dimension of the \emph{state} and the effective dimension of the \emph{organizing structures}: coherent travelling waves, dispersive radiation, slow modulations, invariant tori in near-integrable regimes, metastable manifolds in weakly dissipative regimes, and symmetry-generated families of solutions.  Koopman operator theory reframes reduction and analysis by shifting attention from the nonlinear evolution of states to the linear evolution of \emph{observables} under the flow \cite{koopman1931hamiltonian,mezic2005spectral,budisic2012applied,mezic2013analysis}.  For nonlinear waves, this viewpoint is particularly useful because (i) physically meaningful quantities are often naturally expressed as functionals such as energy, wave action, scattering data, phase/amplitude variables, (ii) translation/phase symmetries can be handled explicitly at the observable level, and (iii) integrable or conjugate-to-linear wave PDEs provide rare analytic ``ground truth'' Koopman spectra that expose algorithmic pitfalls like degeneracy, region-dependent expansions, and observable dependence~\cite{page2018burgers,parker2020isolated,parker2023kdv,nakao2020spectral}. 

\subsubsection{Koopman operators, generators, and eigenfunctionals for wave PDE flows}

Let $u(\cdot,t)$ evolve on a (typically infinite-dimensional) state space $\mathcal{X}$ according to an autonomous nonlinear wave PDE
\begin{equation}
  \partial_t u \;=\; \mathcal{N}(u), 
  \qquad u(\cdot,0)=u_0\in\mathcal{X},
  \label{eq:koop_nlw_pde}
\end{equation}
with flow map (solution operator) $\Phi^t:\mathcal{X}\to\mathcal{X}$ defined by $u(\cdot,t)=\Phi^t(u_0)$.  The Koopman operator semigroup $\{\mathcal{K}^t\}_{t\ge 0}$ acts on complex-valued observables (functionals) $g:\mathcal{X}\to\mathbb{C}$ via composition with the flow:
\begin{equation}
  (\mathcal{K}^t g)(u) \;:=\; g(\Phi^t(u)).
  \label{eq:koop_nlw_def}
\end{equation}
Linearity is immediate:
$\mathcal{K}^t(\alpha g_1+\beta g_2)=\alpha\,\mathcal{K}^t g_1+\beta\,\mathcal{K}^t g_2$,
even though \eqref{eq:koop_nlw_pde} is nonlinear.
Depending on the dynamical setting, one typically considers $\mathcal{K}^t$ as an operator on a function space such as $C_b(\mathcal{X})$ (bounded continuous functionals) or $L^2(\mathcal{X},\mu)$ for an invariant/physical measure $\mu$ \cite{mezic2005spectral,budisic2012applied,mezic2013analysis}.  In measure-preserving, Hamiltonian regimes, $\mathcal{K}^t$ is unitary on $L^2(\mu)$; in dissipative regimes, it is generally a contraction semigroup on suitable spaces.

The infinitesimal generator $\mathcal{L}$ of $\{\mathcal{K}^t\}$ is defined, when the limit exists, by
\begin{equation}
  (\mathcal{L}g)(u)
  \;:=\;
  \lim_{t\to 0}\frac{(\mathcal{K}^t g)(u)-g(u)}{t}.
  \label{eq:koop_nlw_generator_def}
\end{equation}
In finite dimensions, and for an ordinary DE (ODE) of the form $
\dot x=F(x) $,
the Koopman generator reduces to the Lie derivative
$$
\mathcal{L}g(x)=\nabla g(x)\cdot F(x)
$$
For PDE flows, the same object is expressed using a Fr\'echet/G\^ateaux derivative: if $Dg[u]$ denotes the Fr\'echet derivative, then formally
\begin{equation}
  (\mathcal{L}g)[u]
  \;=\;
  Dg[u]\big(\mathcal{N}(u)\big)
  \;=\;
  \int_{\Omega} \frac{\delta g}{\delta u}(x)\,\mathcal{N}(u)(x)\,dx,
  \label{eq:koop_nlw_generator_pde}
\end{equation}
where $\delta g/\delta u$ is the functional derivative with respect to $u$~\cite{nakao2020spectral}.

A Koopman eigenfunctional~\footnote{Notice that
we use the term eigenfunctional here following~\cite{nakao2020spectral}, although elsewhere
the term eigenfunction has been used~\cite{williams2015edmd}.} $\varphi$ with eigenvalue $\lambda\in\mathbb{C}$ satisfies
\begin{equation}
  \mathcal{K}^t\varphi \;=\; e^{\lambda t}\varphi
  \qquad\Longleftrightarrow\qquad
  \varphi(\Phi^t(u)) \;=\; e^{\lambda t}\varphi(u),
  \label{eq:koop_nlw_eig}
\end{equation}
equivalently, for sufficiently regular $\varphi$, $\mathcal{L}\varphi=\lambda\varphi$.  The eigenfunctional coordinate
\begin{equation}
  z(t) := \varphi(u(t)) \quad\Rightarrow\quad \dot z = \lambda z
  \label{eq:koop_nlw_linear_coord}
\end{equation}
evolves linearly.  If one can find a collection $\{\varphi_j\}_{j=1}^r$ that defines an embedding of the relevant invariant set (or a neighbourhood of interest), then the vector of coordinates $z_j(t)=\varphi_j(u(t))$ evolves under a diagonal linear system $\dot z = \Lambda z$ with $\Lambda=\mathrm{diag}(\lambda_1,\ldots,\lambda_r)$; this is the precise sense in which Koopman eigenfunctionals provide global linearizing coordinates when they exist on a set of interest \cite{mezic2005spectral,budisic2012applied}.

For a vector-valued observable $\mathbf{g}:\mathcal{X}\to\mathbb{C}^m$ (for example a discretized field, Fourier coefficients, probe measurements, or reduced features), the Koopman expansion takes the formal form
\begin{equation}
  \mathbf{g}(\Phi^t(u_0))
  \;=\;
  \sum_{j} e^{\lambda_j t}\,\varphi_j(u_0)\,\mathbf{v}_j
  \;+\;
  \int_{\Sigma_{\mathrm{cont}}} e^{\lambda t}\,\varphi(\lambda;u_0)\,\mathbf{v}(\lambda)\,\mathrm{d}\rho(\lambda),
  \label{eq:koop_nlw_kmd}
\end{equation}
where $\mathbf{v}_j\in\mathbb{C}^m$ are Koopman modes associated with $\mathbf{g}$, and the integral represents the contribution of the continuous spectrum through a spectral measure $\rho$ supported on $\Sigma_{\mathrm{cont}}$ \cite{mezic2013analysis,rowley2009spectral}. 

In nonlinear-wave problems this continuous component is often substantial: mixing or chaotic spatiotemporal regimes (for example Kuramoto--Sivashinsky dynamics or strongly forced dispersive waves) may possess broad continuous spectrum, in which case the appropriate object is a spectral measure \cite{korda2020spectral}.

Three structural properties of Koopman eigenfunctionals are especially consequential in wave PDEs:

\emph{(i) Invariants and conservation laws correspond to $\lambda=0$ eigenfunctionals.}
If $I[u(t)]$ is conserved along trajectories (for all $t$ where the solution exists), then
\begin{equation}
  I(\Phi^t(u)) = I(u) \quad\Rightarrow\quad \mathcal{K}^t I = I
  \quad\Rightarrow\quad \mathcal{L}I = 0.
  \label{eq:koop_nlw_invariant}
\end{equation}
This is the operator-theoretic statement that invariants lie in the nullspace of the generator, and it motivates data-driven identification of invariants as approximately $\lambda\approx 0$ eigenfunctionals when the PDE is partially or entirely unknown \cite{budisic2012applied,mezic2013analysis}.
We note in passing that this aspect is also intimately connected
with the characterization of the properties of
the equation and eventually with its potential
integrability, to which we dedicate a considerable
segment of this review below.
To the best of our knowledge, this aspect of the
Koopman operator has not been touched upon sufficiently
in the literature as of yet (i.e., the Koopman 
operator as a potential identifier of conservation
laws and potential integrability).

\emph{(ii) Products of eigenfunctionals are eigenfunctionals.}
If $\varphi_1,\varphi_2$ satisfy \eqref{eq:koop_nlw_eig} with eigenvalues $\lambda_1,\lambda_2$, then
\begin{equation}
  \mathcal{K}^t(\varphi_1\varphi_2)
  =(\mathcal{K}^t\varphi_1)(\mathcal{K}^t\varphi_2)
  =e^{(\lambda_1+\lambda_2)t}\,\varphi_1\varphi_2.
  \label{eq:koop_nlw_product}
\end{equation}
For wave PDEs, this implies that once a small set of ``base'' eigenvalues appears (e.g.\ linear decay rates or base frequencies on an invariant torus), one should generically expect a large family of combination eigenvalues (sums of base eigenvalues), often with severe degeneracy.  Analytic Burgers/KdV examples make this explicit \cite{page2018burgers,parker2020isolated,parker2023kdv}.

\emph{(iii) Eigenfunctionals encode global geometry (basins and invariant manifolds) when defined on a basin.}
For a hyperbolic equilibrium $u_\star$ (in a symmetry-reduced frame if needed), eigenfunctionals associated with unstable eigenvalues can be used to characterize the stable manifold.  In the simplest finite-dimensional setting, if $\varphi_j$ are Koopman eigenfunctions with $\Re(\lambda_j)>0$ associated with unstable directions, then boundedness as $t\to+\infty$ implies $\varphi_j(u)=0$ on the stable manifold, giving the characterization
\begin{equation}
  W^s(u_\star)
  \;\subseteq\;
  \bigcap_{\Re(\lambda_j)>0}\{u\in\mathcal{X}:\ \varphi_j(u)=0\},
  \label{eq:koop_nlw_stable_manifold_levels}
\end{equation}
with equality locally under standard nondegeneracy assumptions \cite{mauroy2016global}.  For dissipative nonlinear waves with multiple outcomes (capture vs.\ transmission, decay to different travelling pulses, distinct pattern states), such level-set characterizations motivate basin and separatrix computation via learned eigenfunctionals.

Relatedly, Koopman eigenfunctionals provide canonical phase/amplitude coordinates near attracting limit cycles and equilibria.  For a stable periodic orbit with fundamental frequency $\omega$, the asymptotic phase $\theta(u)$ satisfies $\theta(\Phi^t(u))=\theta(u)+\omega t\ (\mathrm{mod}\ 2\pi)$, hence the complex phase observable $e^{i\theta(u)}$ is a Koopman eigenfunction with eigenvalue $i\omega$; the isochrons are level sets of $\theta$ \cite{mauroy2016global}.  For a stable equilibrium, level sets of the slowest-decaying eigenfunctional define \emph{isostables}; in PDE contexts these are naturally formulated as level sets of a Koopman eigenfunctional (not an eigenfunction of the state) \cite{nakao2020spectral}.

Finally, symmetries that are ubiquitous in nonlinear waves (translations, rotations, gauge/phase symmetries) act directly on Koopman structure.  If a group $G$ acts on states, $g\cdot u$, and the PDE is equivariant,
$\Phi^t(g\cdot u)=g\cdot \Phi^t(u)$, then Koopman commutes with the induced pullback action on observables, which organizes eigenfunctionals into representation-theoretic subspaces and induces block structure in finite-dimensional approximations \cite{salova2019symmetries}.

\subsubsection{Data-driven and learning-based approximations of Koopman spectra from wave data}

In the basic DMD setting, one takes the observable to be the measured state itself, so that $x_k=u_k$ (or the measured quantity arranged as a state vector), and forms snapshot matrices
\begin{align}
\label{eq:koopman_data_matrix}
X = \begin{bmatrix} x_0 & x_1 & \cdots & x_{m-1}\end{bmatrix}\in\mathbb{C}^{n\times m},\qquad
Y = \begin{bmatrix} x_1 & x_2 & \cdots & x_{m}\end{bmatrix}\in\mathbb{C}^{n\times m}.
\end{align}
More generally, Extended DMD (EDMD) replaces the identity observable by a lifted dictionary $x_k=\mathbf{g}(u_k)\in\mathbb{C}^n$. Thus standard DMD is the identity-observable case, while EDMD uses a richer observable space; see details below.
DMD approximates the action of the Koopman operator on the span of measured observables by solving a finite-dimensional least-squares problem.  Specifically, given snapshot matrices $X, Y$ of Eq.~\eqref{eq:koopman_data_matrix}, DMD seeks the linear operator $A:\mathbb{C}^n\to\mathbb{C}^n$ minimizing
\[
\|Y-AX\|_F^2,
\]
so that $A$ represents the least-squares approximation of the time-$\Delta t$ Koopman action restricted to $\mathrm{range}(X)$.  The minimizer is
\begin{equation}
  A \;=\; YX^{+},
  \label{eq:koop_nlw_dmd_A}
\end{equation}
where $X^{+}$ denotes the Moore--Penrose pseudoinverse, typically computed from a truncated SVD of $X$ in which only the leading $r$ singular values (and corresponding modes) are retained, so that $r$ is the chosen truncation rank used to enforce a rank-$r$ projection \cite{schmid2010dmd,tu2014dmd}.

If $A w_j=\mu_j w_j$, then $\mu_j$ approximate Koopman eigenvalues on the chosen measurement subspace, and the associated continuous-time rates are
\begin{equation}
  \lambda_j \;=\; \frac{1}{\Delta t}\log(\mu_j),
  \label{eq:koop_nlw_dmd_log}
\end{equation}
with the logarithm branch chosen so that the discrete- and continuous-time eigenvalues satisfy $\mu_j=e^{\lambda_j\Delta t}$.  
The DMD reconstruction uses
$x_k\approx \sum_j b_j \mu_j^k w_j$ for coefficients $b_j$ set by the initial condition.  

For wave PDE data, plain DMD is often most effective after (i) restricting to a localized space-time window to avoid mixing distinct local expansions, and/or (ii) symmetry-aware preprocessing, such as comoving-frame alignment or related registration procedures, so that group motion and other extrinsic effects are not misidentified as internal oscillation. More generally, when a single measured observable is not sufficient to represent the relevant dynamics, one is naturally led to richer lifted observables.

Indeed, because nonlinear PDE dynamics rarely close on the span of state coordinates, EDMD lifts the data to a dictionary of nonlinear observables \cite{williams2015edmd}.  Choose a feature map $\psi:\mathcal{X}\to\mathbb{R}^p$,
$\psi(u)=(\psi_1(u),\ldots,\psi_p(u))^\top$,
and fit a matrix $K\in\mathbb{R}^{p\times p}$ such that
\begin{equation}
  \psi(u_{k+1}) \;\approx\; K^\top \psi(u_k).
  \label{eq:koop_nlw_edmd_model}
\end{equation}
With snapshot pairs $\{(u_k,u_{k+1})\}_{k=0}^{M-1}$, define
\begin{equation}
  G=\frac{1}{M}\sum_{k=0}^{M-1}\psi(u_k)\psi(u_k)^\top,\qquad
  A=\frac{1}{M}\sum_{k=0}^{M-1}\psi(u_k)\psi(u_{k+1})^\top,
  \label{eq:koop_nlw_edmd_mats}
\end{equation}
where \(G\) is the instantaneous feature covariance, while \(A\) is the one-step lagged cross-covariance linking \(\psi(u_k)\) to \(\psi(u_{k+1})\). Then, compute
\begin{equation}
  K^\top \;=\; G^{+}A,
  \label{eq:koop_nlw_edmd_K}
\end{equation}
with regularization (e.g.,\ $G+\varepsilon I$) often essential for noisy PDE data.   Eigenpairs $K^\top w_j=\mu_j w_j$ yield approximate Koopman eigenfunctionals $\varphi_j(u)\approx w_j^\top \psi(u)$, while Koopman modes for a chosen vector observable $\mathbf{g}(u)$ are recovered by regression against these coordinates \cite{williams2015edmd,mezic2013analysis}.  The wave-specific issue is not the algebra in \eqref{eq:koop_nlw_edmd_K} but the dictionary design: translation symmetry, localization, and multiscale dispersive content can make naive global polynomial/Fourier dictionaries ill-conditioned or physically misleading unless the dictionary is symmetry-adapted.

A complementary route is to eliminate explicit dictionary construction via \emph{kernel} EDMD/DMD, where one works implicitly in a reproducing kernel Hilbert space (RKHS).  With a kernel $\kappa(u,v)=\langle \psi(u),\psi(v)\rangle$, EDMD computations reduce to linear algebra involving Gram matrices on the data, often improving numerical stability in high-dimensional PDE settings \cite{salova2019symmetries,williams2015edmd}.

\emph{Partial observations and delay coordinates.}
Wave experiments and simulations frequently provide sparse probes rather than full fields.  If one observes $y_k=h(u_k)$, a standard Koopman-compatible approach is delay embedding: form
\begin{equation}
  \mathbf{y}_k = \big(y_k, y_{k-1}, \ldots, y_{k-q}\big)^\top \in \mathbb{R}^{q+1},
  \label{eq:koop_nlw_delay}
\end{equation}
and apply DMD/EDMD to $\mathbf{y}_k$.  Hankel/DMD variants implement this efficiently by building a Hankel matrix
\begin{equation}
  H=
  \begin{bmatrix}
    y_0 & y_1 & \cdots & y_{m-q} \\
    y_1 & y_2 & \cdots & y_{m-q+1} \\
    \vdots & \vdots & \ddots & \vdots \\
    y_q & y_{q+1} & \cdots & y_{m}
  \end{bmatrix},
  \label{eq:koop_nlw_hankel}
\end{equation}
and performing DMD on (a low-rank SVD truncation of) $H$.
In ergodic settings, Hankel/DMD constructions can be placed on firm theoretical footing as approximations of Koopman spectral objects \cite{arbabi2017koopman}.  The HAVOK formulation refines this idea by fitting a linear model in the leading delay coordinates and interpreting the remaining coordinates as an intermittent forcing term \cite{brunton2017havok}.  For nonlinear-wave data, this is practically relevant whenever (i) only a handful of spatial probes are available or (ii) translation symmetry makes full-field snapshots expensive or redundant.

\emph{Beyond point spectra: spectral measures and continuous spectrum.}
When the Koopman operator has substantial continuous spectrum, a finite list of DMD eigenvalues is not, in general, a stable object.  A robust alternative is to target the spectral measure of an observable.  In a measure-preserving ergodic setting with $g\in L^2(\mu)$, define the (discrete-time) correlation moments
\begin{equation}
  m_k := \langle g, \mathcal{K}^k g\rangle_{L^2(\mu)},
  \qquad k\in\mathbb{Z},
  \label{eq:koop_nlw_moments}
\end{equation}
which satisfy
$m_k=\int_{0}^{2\pi} e^{ik\theta}\,d\mu_g(\theta)$ for a spectral measure $\mu_g$ on the unit circle.  Korda, Putinar, and Mezi\'c develop a data-driven procedure to reconstruct fine spectral structure (including continuous components) from finitely many moments $m_k$ computed along a single trajectory \cite{korda2020spectral}.  This ``spectral measure'' viewpoint is well aligned with broadband nonlinear wave regimes where one expects spectral densities rather than isolated eigenvalues.

\emph{Learning Koopman-invariant feature spaces.}
A recurring obstacle in wave PDEs is that the ``right'' observables are rarely obvious.  Learning-based approaches address this by optimizing a feature map $\Psi_\theta(u)$ and a linear operator $K$ so that
\begin{equation}
  \Psi_\theta(\Phi^{\Delta t}(u))
  \;\approx\;
  K\,\Psi_\theta(u),
  \label{eq:koop_nlw_learned_lift}
\end{equation}
typically via least squares losses on pairs $(u_k,u_{k+1})$ plus regularization that encourages low rank or interpretability.  Neural approaches that explicitly learn Koopman-invariant subspaces \cite{takeishi2017learning} and deep learning approaches aimed at discovering Koopman eigenfunction coordinates \cite{lusch2018deep} provide practical mechanisms for lifting high-dimensional PDE snapshots into approximately Koopman-invariant coordinates, which can then be used for forecasting, estimation, and reduced modeling.  In a nonlinear-wave review, the key point is methodological, in that the success of Koopman computations hinges on whether the observable class can represent the eigenfunctionals relevant to the wave physics (scattering data, phase/amplitude variables, symmetry-reduced coordinates), and learning provides a systematic way to search that class.}

We will return to the topic of the Koopman operator
and its applications with a more specific bend toward
nonlinear wave problems in section~\ref{sec_appK}.

\section{Applications of PINNs}
\label{sec3}

\subsection{PINNs for learning nonlinear dynamics on lattices}

One of the recent examples of the application of 
PINNs has been in the consideration and potential discovery
of 1D nonlinear dynamical lattices consisting of a finite number
($N$) of nodes in the work of~\cite{SAQLAIN2023107498}.
Such lattices have been one of the workhorses of
nonlinear wave theory, given their broad 
relevance to a number of settings~\cite{Aubry06,Flach:2008,FPUreview,pgk:2011}.
Among the many relevant applications, we mention
atomic BECs in 
optical lattices~\cite{Morsch}, 
optical waveguide arrays~\cite{moti}, micromechanical oscillator arrays~\cite{sievers} %
 nonlinear electrical circuits~\cite{remoissenet}, engineered granular (metamaterial) crystals~\cite{yuli_book,granularBook}, antiferromagnetic crystals~\cite{lars3}, and superconducting
settings of Josephson-junction ladders~\cite{alex,alex2}.

The relevant models that were considered for the 
(real or complex, $n=1,\dots,N$ node) field $u_n(t)$
included, e.g.,  
the discrete $\phi^{4}$ model~\cite{p4book}
\begin{align}
\ddot{u}_{n} = C(u_{n+1} + u_{n-1} - 2u_n) + 2(u_n - u_n^3), \quad u_{n}\in\mathbb{R},
\label{dphi4}
\end{align}
where an overdot stands for the temporal derivative of $u_{n}$, and $C=1/h^{2}(>0)$ is 
the coupling constant (and $h$  the lattice spacing) between adjacent nodes. 
Also relevant to this type of study was 
the discrete SG (DsG)~\cite{SGbook}, 
often also referred to as the Frenkel-Kontorova 
model~\cite{braun1998}:

\begin{align}
\ddot{u}_{n} = C(u_{n+1} + u_{n-1} - 2u_n) -\sin{(u_n)}, \quad u_{n}\in\mathbb{R}.
\label{dsG}
\end{align}
This model similarly to the discrete $\phi^4$ showcased
the existence of kink (and antikink) solutions.

Another prototypical model that has been found
to be physically relevant consisted of the discrete 
NLS (DNLS) equation~\cite{kevrekidis2009dnls}, in particular with a focusing  nonlinearity:

\begin{align}
i\dot u_n = -C(u_{n+1}+u_{n-1}-2u_n) -|u_n|^2u_n, \quad u_{n}\in\mathbb{C}.
\label{dnls}
\end{align}

Finally, one more key model of broad interest, including
the important feature of breaking the Hamiltonian
character was the discrete, complex 
Ginzburg-Landau (DCGL) equation:
\begin{align}
\dot u_n = (1+i)C(u_{n+1} + u_{n-1} - 2u_n) - (1-i) |u_n|^2u_n + u_n, \quad u_{n}\in\mathbb{C},
\label{dcgl}
\end{align}
with a cubic nonlinearity.

In the work of~\cite{SAQLAIN2023107498}, a number
of relevant modifications were introduced in connection 
to the regular PINNs.  In particular, for the case of
real dynamical variables $\bu(t)\in \R^N$, the PINN $\hat{\bu}: \R\to\R^N$ only involved time $t$ as the input.
This was mapped through an $L$-layer fully-connected neural network to the output forming an $N$-dimensional vector 
$\hat{\bu}(t) = (\hat{u}_1(t),\hat{u}_2(t),\dots, \hat{u}_N(t))\in\R^N$. Since the specific right hand
side of the nonlinear model $\N(u_1, \dots, u_N)$ was
not known,  an overcomplete library $\mathrm{Lib}= \{D_\alpha\}_{\alpha\in A}$ of shift-invariant 
discrete spatial operators was utilized. The
latter was selected so as to incorporate both the 
linear 
couplings between nearest neighbours, and 
the different forms of nonlinear contributions.

The unknown operator $\N:\R^N\to\R^N$ is then considered
to be a linear combination $\N = \sum_{\alpha\in A}\lambda_\alpha D_\alpha$ 
of elements $D_\alpha$ in the library, and the 
scope of the relevant effort was to identify the expansion coefficients $\bm{\lambda} = (\lambda_\alpha)_{\alpha\in A}$,
i.e., to solve the associated inverse problem by minimizing the loss function
\begin{align}
    \CL(\btheta, \bm{\lambda}; \T_\N, \T_{\bm{f}}) \coloneqq w_\N\CL_\N(\btheta, \bm{\lambda};\T_\N) + w_{\bm{f}}\CL_{\bm{f}}(\btheta, \bm{\lambda};\T_{\bm{f}}).
\end{align}
Here
\begin{align}
  & \CL_\N(\btheta, \bm{\lambda}; \T_\N) = \frac{1}{|\mathcal{T}_\N|}\sum_{t\in\T_\N}\left|\dot{\hat{\bu}}(t;\btheta)%
  -\sum_{\alpha\in A}\lambda_\alpha D_\alpha \hat{\bu}(t; \btheta) \right|^2,\\
  & \CL_{\bm{f}}(\btheta, \bm{\lambda}; \T_{\bm{f}}) = \frac{1}{|\mathcal{T}_{\bm{f}}|}\sum_{t\in\T_{\bm{f}}}\left|\hat{\bu}(t;\btheta)%
  -\bm{f}(t; \btheta) \right|^2,
\end{align}
with $\T_\N$, $\T_{\bm{f}}$, respectively, being subsets of $[0,T]$ representing the training 
collocation points.  It is, in particular, at these
points that the ODEs residual and the discrepancy between $\hat{\bu}$ and the observed 
$\bm{f}$ was minimized. 

The initial value problem was solved with a suitable
integrator, e.g., 4th order Runge-Kutta for the case
of the $\phi^4$ model that we will detail below.
As relevant initial conditions, a travelling kink soliton 
was used for the $\phi^4$ data generation. The DeepXDE
library of~\cite{lu2021deepxde} was used and appropriately
modified in order to consider a first order system
(e.g., for positions and velocities in the $\phi^4$ case),
with $t$ as the relevant input variable.  The neural network used in all the experiments considered
only interior nodes in the relevant loss function.
Moreover, the associated architecture involved fully-connected networks  of three hidden 
layers with 40 neurons each, and utilizing a $\tanh$ activation function.

\begin{figure}[htbp]
    \centering
      \subfloat[]{
    \includegraphics[width=7.2cm,height=5cm]{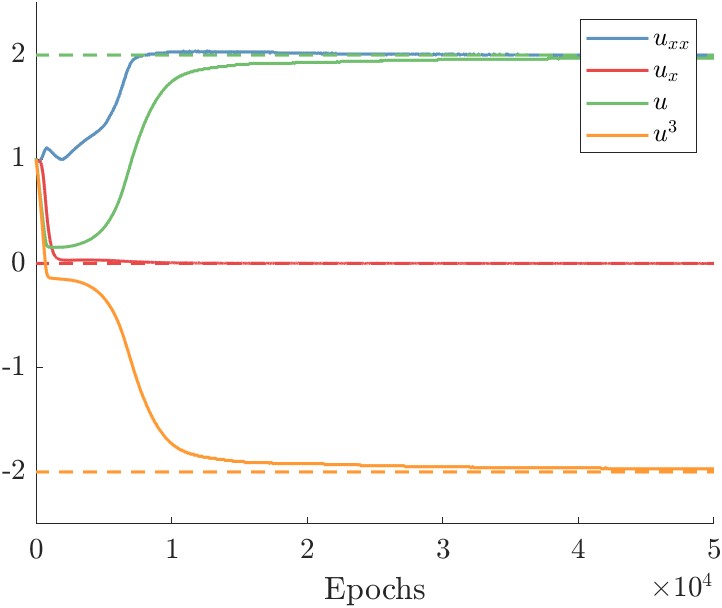}
  }
  \hfill
        \subfloat[]{
    \includegraphics[width=7.2cm,height=5cm]{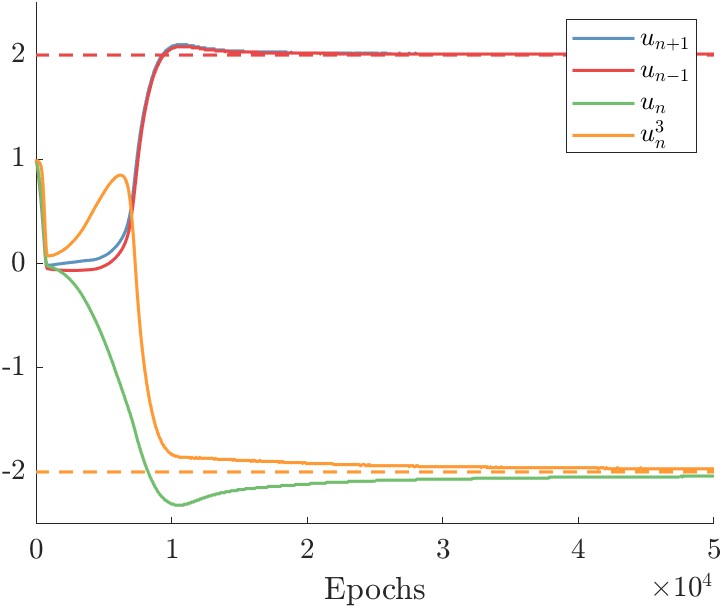}
  }\\
  
          \subfloat[]{
    \includegraphics[width=7.2cm,height=5cm]{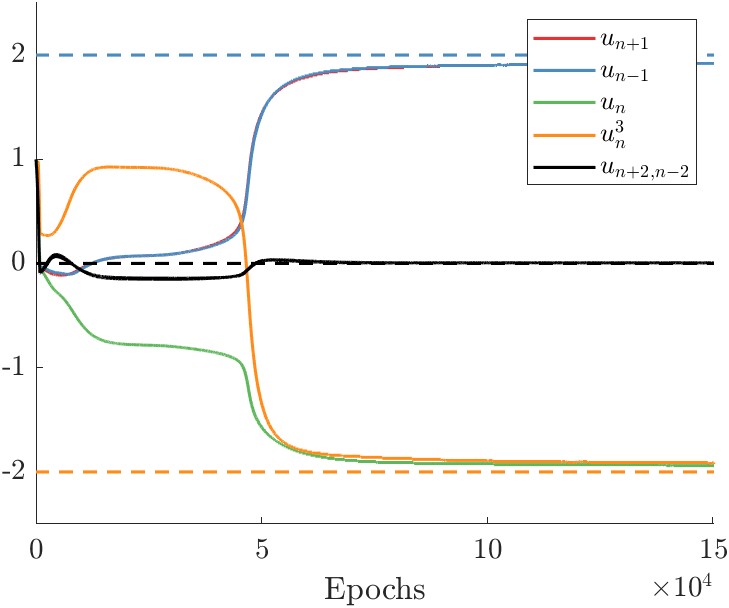}
  }
  \hfill
          \subfloat[]{
    \includegraphics[width=7.2cm,height=5cm]{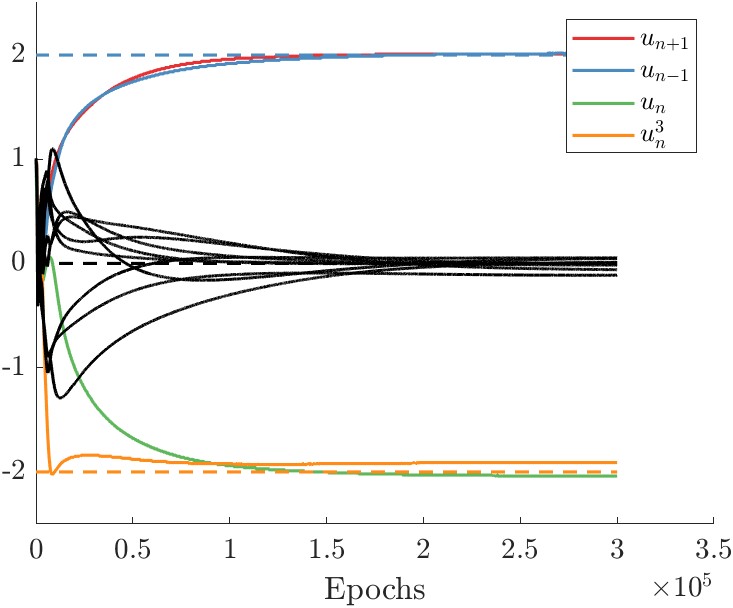}
  }
	\caption{Discrete $\phi^{4}$ model numerical results (cf equation~\eqref{dphi4}) for a coupling of $C=2$, The library $\mathrm{Lib}^{(1)}$ of equation~\eqref{dphi4_lib1} was considered in panel (a) with the solid blue, red, green and yellow lines corresponding to the discrete representation of the second and first derivative, as well as $u$, and $u^{3}$, respectively. The numerical results obtained by using the library $\mathrm{Lib}^{(2)}$ [cf equation~\eqref{dphi4_lib2}] are presented in panel (b) where solid blue, red, green, and yellow depict the $u_{n+1}$, $u_{n-1}$, $u_{n}$, and $u_{n}^{3}$, respectively. Panels (c) and (d) utilized the libraries of equations~\eqref{dphi4_lib3} and~\eqref{dphi4_lib4}, respectively (with the same notational conventions). The solid black lines therein correspond to (c) the terms $u_{n\pm2}$, and (d) to all the other cubic terms. It is relevant to highlight that the dashed lines represent the reference values for the coefficients. Reproduced from~\cite{SAQLAIN2023107498}. CC BY 4.0.}
    \label{fig:exps_dphi4}
\end{figure}

In what follows, we report the prototypical results
from the $\phi^4$ case considered in~\cite{SAQLAIN2023107498},
which were representative of the relevant findings also
for the DsG, as well as for the DNLS and the DCGL models
reported in the same publication. 
In the case of figure~\ref{fig:exps_dphi4}(a), a library
set of terms including
\begin{align}
\mathrm{Lib}^{(1)}=\Big\{\left(u_{n+1}+u_{n-1}-2u_n\right),\left(u_{n+1}-u_{n-1}\right)/2,u_n,u_n^3\Big\},
\label{dphi4_lib1}
\end{align}
was utilized, inspired by the continuum variant of the 
$\phi^4$ model. 
It can be clearly seen that the PINN 
learns the correct coefficients, leading the 
existing coefficients (including the nonlinearity and
the second difference one) to acquire their expected
values, while the ``discrete derivative'' term is
accurately recognized as featuring a vanishing prefactor.

Extending this perspective in figures~\ref{fig:exps_dphi4}(b)-(d), 
 a more ``inherently discrete'' approach to the relevant
problem was followed and an expanded library function 
choice was made. The relevant libraries were of the form. 
\begin{align}
\mathrm{Lib}^{(2)}=\Big\{u_{n+1},u_{n-1},u_{n},u_{n}^{3}\Big\},
\label{dphi4_lib2}
\end{align}
\begin{align}
\mathrm{Lib}^{(3)}=\Big\{\mathrm{Lib}^{(2)},u_{n+2},u_{n-2}\Big\}, 
\label{dphi4_lib3}
\end{align}
and
{\small
\begin{align}
\mathrm{Lib}^{(4)}=\Big\{\mathrm{Lib}^{(2)}, u_{n+1}^2u_n,  %
u_{n-1}^2u_n,u_{n+1}u_{n-1}u_n, u_{n+1}^2u_{n-1}, u_{n-1}^2u_{n+1}, %
u_n^2u_{n+1}, u_n^2u_{n-1}, u_{n+1}^3,  u_{n-1}^3\Big\}.
\label{dphi4_lib4}
\end{align}
}
Among these,
$\mathrm{Lib}^{(2)}$ was deemed to be the simplest example 
encompassing the principal ingredients of the 
model at hand. 
$\mathrm{Lib}^{(3)}$ constituted an extension of $\mathrm{Lib}^{(2)}$ encompassing the next-nearest neighbours,
namely, $u_{n\pm2}$ are appended therein. Finally, $\mathrm{Lib}^{(4)}$ further expanded on the possibilities of the cubic
nonlinear term
including several options towards
the cubic nonlinearity of the model.

In all the cases depicted in figures~\ref{fig:exps_dphi4}(b)-(d),
the nonlinear dynamical lattice model was accurately 
``discovered''. This entails both all the enclosed terms
being identified correctly as ``participating'' in the model
and with the right prefactors, as well as ``extraneous''
terms reaching a converged vanishing prefactor, thus revealing
their absence from the model dynamics. 
That being said, features such as the role of symmetries
(that will be further explored in what follows through
structure-preserving PINNs (S-PINNs)) was found to be of importance
here. More specifically, in the case where the libraries
contained quadratic or quartic terms, data augmentation and the
fact that $-u$ is a solution if $u$ is a solution were 
used in order for the model to converge to the appropriate
coefficients. 

\subsection{GNNs learn short and long range interactions on lattices}

While for many of the tasks of interest, using
``vanilla'' feedforward neural networks will suffice,
for specific tasks, such as the discovery of complex
interaction patterns between elements (particles)
in lattice systems, it has been argued that more
specialized neural networks, such as, e.g., graph
neural networks, may offer superior performance~\cite{geng2024separable,PGKGeng}.
Indeed, it has been argued therein that
since data that exhibits relationships between elements can be modelled as a graph, GNNs are a rather natural choice. Here, elements are named graph nodes, and the relationships between them are 
naturally represented by edges. Accordingly, the graph can serve as a systematic representation of lattice (nonlinear)
dynamical systems. In that vein, GNNs are rapidly emerging as a new field for studying 
nonlinear dynamical lattices.

The paradigms explored initially in~\cite{geng2024separable}
and subsequently in~\cite{PGKGeng} 
concerned many degrees of freedom and complex interactions.
It was in such settings where
GNNs were found to exhibit a proficient ability to accurately identify key information around nodes, ultimately leading
to an improved accuracy of the model discovery.  
This was inspired also by the
 outstanding performance of GNNs in some classic systems, e.g., in the works of~\cite{sanchez2019hamiltonian,bishnoi2023learning}.
Indeed,  lattice systems such as gravitationally 
interacting celestial bodies,  can be 
formulated on the basis of fully connected graphs, 
since they involve
interactions  between all particles.
This may allow the capturing of even
long-range interactions~\cite{laskin2006nonlinear}.

The approach of~\cite{geng2024separable,PGKGeng}
involved the representation of relationships 
in the form of 
a graph $\mathcal{G}=(\mathcal{V}, \mathcal{E})$ with $\mathcal{V}=\{v_1\cdots v_N\}$ being the set of nodes and $\mathcal{E}=\{e_{ij}\}$  the set of (directed) 
edges between nodes. The method then consisted
of two parts, with the first one adjusting the weight of the edge through trajectory data to extract the underlying interaction between particles, while the second part 
utilized the learned interactions towards the more accurate and effective trajectory prediction. Moreover,
the learning was designed to take place in two
components, a potential energy learning part ($V$-net)
and a kinetic energy learning part ($T$-net).
Each of these involved a $K$-layer GNN
updating nodes (for both the $T$- and $V$-nets)
and edges (only for the $V$-net). 

Then, assuming that the Lagrangian of the model could
be written as the difference of the kinetic energy
$T_\theta$ minus the potential energy 
${V_\theta}$, the relevant loss function for the
first of the two parts indicated above (the 
structural learning) can be written as:
\begin{equation} \label{loss_1}
\mathcal{L} = \mathcal{L}_{pred}+\gamma \mathcal{L}_{GL},
\end{equation}
where
\begin{equation}
\mathcal{L}_{pred}=\left\| \frac{\partial{T_\theta}}{\partial \mathbf{p}}-\frac{d\mathbf{q}}{dt}\right\|_2+\left\| -\frac{\partial {V_\theta}}{\partial \mathbf{q}}-\frac{d\mathbf{p}}{dt}\right\|_2.
\end{equation}
Also, here the graph learning loss $\mathcal{L}_{GL}$ we use is in the form:
\begin{equation}
\mathcal{L}_{GL} =  \|\mathbf{\mathbf{\alpha}}\|_F^2,
\end{equation}
where $\| \cdot\|_F$ denotes the Frobenius norm of a matrix.
The choice of the hyperparameter
$\gamma \in \mathbb{R}$ balances the two terms.
A sparse representation of the graph is sought, to the
degree possible, through
the graph learning loss.
 Here, $\mathbf{\alpha}=\{\mathbf{\alpha}_{i,j}\}$  
 represents a parameter matrix initialized by a neural network acting on each edge and adaptively labelling the strength of the edges through the training process. 
It is practically the key to learning the graph structure
in the $\alpha$-SGHN (i.e., separable
graph Hamiltonian network) model~\cite{PGKGeng}.
After the update has been completed, there is a 
prediction stage, where  the predicted trajectory
is obtained from the network as:
$(\hat{\mathbf{q}}^t,\hat{\mathbf{p}}^t)$,
through integration according to 
\begin{equation}\label{pred}
(\hat{\mathbf{q}}^t,\hat{\mathbf{p}}^t)=(\mathbf{q}^0,\mathbf{p}^0)+\int_{t_0}^t \alpha\text{-}\mathrm{SGHN} dt,
\end{equation}
with suitable initial values 
$(\mathbf{q}^0,\mathbf{p}^0)$. Here the loss is defined
based on the accuracy of the prediction of the trajectory,
i.e., $\mathcal{L} = \mathcal{L}_{pred}$.

\begin{figure*}%
\centering
\includegraphics[scale=0.55]{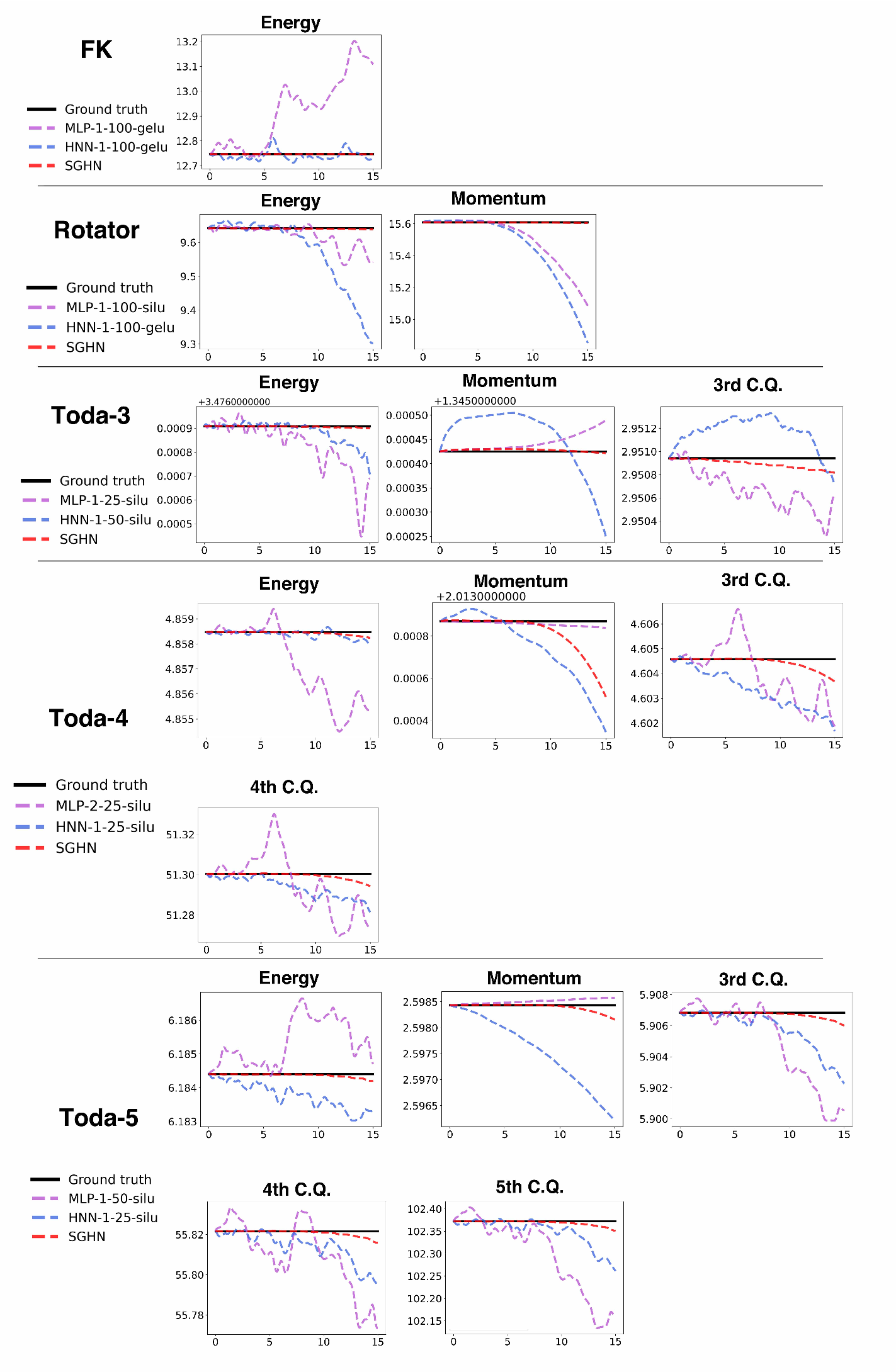}
\caption{This figure, illustrates evolution over time of the average true conservation values and the average predicted values of 20 samples for the different networks. C.Q. represents conserved quantity. Reproduced from~\cite{PGKGeng}. CC BY 4.0.}\label{c-frt}
\end{figure*}

The relevant methodology was explored in multiple
test examples in~\cite{PGKGeng}. These included the Frenkel-Kontorova lattice (bearing one conserved quantity), with the Hamiltonian:
\begin{equation}
H=\sum_{i=1}^N \Bigg(\frac{p_i^2}{2}+ \frac{(q_{i+1}-q_i)^2}{2}+1-\cos(q_i)\Bigg),	
\end{equation}
the rotator lattice (which
has two conserved quantities) with Hamiltonian:
\begin{equation}
H=\sum_{i=1}^N \Bigg(\frac{p_i^2}{2}+ \frac{(q_{i+1}-q_i)^2}{2}+1-\cos(q_{i+1}-q_i)\Bigg),	
\end{equation}
and the Toda lattice (which
is integrable and bears as many conserved quantities
as the number of degrees of freedom) with Hamiltonian:
\begin{equation}\label{Toda}
H=\sum_{i=1}^N \Bigg(\frac{p_i^2}{2}+\exp(q_i-q_{i+1})\Bigg).
\end{equation}
The evolution over time of the different conserved quantities
of these models in the case of the evolution prediction
from the $\alpha$-SGHN model, in comparison with the
standard multilayer perceptron (denoted as MLP) and the
so-called Hamiltonian neural network (denoted as HNN)~\cite{greydanus2019hamiltonianneuralnetworks}
is shown in figure~\ref{c-frt}. While there may still be
some room for improvement in higher order conserved 
quantities, generically the $\alpha$-SGHN model was
found to do a consistently better job in associated
conservation laws.

\subsection{S-PINNs}

As explained in section~\ref{sec:dl_of_functions}, PINNs aim to integrate physical principles into the learning process by minimizing losses derived from the underlying DEs. While this approach has shown success across a range of problems, standard PINNs often fail to capture deeper structural properties of the system, such as symmetries and conservation laws, which are essential for accurately modeling nonlinear dynamics. These limitations are particularly pronounced in nonlinear wave phenomena, where spatio-temporal symmetries, periodicity, and localization critically influence the solution dynamics.

To address these challenges, recent work has focused on structure-preserving variants of PINNs that explicitly encode such invariants into the architecture. A notable example is the development of S-PINNs~\cite{zhu2022neural}, which are motivated by the special structures intrinsic to solutions of nonlinear dynamical lattices. Consider, for instance, the completely integrable Ablowitz--Ladik (AL) model~\cite{doi:10.1063/1.523009,doi:10.1063/1.522558,kevrekidis2009dnls}, given by
\begin{align}
\label{eq:AL_before_separation}
\ii\dot{\Psi}_n+(\Psi_{n+1}-2\Psi_n+\Psi_{n-1}) %
+(\Psi_{n+1}+\Psi_{n-1})|\Psi_n|^2=0,
\end{align}
where $\Psi_n(t):\mathbb{Z}\times\mathbb{R}\to\mathbb{C}$ denotes the complex wavefunction at lattice site $n\in\mathbb{Z}$ and time $t\in\mathbb{R}$, and $\ii=\sqrt{-1}$. Using the ansatz
\begin{equation}
\Psi_n=\psi_ne^{2\ii q^2t},
\label{eq:Sep_ansatz}
\end{equation}
with background amplitude $q^2$, equation~\eqref{eq:AL_before_separation} transforms into
\begin{align}
\label{eq:AL}
\ii\dot{\psi}_n+(\psi_{n+1}-2\psi_n+\psi_{n-1}) %
+(\psi_{n+1}+\psi_{n-1})|\psi_n|^2-2q^2\psi_n=0.
\end{align}
Hereafter we set $q\equiv1/\sqrt{2}$ for simplicity.

Among the exact solutions of equation~\eqref{eq:AL}, one notable example is the Kuznetsov--Ma (KM) soliton~\cite{Ankiewicz2010,doi:10.1063/1.4961160}, a discrete solution that is temporally periodic and spatially localized:
\begin{align}
\label{eq:ma}
\psi(n,t):=\psi_n(t)=\frac{1}{\sqrt{2}}\frac{\cos(\omega t+\ii\theta)+ %
G\cosh(rn)}{\cos(\omega t)+G\cosh(rn)},
\end{align}
where the period is $T=2\pi/\omega$, and the parameters $\theta$, $r$, and $G$ are determined by $\omega$ through
\begin{align}
\theta=-\arcsinh(\omega),\qquad r=\arccosh\left(\frac{2+\cosh(\theta)}{3}\right),\qquad G=-\frac{\omega}{\sqrt{3}\sinh(r)}.
\end{align}
Besides temporal periodicity, a key feature of the KM soliton is its spatio-temporal parity symmetry:
\begin{align}
\label{eq:parity_symmetry}
\psi(n,-t)=\psi(n,t),\qquad \psi(-n,t)=\psi(n,t).
\end{align}
which reflects the underlying parity and time-reversal invariance of the AL model itself.

\begin{figure}[h]
  \centering
  \subfloat[Exact KM soliton\label{fig:ma_exact}]{
    \includegraphics[width=0.3\textwidth]{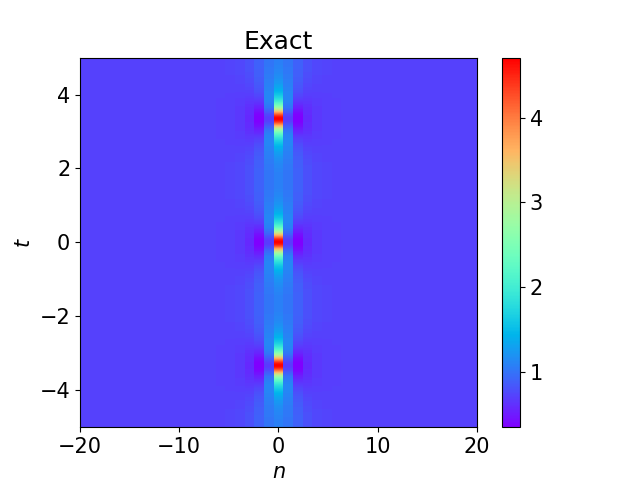}
  }
  \hfill
  \subfloat[PINN\label{fig:ma_original_pinn}]{
    \includegraphics[width=0.3\textwidth]{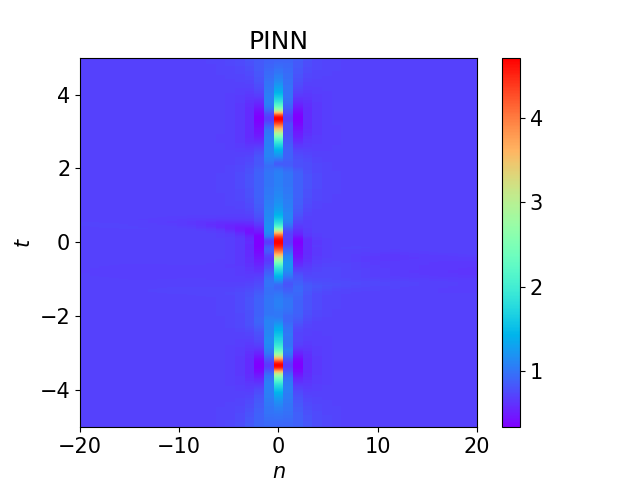}
  }
  \hfill
  \subfloat[S-PINN\label{fig:ma_spinn}]{
    \includegraphics[width=0.3\textwidth]{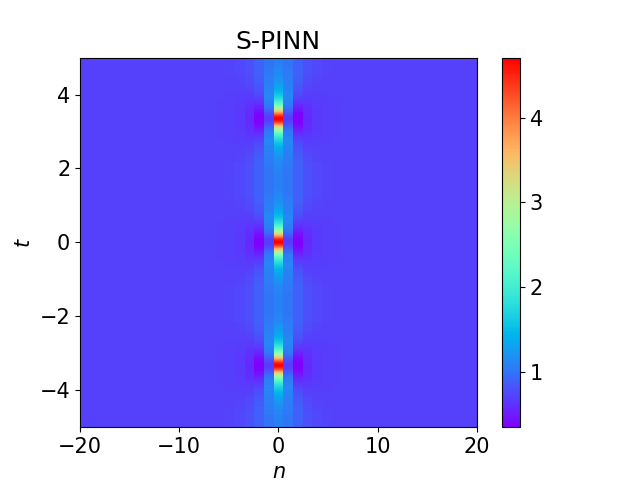}
  }

  \vspace{0.5em}

  \subfloat[$\psi(n,t)$ for $t=-0.67$\label{fig:ma_t_fixed}]{
    \includegraphics[width=0.3\textwidth]{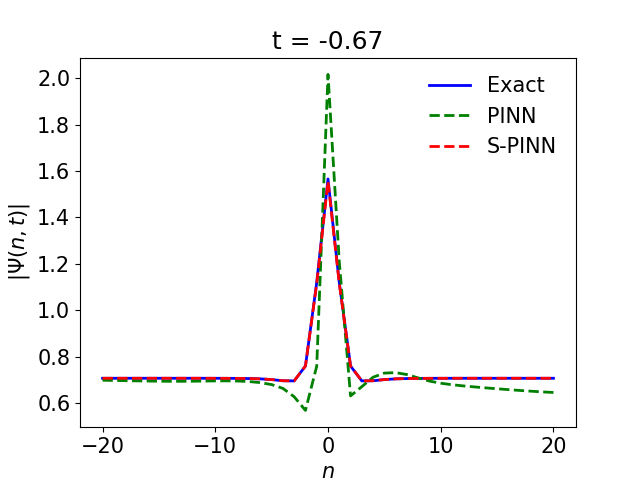}
  }
  \hfill
  \subfloat[$\psi(n,t)$ for $n=0$\label{fig:ma_x_fixed}]{
    \includegraphics[width=0.3\textwidth]{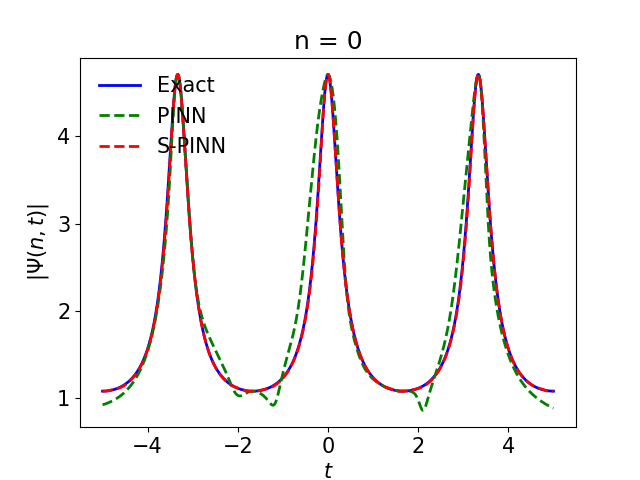}
  }
  \caption{Numerical results, for the KM soliton using PINN and S-PINN. Top panels show the spatio-temporal evolution of the amplitude $|\psi(n,t)|$ for the exact solution (a), PINN (b), and S-PINN (c). Bottom panels show the spatial profile at $t=-0.67$ and the temporal evolution at $n=0$. Solid blue lines indicate the exact solution, while dashed green and red lines correspond to PINN and S-PINN, respectively. Standard PINN fails to capture the time-periodicity and spatio-temporal parity symmetry~\eqref{eq:parity_symmetry}, whereas S-PINN accurately preserves both. Reprinted from~\cite{zhu2022neural}, Copyright (2022), with permission from Elsevier.}
  \label{fig:acc_full_space_ma}
\end{figure}

Zhu et al~\cite{zhu2022neural} proposed a principled approach to encode such physical structures directly into the neural network architecture. By enforcing spatio-temporal parity symmetry~\eqref{eq:parity_symmetry} and temporal periodicity within the network design, they introduced a S-PINN that significantly outperformed conventional PINNs. As illustrated in figure~\ref{fig:acc_full_space_ma}, S-PINNs produce markedly more accurate reconstructions of the KM soliton across both spatial and temporal domains. Standard PINNs, by contrast, fail to respect the symmetry and periodicity of the underlying solution, leading to substantial inaccuracies. This highlights the critical role of embedding structural priors when modeling nonlinear wave phenomena with neural networks.

\subsection{PINNs for scale- and translational-invariant wave solution identification: the Burgers equation case example}

One of the aspects of the usefulness of PINNs that we also briefly touch upon concerns their ability to dynamically evolve equations, including potentially along a group orbit (e.g. of translation or of self-similar rescaling), leading to the identification of potential stationary---or, for that matter, dynamical---states in such frames. Here, we provide a prototypical example along this vein, from the recent work of~\cite{kavousanakis2025symmetrypinns}, concerning the self-similar and moving evolution of a wave in the well-known Burgers equation, inspired from the earlier example of~\cite{rowley2003reduction}.

Consider the one-dimensional Burgers equation, also considered in the realm of PINNs in~\cite{shahab2025neuralnetworksbifurcationlinear}:
\begin{equation}\label{eq:burgers}
\partial_tu=\nu\partial_{xx}^2u-u\partial_xu\equiv\mathcal D_x(u).
\end{equation}
The value of viscosity for this example has been set to $\nu=0.025$. Using scaling arguments starting from an ansatz of the form
\begin{equation}
u(x,t)=Bw\left(\frac{x-c}{A}\right)\equiv Bw(y),
\end{equation}
one infers that $B=1/A$.

\begin{figure}[ht!]
    \centering
    \subfloat[]{\includegraphics[width=0.33\linewidth]{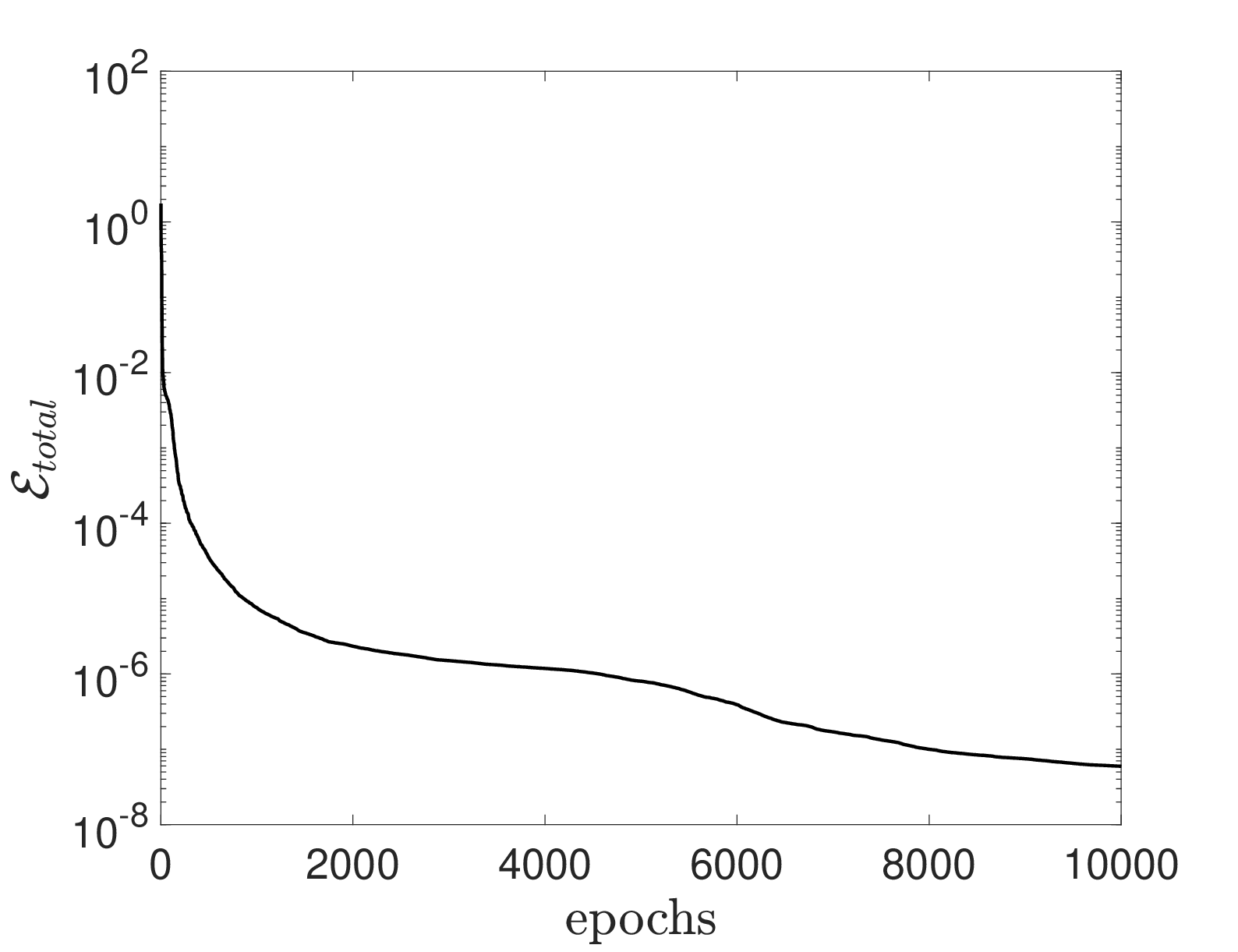}\label{fig:burgersMN_b}}
    \subfloat[]{\includegraphics[width=0.33\linewidth]{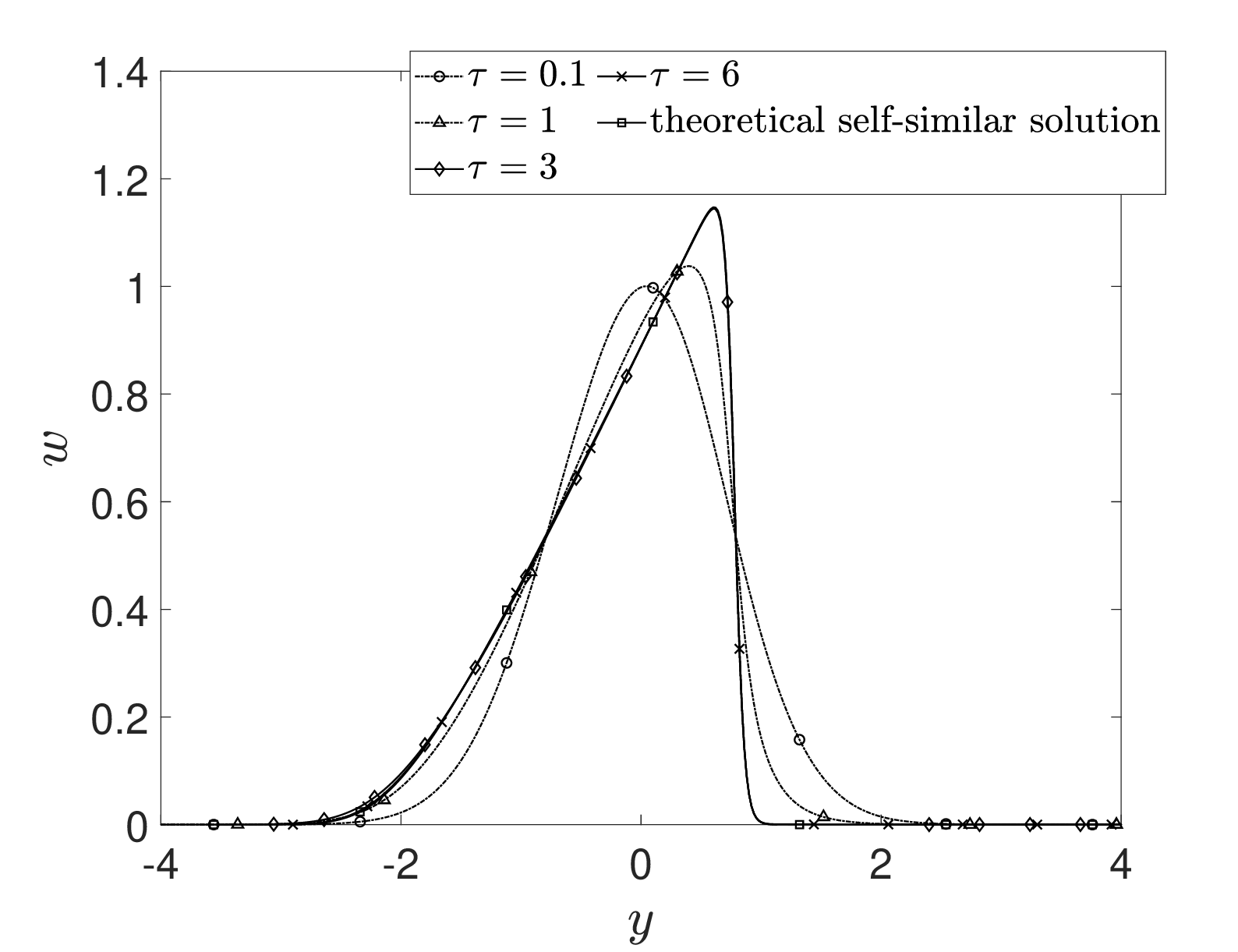}\label{fig:burgersMN_c}}
    \subfloat[]{\includegraphics[width=0.33\linewidth]{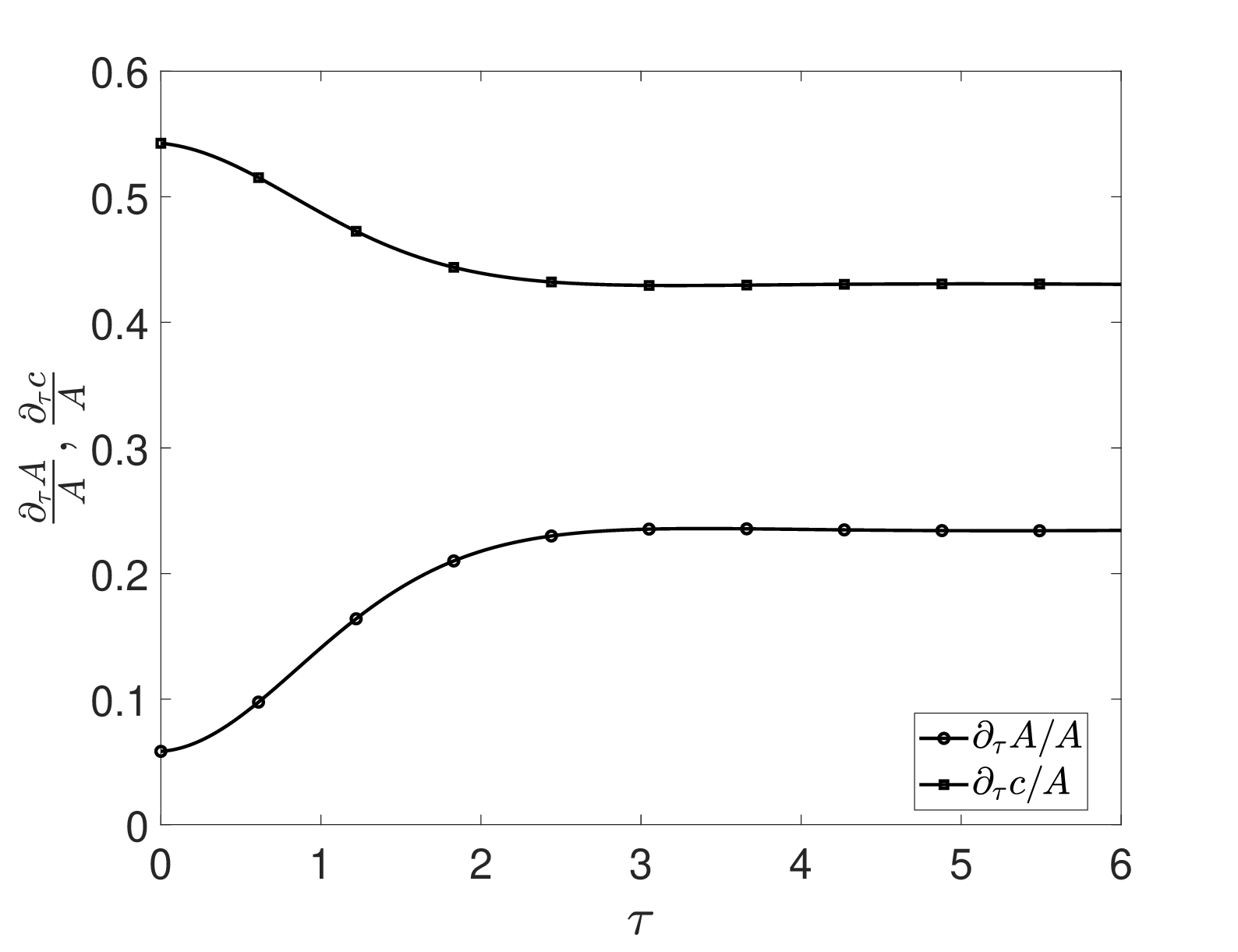}\label{fig:burgersMN_d}}
    \caption{PINN prediction of the self-similar dynamics for the Burgers equation (a) Convergence of the total loss $\mathcal E_{total}$ during training for the rescaled/co-moving Burgers equation. (b) Snapshots of rescaled PINN predicted solution $w$ at different $\tau$ values. The converged self-similar solution $w$ (at $\tau\approx3$) and the analytically predicted self-similar solution~\cite{whitham1999linear} are practically coincident. (c) Evolution of the scaling rates $\partial_\tau A/A$ and $\partial_\tau c/A$ over rescaled time $\tau$. For sufficiently long $\tau$ values ($\tau>3$), we can observe the convergence of the rescaled solution to a stationary, self-similar profile. Reproduced from~\cite{kavousanakis2025symmetrypinns}. CC BY 4.0.}
    \label{fig:burgersMN}
\end{figure}

With this scaling, one obtains:
\begin{equation}
\mathcal D_x\left(\frac{1}{A}w\left(\frac{x-c}{A}\right)\right)=A^{-3}\mathcal D_y(w).
\end{equation}
This, in turn, inspires the dynamic scaling choice of the form:
\begin{equation}
u(x,t)=\frac{1}{A(\tau)}w\left(\frac{x-c(\tau)}{A(\tau)},\tau(t)\right),
\end{equation}
incorporating a rescaling of time $\tau=\tau(t)$, such that $d\tau/dt=1/A^2$, which leads the original Burgers equation to be reshaped as:
\begin{equation}\label{eq:burgersMN}
\partial_\tau w=\nu\partial_{yy}^2w-w\partial_yw+\frac{\partial_\tau A}{A}(w+y\partial_yw)+\frac{\partial_\tau c}{A}\partial_yw,\qquad \textrm{with }y=\frac{x-c}{A}.
\end{equation}
The two $\tau$-dependent unknown quantities (as they need to be to ensure the factoring out of the symmetries), $\partial_\tau A/A$ and $\partial_\tau c/A$ are determined by imposing two template constraints, as described in the work of~\cite{kavousanakis2025symmetrypinns,rowley2003reduction}. Each of these conditions eliminates one of the corresponding symmetries/degeneracies, enabling the identification of a unique solution upon application of the PINN methodology.
The conditions arise, e.g. from minimizing
\begin{equation}
E\equiv\int_{y_{min}}^{y_{max}}\left(w-\frac{1}{A}T\left(\frac{y-c}{A}\right)\right)^2dy,
\end{equation}
evaluated at $A=1$ and $c=0$ with $T(y)$ being a chosen template function.
This results in the constraints:
\begin{equation}\label{eq:burgerstemplate}
\int_{y_{min}}^{y_{max}}(w-T)(T+y\partial_yT)dy=0\qquad\textrm{and}\qquad\int_{y_{min}}^{y_{max}}(w-T)\partial_yTdy=0.
\end{equation}
The specific $T(y)$ chosen in~\cite{kavousanakis2025symmetrypinns} was $T(y)=\exp(-y^2)$.

The above theoretical formulation of equation~\eqref{eq:burgersMN} was solved using PINNs in the domain $y\in[-6,6]$ with zero-flux boundary conditions at both ends in~\cite{kavousanakis2025symmetrypinns}. A similar proposal for the solution of fluid mechanical models---such as variants of the Euler equation---bearing (asymptotic) self-similarity was made earlier in the work of~\cite{wang2023asymptotic}.
The initial condition is $w(y,0)=\exp(-y^2)$.
For this implementation, the neural network $\mathcal N_w(y,\tau)$ consisted of three hidden layers with 20 neurons each, with a hyperbolic tangent activation function.
An auxiliary network $\mathcal N_p(\tau)$ was used in order to predict $\partial_\tau A/A$ and $\partial_\tau c/A$ and contained a single hidden layer with 4 neurons.
The neural network $\mathcal N_w(y,\tau)$ was trained using about 180 000 collocation points in the $y-\tau$ domain.
Training was performed using the L-BFGS optimizer and the corresponding convergence of the total loss $\mathcal E_{total}$ is shown in figure~\ref{fig:burgersMN}.
Upon the spatio-temporal evolution, the initial data was found to steepen into a sharp front, eventually leading to a stationary state in the renormalized frame of variables $(y,\tau)$; see also~\cite{rowley2003reduction} and figure~\ref{fig:burgersMN_b}.
A comparison of the resulting self-similar solution
with the analytical self-similar solution of the Burgers equation~\cite{whitham1999linear}:
\begin{equation}\label{eq:burgersanalytic}
u(x,t_0)=\sqrt{\frac{\nu}{\pi t^*}}\frac{[\exp(A^*/(2\nu))-1]\exp[-(x-c^*)^2/(4\nu t^*)]}{1+[\exp(A^*/(2\nu))-1]/2\cdot\operatorname{erfc}((x-c^*)/\sqrt{4\nu t^*})},
\end{equation}
for a suitable choice of $A^*$, $c^*$ and $t^*$, revealed the accuracy of the obtained result as shown in figure~\ref{fig:burgersMN_c}. In turn, the evolution of $\partial_\tau A/A$ and $\partial_\tau c/A$ over rescaled time $\tau$ is shown in figure~\ref{fig:burgersMN_d}, converging to the asymptotic values of the relevant quantities.

\subsection{Lax pair informed neural networks}\label{section:LLPN}
The recent work of~\cite{pu2024lax} shows how Lax pairs can be used to assist in the training of neural networks to solve integrable systems. A Lax pair of operators, denoted by $(L(t),P(t))$ and depending on time $t$, consists of operators acting on some fixed Hilbert space that satisfy the Lax equation
\begin{equation}
\frac{dL}{dt}=[L,P]:=LP-PL.
\label{eq:lax-eq}
\end{equation}
This solvability condition can sometimes amount to an integrable nonlinear partial (or lattice) DE. For example, it is easy to verify that the KdV equation
\begin{equation}\label{eq:KdV}
u_t-6uu_x+u_{xxx}=0
\end{equation}
has the Lax Pair
\begin{equation}
L=-\partial_x^2+u,\qquad P=4\partial_x^3-3(u\partial_x+\partial_xu).
\end{equation}
It is well-known that for operators satisfying the Lax equation, one finds the spectral operator $L(t)$ evolves by a similarity transformation:
$$
L(t)=Q(t)L(0)Q(t)^{-1},\qquad \dot Q=-PQ.
$$
Therefore, the spectrum of $L(t)$ is preserved for all $t$. This is one of the algebraic hallmarks of integrability, that is, the invariants of motion are the spectral invariants of $L$ (e.g. in finite dimensions the traces $\operatorname{tr}(L^n)$).

In the PDE setting, $L$ and $P$ are typically differential operators acting on a spectral function $\psi(x,t)$ through the overdetermined system
\begin{equation}
L(u)\psi=\lambda\psi,\qquad \psi_t=-P(u)\psi,
\label{eq:lax-pde}
\end{equation}
where $\lambda$ is the so-called spectral parameter and $u(x,t)$ is the physical field corresponding to original PDE dynamics. The compatibility condition $\psi_{xxt}=\psi_{txx}$ yields equation~\eqref{eq:lax-eq} and hence the nonlinear PDE for $u$.

Another particularly useful viewpoint, when $x\in\mathbb{R}$, comes from rewriting the Lax pair in terms of auxiliary linear problems for the wavefunction $\Phi(x,t)$:
\begin{equation}
\Phi_x=U(u,\lambda)\Phi,\qquad \Phi_t=V(u,\lambda)\Phi,
\label{eq:aux-lin}
\end{equation}
where $U$ and $V$ are matrix- (or operator-) valued functions of the field $u$ and the spectral parameter $\lambda$. In this formulation, $U$ plays the role of a spatial connection and $V$ a temporal connection on a trivial vector bundle over the $(x,t)$-plane. The requirement that these two linear equations be compatible, $\Phi_{xt}=\Phi_{tx}$, is equivalent to the flatness of the associated connection
\begin{equation}
U_t-V_x+[U,V]=0.
\label{eq:zero-curvature}
\end{equation}
Equation~\eqref{eq:zero-curvature} is known as the zero curvature condition or Zakharov--Shabat equation. It encodes the original nonlinear PDE in the vanishing of the curvature of a connection depending on the spectral parameter, thereby tying integrability to the machinery of gauge theory.

With this context in mind, a Lax Pair Informed Neural Network (LPNN) can be built depending on the formulation of Lax integrability used. More specifically, in this approach one seeks to embed the Lax pair structure into a SciML model for $u(x,t)$. Just as with PINNs and all of the previous frameworks presented in this section, one regresses in the sense of $L^2$, alongside boundary and initial data, to learn the trainable parameters that furnish the original PDE variable. The key difference here is that LPNNs are structure preserving in that they also learn the spectral function $\psi$. This is the essence of what~\cite{pu2024lax} calls LPNN version 1 (LPNN-v1). LPNN version 2 (LPNN-v2) considers, in tandem with version 1, an additional loss term that incorporates the zero-curvature condition, either given by the Zakharov--Shabat or auxiliary linear problem depending on the context.

For a concrete example, consider the KdV equation which has a spectral function $\psi$ that satisfies
$$
\psi_{xx}=(u-\lambda)\psi,\qquad \psi_t=-u_x\psi+(4\lambda+2u)\psi_x.
$$
Since the KdV equation can be derived from the compatibility condition $\psi_{xxt}-\psi_{txx}=0$, this compatibility condition can be reformulated into a standard loss function for LPNN-v1. Thus, with appropriate boundary and initial data, the basic idea of LPNN-v1 here is, for a fixed $\lambda$, to parametrize $u(x,t)$ and $\psi(x,t)$ with a feedforward fully connected neural network and to regress until the compatibility condition $\psi_{xxt}-\psi_{txx}=0$ and KdV equation~\eqref{eq:KdV} are approximately satisfied.

As a benchmark, the work of~\cite{pu2024lax} compares LPNN-v1 with a vanilla PINNs approach; see~\cite{pu2024lax} for the computational details. More specifically, it is reported that the LPNN-v1 error, in the sense of $L^2$, is about twice that of PINNs, yet the training time is about four times less. The work of~\cite{pu2024lax} similarly goes on to study the Camassa--Holm, and the Kadomtsev--Petviashvili, a two-dimensional generalization of the KdV equation, reporting, more or less, similar results.

LPNN-v2 seems to show more promise in that it consistently outperforms PINNs in the resulting accuracy. Perhaps most noteworthy is their result for the modified KdV (mKdV) equation
$$
u_t+6u^2u_x+u_{xxx}=0.
$$
Here, the auxiliary linear problem involves the following matrices
$$
U=\begin{bmatrix}\lambda&u\\-u&-\lambda\end{bmatrix},\qquad
V=\begin{bmatrix}-4\lambda^3-2u^2\lambda&-4u\lambda^2-2u_x\lambda-2u^3-u_{xx}\\4u\lambda^2-2u_x\lambda+2u^3+u_{xx}&4\lambda^3+2u^2\lambda\end{bmatrix}.
$$
LPNN-v2 simply parametrizes the complex two by one vector $\Phi$, along with $u$, using a neural network and trains until the compatibility condition $\Phi_{xt}-\Phi_{tx}=0$ is approximately satisfied. The work of~\cite{pu2024lax} reports that LPNN-v2 achieves an accuracy eight times that of PINNs. They further go on to adapt LPNN-v2 to the SG, NLS, and Short-Pulse equations, in all cases outperforming PINNs in terms of accuracy.

We comment that LPNNs, as used by~\cite{pu2024lax} are an example of function learning. Instead of learning the function satisfying a PDE $\partial_tu=\mathcal F(u)$ and associated spectral functions directly, one may adapt to learn operator-valued maps
$$
u\mapsto L_\theta(u),\qquad u\mapsto P_\varphi(u),
$$
with trainable parameters $\theta,\varphi$, constrained so that the evolution predicted by the network satisfies
\begin{equation}
\frac{d}{dt}L_\theta(u(t))\approx[L_\theta(u(t)),P_\varphi(u(t))].
\label{eq:lpnn-constraint}
\end{equation}
In an LPNN discretization, $L_\theta$ could be represented by a symmetric banded matrix with main diagonal given by $u$, and $P_\varphi$ by a skew-symmetric matrix encoding the learned differential stencil. However the discretization is performed, learning $L_\theta$ and $P_\varphi$ can be facilitated by either a DeepONet or FNO as discussed in section~\ref{section:OLearn}.

Besides outperforming PINNs in terms of accuracy, there are a few advantages to consider going forward with deep learning integrable and nearly-integrable dynamics using LPNNs.
\begin{itemize}
\item \emph{Structure preservation:} Spectral invariants are (approximately) preserved by construction, improving long-term stability of learned dynamics.
\item \emph{Integrability bias:} By constraining to a (possibly learned) Lax form, the network embeds a strong prior aligned with integrable or near-integrable physics.
\item \emph{Interpretability:} After learning $L_\theta$ and $P_\varphi$ there may be opportunities to recover hidden symmetries, conserved quantities, or other approximate integrable structures.
\end{itemize}

\section{Discovering dynamical systems and reduced order models from dispersive PDE}
\label{sec4}

\subsection{Use of SINDy to develop moment equation reductions stemming from nonlinear dispersive PDEs}

Often in the context of nonlinear wave equations, it is possible
to infer effective information about the system's dynamics
(e.g. its centre of mass, its variance and the associated
wavefunction width, the kurtosis etc) through the consideration of
the so-called moments~\cite{victor_theory,Victor_NLS_equation}.
Indeed, in these (and related) works,
the technique of moment methods was demonstrated as being
particularly useful in exploring the dynamics of NLS models
and their effective dynamics including the examination of
collapse type phenomena and their potential prevention.
Such methods were not only used in themes from
optics to atomic BECs~\cite{Kevrekidis2000Parametric}, but they
were also extended to models of Fisher-KPP type with 
potential applications to the dynamics of brain tumours~\cite{BELMONTEBEITIA20143267}. 

Given that these moment equations are effective ODEs
that may {\it or may not} close at the level of analytical
considerations, a natural use of data-driven methods 
is to potentially leverage, e.g. the SINDy approach, or other
ones such as Neural ODE~\cite{chen2018neural}, in order to identify the relevant
ODEs in either one of these cases, and to compare the
results of training data with possible testing evolution
data to examine the potential discovery of known---or,
more excitingly unknown---such moment relations.
Such approaches were recently considered in the work of~\cite{suyang}.
Here, we will consider some of the basic examples of the latter
work and comment on the extensions and further possibilities
(as well as on more recent work) along this vein.

Our relevant starting point for the purposes of this
discussion will be a prototypical NLS model with a parabolic
trap (of wide relevance to atomic BECs~\cite{becbook2})
in the form of:
\begin{equation}
\label{eq:nls_harmonic}
    iu_t = -\frac{1}{2}u_{xx} + \frac{1}{2}x^{2}u + g\left(\left|u\right|^{2}, t\right)u,
\end{equation}
where $g\left(|u|^{2},t\right)$ denotes the nonlinearity.
It is well known from the theory of~\cite{victor_theory}
that relevant moment quantities in such a setting
are of the form:
\begin{align}
    I_{k}(t) &= \int_{\mathbb{R}} x^{k}|u(x,t)|^{2}dx, 
    \label{eq:def_I}\\
    V_{k}(t) &= 2^{k-1}i\int_{\mathbb{R}}  x^{k}\left(u(x,t)\frac{\partial \bar{u}(x,t)}{\partial x} - \bar{u}(x,t)\frac{\partial u(x,t)}{\partial x}\right)dx, 
    \label{eq:def_V}\\
    K(t) &= \frac{1}{2}\int_{\mathbb{R}}  \left|\frac{\partial u(x,t)}{\partial x}\right|^{2} dx, 
    \label{eq:def_K}\\
    J(t) &= \int_{\mathbb{R}}  G\left(\rho(x, t), t\right)dx = \int_{\mathbb{R}}  G\left(\left|u(x, t)\right|^2, t\right)dx.
    \label{eq:def_J}
\end{align}
Here $\bar{u}$ denotes the complex conjugate of $u$, while
$\rho(x, t) = \left|u(x, t)\right|^{2}$. Also, $G=G(\rho, t)$ is a function such that $\frac{\partial G}{\partial \rho}(\rho, t) = g(\rho, t)$. The definitions of \eqref{eq:def_I}-\eqref{eq:def_J} can be intuitively 
interpreted. Indeed, the first moment $I_1(t)$ 
represents the centre of mass of the probability density $\rho = |u|^2$ (when the latter is normalized). 
$I_k$ represent higher moments associated with 
this density distribution such as its variance,
while  $V_k$ reflects the moments associated
with the momentum density (the latter is the quantity
in the corresponding parenthesis in the right hand
side of the $V_k$ definition). $K$ is drawn from  
the kinetic part of
the Schr{\"o}dinger problem energy, while $J$
from the nonlinear part of the corresponding
energy.

We now focus on a couple of concrete (simple) examples of 
associated moment dynamics. For instance, 
the moments $I_1$ and $V_0$ satisfy
    \begin{align}
    \label{eq:first_ME}
    \left\{
        \begin{aligned}
            \frac{dI_1}{dt} &= V_0,\\
            \frac{dV_0}{dt} &= -I_1.
        \end{aligned}\right.
    \end{align}
This is an intuitive analytical finding suggesting
that the centre of mass of the density distribution 
behaves as a harmonic oscillator inside a parabolic
trap. Interestingly, this happens irrespectively of
the nonlinearity $g(\rho, t)$.

In the work of~\cite{suyang}, the authors constructed
data matrices $\bfX^{(0)} = [\mathbf{I}_1^{(0)}, \mathbf{V}_0^{(0)}] \in \R^{N \times 2}$ and $\bfX^{(1)} = [\mathbf{I}_1^{(1)}, \mathbf{V}_0^{(1)}]\in \R^{N \times 2}$ from numerically solving the PDE~\eqref{eq:nls_harmonic} with two distinct suitably localized
initial conditions in the form of Gaussian and sech-like
pulses. Enlarging the relevant dataset using
$\bfX = [\bfX^{(0)\top}, \bfX^{(1)\top}]^\top \in \R^{2N \times 2}$
to avoid overfitting, it was found that SINDy predicts the
correct dynamics that match equation~\eqref{eq:first_ME}, even when
deploying a large 
library $\mathbf{\Theta}_{\deg\le n}(\bfx)$ for $n$ up to 16, i.e.,
it yields:
\begin{equation}\label{eq: SINDy output for n = 16}
\left\{
    \begin{aligned}
        \frac{dI_1}{dt} &= 1.000V_0,\\
        \frac{dV_0}{dt} &= -1.000I_1,
    \end{aligned}\right.
\end{equation}
where the coefficients are rounded to three decimal places.
This suggested the potential usefulness of the concatenation
of data stemming from different time series.

\begin{figure}[t]
\begin{center}
\includegraphics[width=0.85\linewidth]{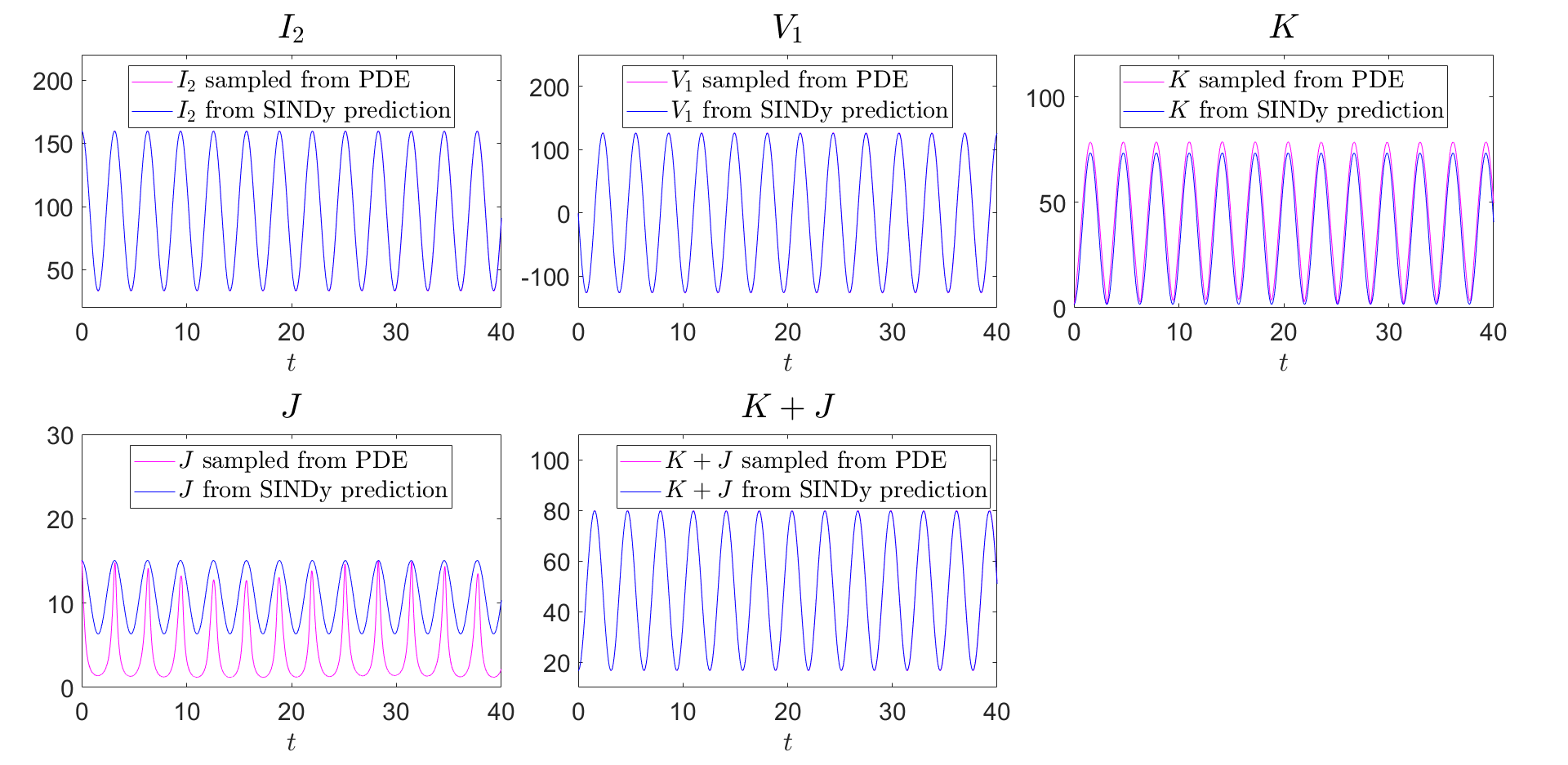}
\end{center}
\caption{Comparison of the moment evolutions of $[I_2,V_1,K,J,K+J]$ between SINDy and the ground truth. The training occurs for SINDy only on the selected moments $\bfx=[I_2,V_1,K,J]$, where a closure does not exist, using a linear library $\mathbf{\Theta}_{\deg=1}(\bfx)$. Interestingly, SINDy captures the correct dynamics for $E=K+J$, although not so the individual ones of $K$ and $J$. Reproduced from~\cite{suyang}. CC BY 4.0.}\label{fig:ex_3_sindy_lib_1}
\end{figure}

Another example considered in the work of~\cite{suyang} involved
the case where the nonlinearity $g(\rho, t) = g(\rho)$ is time-independent and given by $g(\rho) = g_0\rho^{2}$, where $g_0\in \R$ is a constant. Interestingly, in this case, the dynamics of the
moments $I_2$, $V_1$, $K$ and $J$ is not closed, yet it becomes closed under a suitable coordinate transformation, namely $E = K+J$:
  \begin{align}
    \label{eq:third_ME}
    \left\{
    \begin{aligned}
        \frac{dI_2}{dt} &= V_1,\\
        \frac{dV_1}{dt} &= 4E - 2I_2,\\
        \frac{dE}{dt} &= -\frac{1}{2}V_1,
    \end{aligned}\right.
    \end{align}
In this case, too, a diverse set of localized single and multihump
initial data was evolved and the corresponding data was
concatenated for the selected moments  \(\bfx = [I_2, V_1, K, J]\),
to examine the outcome of SINDy in this setting. When trying 
a polynomial {\it linear} library to discover the equations for
these moments, the following system was obtained:
\begin{align}
\label{eq:ex_3_sindy_raw_lib_1}
\left\{
\begin{aligned}
    \frac{dI_2}{dt} &= 1.000V_1,\\
    \frac{dV_1}{dt} &= -2.000I_2 + 4.000K + 3.998J,\\
    \frac{dK}{dt} &= -0.569V_1,\\
    \frac{dJ}{dt} &= 0.069V_1.
\end{aligned}\right.
\end{align}
When simulating this system and comparing it with the ground
truth, it was found that it does not capture the correct dynamics
for $K$ and $J$ for which there is no closed system;
see, in particular, figure~\ref{fig:ex_3_sindy_lib_1}.

However, when the ODE for $K$ was added to that for $J$, the
resulting system is found to be:
\begin{align}
 \label{eq:ex_3_sindy_transformed_lib_1}
\left\{
\begin{aligned}
    \frac{dI_2}{dt} &= 1.000V_1,\\
    \frac{dV_1}{dt} &= -2.000I_2 + 4.000K + 3.998J,\\
    \frac{d\left(K+J\right)}{dt} &= -0.500V_1.
\end{aligned}\right.
\end{align}
which matches well the relevant ground truth.
This is also suitably reflected in the last panel
of figure~\ref{fig:ex_3_sindy_lib_1}, where the two
relevant time series are added, capturing very
satisfactorily the ground truth result. 

Some key additional findings of the above work
were that the potential use of quadratic libraries
could be problematic for the above case which only
closes for the quantity $K+J$ (but not individually
for $K$ or/and $J$). In particular, in such a setting
overfitting issues arose, leading to ODE models
which are proximal but not identical to the theoretically
expected ones. This type of issue also arose in earlier
studies such as those of~\cite{Champion2019} and
especially~\cite{kutz1}. Indeed, in such settings
where the selected library may be ``richer'' than the terms
anticipated, it often turns out to be the case that the
resulting ``discovered'' model will not be the one 
theoretically expected. This is because the ``wealth''
of available nonlinear dependent variable combinations
within the libraries may yield better data approximations.
However, the key question of how well such models generalize
is, in principle, unclear beyond the relevant training data.

Another vein that was pursued in this context is that of
{\it identification} of a coordinate transformation
in the form of $\widetilde{\bfy} = \widetilde{\mathbf{A}}^\top \bfx$,
where the transformation matrix $\mathbf{A} \in \R^{4 \times 3}$
(and is not unique). The set of $\mathbf{A}$ satisfying the constraint $\mathbf{A}^\top\mathbf{A}=\mathbf{I}_{3\times 3}$ is called the \textit{Stiefel manifold} \cite{edelman1998geometry,absil2008optimization}.
Thus, in~\cite{suyang}, a concurrent minimization seeking to 
identify $\mathbf{A}$, along with identifying the optimal, sparse
set of resulting DEs was pursued and
was found to accurately obtain the dynamics of \([I_2, V_1, K, J]\).

Finally, the method was also sought to be used in scenarios
where no closed-form moment equations were known to exist.
As a concrete example of this kind, the non-autonomous case
of the time-dependent nonlinearity was utilized
\begin{align}
\label{eq:unclosed_nonlinearity}
    g(\rho, t) = \left(\sin(t) + 2\right)\left|\rho\right|^{2},
\end{align}
for different initial conditions and the dynamics of the moments
 \(\bfx = [I_2, V_1, E]\), where $E = K+J$, was monitored for 
 all the relevant cases accordingly. The result was found to be
 quite encouraging both for cases of simpler, bounded, oscillatory
 dynamics, as well as for ones of apparent unstable growth. What 
 was seen in the relevant results of~\cite{suyang} was that the
 proximity between the ground truth and the SINDy-based outcomes
 extended well past the training time interval of $T=20$ and 
 the matching between the two was qualitatively (and even
 semi-quantitatively) very good, even up to times about 5 times
 as large.

\subsection{Use of SINDy to discover soliton effective particle dynamics}

In the above section, we saw how to leverage sparse methods
for the identification of nonlinear dynamics via ODEs
in the context of moments of the original distribution
in nonlinear Schr{\"o}dinger type models.
Another vein of derivation of effective ODEs that has
been extremely popular in the physical literature
has been the extraction of ordinary differential
Equations for the features of solitary waves (and
their interactions) via the so-called 
variational approximation~\cite{Malomed2002}.
In this class of methods, one approximates the profile
of the wave field via a solitary wave or a collection of
solitary waves, potentially adding them ---notably, in the
case of bright solitary waves--- (or, e.g. 
multiplying them in the case of dark solitons).

Typically, this effective solitary wave ``manifold'' is 
characterized  by the position of the wave
centre (and the corresponding canonically conjugate
variable of the velocity), but additionally other parameters may come
into play including the amplitude and the width of the
wave, or possibly factors involving its phase, such as
the so-called chirp~\cite{Malomed2002,CarreteroGonzalez2024}.
The relevant Ansatz (i.e. attempted wave description)
is inserted typically in the Lagrangian of the model
and then Euler--Lagrange equations are extracted for
the solitary wave parameters. In a Hamiltonian
system, these come in pairs of conjugate variables
and essentially represent a nonlinear, wave-dynamics 
motivated, low-dimensional projection of the infinite-dimensional
PDE dynamics on the ``soliton manifold''. Naturally, also,
such representations fail to capture radiation features.
The latter can result, e.g. from the potential 
non-integrability of the models and, possibly, from the
interaction between solitary waves or between one of them
and the (potentially) spatially heterogeneous landscape.
As long as the dynamics does not produce considerable
radiation (and the interactions are not dramatic enough
to drastically change the character of the solitary waves),
this approximation can provide a meaningful and interpretable
characterization of the soliton dynamics and interactions.

It is also worthwhile to add here that although the variational
method is inherently Hamiltonian in the vast majority of
its considerations, non-conservative generalizations thereof
can be formulated based, e.g. on the work of~\cite{Galley}.
Indeed, relevant applications to nonlinear wave systems
have been considered in~\cite{KevGalley} and have also been
used for the examination of the temporal dynamics of 
cavity solitons in the experimentally relevant example
of~\cite{RossiChandramouliCarreteroKevrekidis2024}.

It is in the above spirit of the variational approximation 
that the recent work of~\cite{YangChenZhuKevrekidis2025} is seeking to identify the sparse dynamics of solitary waves for both the
case of dark and that of bright solitary waves. We
examine here, for proof-of-principle purposes, the case
of the dark solitons which is structurally somewhat simpler
due to the adequacy of a description merely involving
the positions and associated velocities of the solitary
waves. Indeed, it is known from classical results 
on the variational approximation of dark solitons~\cite{KivsharKrolikowski1995} that they 
feature an exponential tail--tail interaction in one
spatial dimension. It is similarly known from the
central work of~\cite{Busch} (and its generalization
for arbitrary, slowly-varying potentials in~\cite{konotopita})
that dark (and grey) solitary waves are subject to a 
restoring force which amounts to one half of the gradient
of the external confining potential in which they move. 

The combination of the above two effects led to the 
characterization of the motion of dark solitons
and of their interactions in the context of
parabolic traps of the form
\begin{equation}\label{eq: Parabolic trap potential}
    V_{\text{MT}}\left(x\right) = \frac{1}{2}\Omega^{2}x^{2}.
\end{equation}
relevant to atomic Bose--Einstein 
condensates~\cite{becbook2}. Then, for a sequence
of dark solitary waves inside the trap, variational
approximations have been brought to bear in the work 
of~\cite{Coles_2010} to give rise to the effective model:
\begin{equation}\label{eq: dynamics when parabolic potential is in}
    \ddot \xi_{i} = 8\exp\left(2\left(\xi_{i-1} - \xi_{i}\right)\right) - 8\exp\left(2\left(\xi_{i} - \xi_{i+1}\right)\right) - \frac{\Omega^{2}}{2}\xi_{i}.
\end{equation}

In the work of~\cite{YangChenZhuKevrekidis2025}, this type
of variational prediction was sought to be verified 
firstly for a single dark solitary wave in a parabolic
trap and then for solitary waves both inside, as well
as without a trap and even for four such solitary waves
(again, with and without a trap). We briefly 
survey some representative results from this
effort. In a parabolic trap of frequency 
$\Omega = 0.025$, using the time series of
the positions and velocities of the dark soliton
extracted from the PDE simulation, SINDy was
used to predict the relevant motion from a library
using monomials (up to 10th order). The relevant result
was:
\begin{equation}\label{eq: dynamics of a single DS}
    \ddot\xi = -0.00031287\xi.
\end{equation}
which is very proximal to the theoretically predicted
prefactor of the linear term (which should be
$\Omega^2/2$ in accordance to~\cite{Busch,konotopita}).
Here, SINDy with control via the so-called FROLS
(i.e. Forward Regression Orthogonal Least Squares) optimizer which is a greedy method, has been used to extract the relevant result.

\begin{figure}[t!]
    \centering
    \includegraphics[width=1\linewidth]{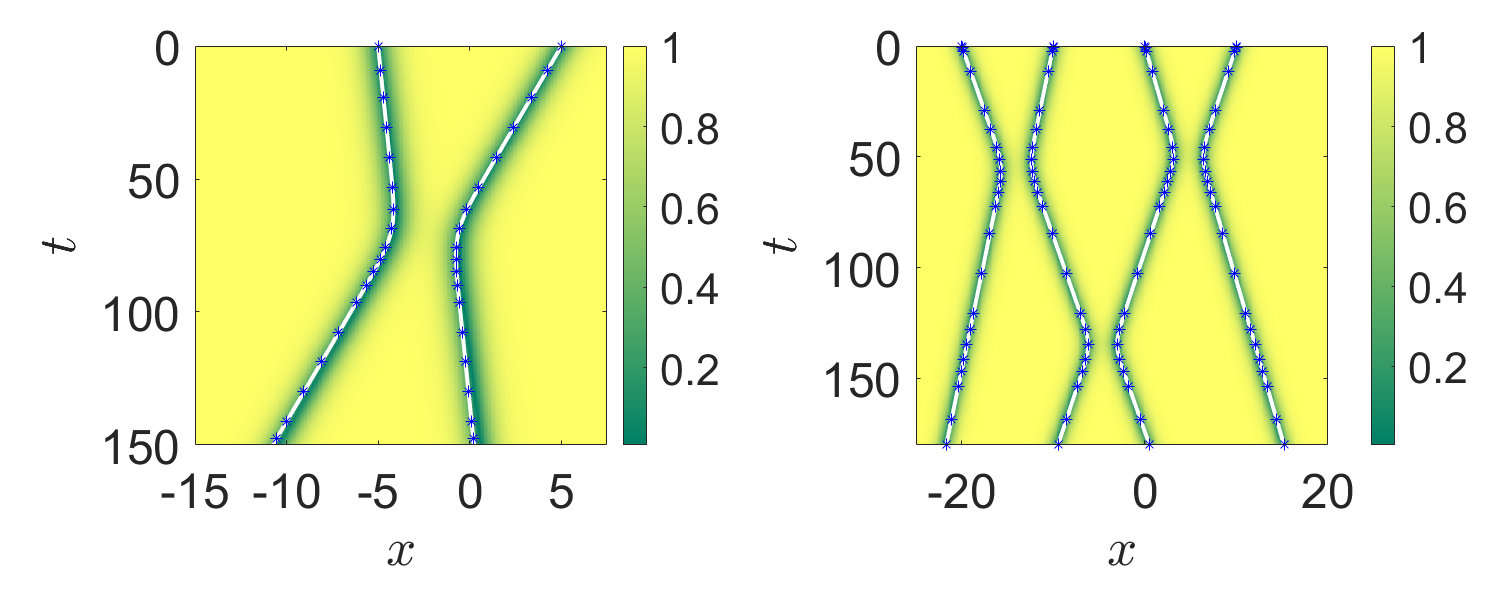}
    \caption{The spatio-temporal dark-soliton interaction dynamics in the absence of a parabolic trap: two dark solitary waves can be discerned on the left, and four on the right, interacting once before they indefinitely separate. The underlying contour map represents the PDE, while white solid curves and blue asterisks refer, respectively, to the theoretical prediction of the soliton positions based on the variational approximation (for $\Omega=0$) and the SINDy prediction, respectively. Reproduced from~\cite{YangChenZhuKevrekidis2025}. CC BY 4.0.}
    \label{fig:dark soliton positions comparison}
\end{figure}

On the other hand, for the case of two dark solitons in the 
presence of the trap, additionally, an exponential term
associated with the soliton interaction, as well as monomials
up to 10th order were used in the library and the relevant
SINDy prediction was:
\begin{equation}\label{eq: SINDy prediction for 2 ds with parabolic trap}
    \begin{aligned}
        \ddot \xi_1 &= -7.757\exp\left(2\left(\xi_1 - \xi_2\right)\right) - 0.00031635\xi_1,\\
        \ddot \xi_2 &= 7.803\exp\left(2\left(\xi_1 - \xi_2\right)\right) - 0.00031674\xi_2.
    \end{aligned} 
\end{equation}
The same approach was also attempted for the case of
4 solitary waves in which case the SINDy prediction
was found to be:
\begin{equation}\label{eq: dynamics for 4 ds with parabolic trap}
    \begin{aligned}
        \ddot \xi_1 &= -7.950\exp\left(2\left(\xi_1 -\xi_2\right)\right) - 0.00031272\xi_1,\\
        \ddot \xi_2 &= 8.420\exp\left(2\left(\xi_1 - \xi_2\right)\right) - 7.506\exp\left(2\left(\xi_2 - \xi_3\right)\right),\\
        \ddot \xi_3 &= 7.522\exp\left(2\left(\xi_2 - \xi_3\right)\right) - 8.406\exp\left(2\left(\xi_3  -\xi_4\right)\right),\\
        \ddot \xi_4 &= 7.900\exp\left(2\left(\xi_3 - \xi_4\right)\right) - 0.00031206\xi_4.
    \end{aligned}
\end{equation}
In all of these cases, SINDy can be seen to be quite
successful towards retrieving the dark soliton
dynamics. A pictorial representation of such outcomes
in the context of two and four dark solitary waves without  
a trap is given in figure~\ref{fig:dark soliton positions comparison}.

Nevertheless, it is interesting to point out that
while the data-driven variational approximation 
yields results quite proximal to the expected
ones, there are some deviations. These stem for
instance, from the fact that the force prefactors
pairwise between the solitons are not found to
be equal and opposite. Here, a hard-wired implementation
of Newton's 3rd law (regarding action and equal-opposite
reaction) is relevant towards a more physically meaningful result,
as was subsequently shown in~\cite{YangChenZhuKevrekidis2025}. Moreover, one can see that
a greedy method may select to weigh more on the exponential
dynamics rather than introduce a parabolic trap term
in the intermediate solitary waves such as $\xi_2$ and
$\xi_3$ (as is, in principle, physically relevant). There are numerous such features worth considering.
Perhaps, even more notably, in the study of~\cite{YangChenZhuKevrekidis2025}, it was found to be
considerably more difficult to capture the 
dynamics of bright solitons, especially in light of the
numerous degrees of freedom (potentially 4 or 6) that each
solitary wave might bear. This suggests the perhaps
anticipated feature that if the regression within the 
``soliton manifold'' is performed at the level of a
well-physically-informed and sufficiently low dimensional
manifold, there is a very good chance of extracting
meaningful, interpretable results. For high dimensional
manifolds, potentially encompassing multiple features
at different scales (e.g. the individual soliton motion 
and their exponential or modulated exponential tail--tail
interaction), it becomes considerably harder to 
accurately retrieve, and more importantly discover
in cases where they are unknown, the appropriate
solitary wave dynamical and interaction equations.

{
\section{Autoencoder-based model reductions}
Reduced-order modeling (ROM) for nonlinear waves often begins with the observation that many wavefields concentrate near low-dimensional sets: coherent-structure manifolds, weakly interacting multi-soliton families, modulated wave trains, or attractors of dissipative wave equations.  Classical ROM approaches (POD/Galerkin~\cite{sirovich1987turbulence,Holmes_Lumley_Berkooz_1996}, balanced truncation~\cite{moore2003principal}, inertial manifolds~\cite{foias1988inertial}, collective coordinates) typically impose \emph{linear} trial subspaces by representing the solution as a linear combination of a fixed set of global basis functions learned from data or derived from the governing equations. While effective in many regimes, such linear approximations can become inefficient when the solution manifold is strongly curved, or when the Kolmogorov $n$-width---defined as the smallest achievable worst-case error when approximating the solution set by an $n$-dimensional linear subspace---
decays slowly, motivating the use of nonlinear reduced representations~\cite{rowley2000reconstruction,ohlberger2013nonlinear,Peherstorfer2022breaking,hesthaven2026nonlinear}.

Autoencoders offer a data-driven approach to learning \textit{nonlinear} coordinate charts and trial manifolds, by mapping high-dimensional states to a low-dimensional latent space and reconstructing them in the original ambient space~\cite{hinton2006reducing,7799153,lee2020manifold,fukami2020convolutional,simpson2021machine,agostini2020exploration,murata2020nonlinear,eivazi2020deep,maulik2021reduced,BUCHFINK2024134299,KIM2022110841,GRUBER2022114764}. VAEs extend this framework by equipping the latent space with a probabilistic structure, which promotes regularity, enhances robustness to noise, and enables sampling and interpolation via a generative map~\cite{akkari2022bayesian,simpson2024vprom}. In regimes where the solution manifold is strongly curved, or where the Kolmogorov $n$-width decays slowly for linear subspaces, such nonlinear manifold-based ROMs can be substantially more expressive. In what follows, we explain how autoencoders (through an offline snapshot training stage and an online reduced dynamics identification stage) can be applied to the reduced-order modeling of nonlinear wave equations.
{

An autoencoder consists of an encoder $E_\theta:\mathbb{R}^{N}\to\mathbb{R}^{r}$ and a decoder $D_\theta:\mathbb{R}^{r}\to\mathbb{R}^{N}$, with the latent dimension $r\ll N$. The encoder $E_\theta$ maps the high-dimensional discretized wave field to a low-dimensional latent representation, while the decoder lifts this latent representation back to the ambient space. During the offline snapshot training stage, the autoencoder is trained to minimize the reconstruction error over snapshot data $\{u_k\}_{k=1}^{M}$:
\begin{equation}
\min_\theta \;\sum_{k=1}^{M}\big\|u_k - D_\theta(E_\theta(u_k))\big\|^2
\;+\;\text{(regularization/constraints)}.
\label{eq:ae_recon_loss}
\end{equation}
When successfully trained, the decoder defines a nonlinear trial manifold
\[
\mathcal{M}_r \;=\;\{D_\theta(z): z\in\mathbb{R}^{r}\}
\]
that approximates the solution set.  For nonlinear waves, $\mathcal{M}_r$ can implicitly encode translations, phase shifts, and amplitude modulations if the training set includes those variations; however, this also means that the learned coordinates may entangle symmetry directions with intrinsic dynamics unless one explicitly builds equivariance or performs symmetry reduction beforehand.

Once an autoencoder is trained offline, reduced dynamics on the nonlinear trial manifold can then be identified online. When the governing equation $\dot{u}=F(u)$ is available (the intrusive ROM setting), the dynamics on $\mathcal{M}_r$ are obtained by projecting the vector field onto the decoder tangent space~\cite{lee2020manifold}.  Specifically, writing $u\approx D_\theta(z)$ and denoting the Jacobian $J_\theta(z)=\partial D_\theta(z)/\partial z\in\mathbb{R}^{N\times r}$, we seek $\dot{z}$ so that $J_\theta(z)\dot{z}\approx F(D_\theta(z))$.
A natural choice is the least-squares tangent projection:
\begin{equation}
\dot{z}
\;=\;
\arg\min_{w\in\mathbb{R}^{r}}\;
\big\|J_\theta(z)w - F(D_\theta(z))\big\|^2
\;=\;
J_\theta(z)^{+}\,F(D_\theta(z)),
\label{eq:manifold_galerkin}
\end{equation}
where $J_\theta(z)^{+}$ denotes a (possibly regularized) pseudoinverse.  This is the nonlinear-manifold analogue of Galerkin projection, and it reduces to standard POD--Galerkin when $D_\theta(z)=\Psi z$ is linear.

For time-discretized systems, one can similarly define a manifold least-squares Petrov--Galerkin (LSPG) update by minimizing the time-step residual in the full space over the reduced coordinates \cite{lee2020manifold}.  For example, with implicit Euler and step size $\Delta t$, define the residual
\[
R(z_{n+1};z_n)
=
D_\theta(z_{n+1})-D_\theta(z_n)-\Delta t\,F(D_\theta(z_{n+1})),
\]
and compute
\begin{equation}
z_{n+1}
\;=\;
\arg\min_{z\in\mathbb{R}^{r}} \;\|R(z;z_n)\|^2.
\label{eq:manifold_lspg}
\end{equation}
These intrusive constructions are attractive for nonlinear waves when $F$ is known and one wants a \emph{structure-aware} reduced integrator, but they require differentiating through $D_\theta$ and evaluating $F$, which can be expensive for PDE discretizations.

In non-intrusive settings (or when one aims to learn effective dynamics directly from data), one instead learns a reduced evolution map on latent coordinates, e.g.\ $z_{k+1}=G_\phi(z_k)$ or $\dot{z}=g_\phi(z)$, and trains with a combined reconstruction/prediction objective:
\begin{equation}
\min_{\theta,\phi}\;
\sum_{k}\|u_k-D_\theta(E_\theta(u_k))\|^2
\;+\;
\alpha \sum_{k}\|E_\theta(u_{k+1})-G_\phi(E_\theta(u_k))\|^2,
\label{eq:ae_latent_dynamics_loss}
\end{equation}
with $\alpha$ setting the accuracy trade-off.  Sparse-model variants replace $G_\phi$ by a parsimonious library model (e.g.\ SINDy) to improve interpretability and long-time stability.  A representative example is the SINDy-autoencoder approach \cite{Champion2019}, which simultaneously learns coordinates and sparse governing equations in latent space by penalizing the mismatch between $\dot{z}$ and a sparse library expansion.

\subsection{SHallow REcurrent Decoder architectures (SHRED)-ROM for partially observed wave fields}\label{sec:autoencoders_shredrom}

Autoencoder ROMs are typically trained and evaluated using (at least some) full-state snapshots $u_k$. In many nonlinear-wave experiments, however, only sparse sensor measurements are available during deployment:
\begin{equation}
s_k = C u_k \in \mathbb{R}^{N_s},\qquad N_s\ll N,
\label{eq:sensors}
\end{equation}
where $C$ is a sampling operator that selects point values or local averages.  This creates a fundamental challenge: even if an encoder $E_\theta$ is available, it cannot be applied online because the full state $u_k$ is not observed. Instead, one must infer reduced coordinates, and possibly full-state reconstructions, directly from sensor time series.

To address this partial-observation setting, encoder--decoder pipelines must be adapted to operate on sparse measurements. SHRED and their reduced-order modeling extension SHRED-ROM provide a computationally efficient framework for learning such mappings from sensor data to latent states and reconstructed fields~\cite{williams2024shred,tomasetto2025shredrom}. The key idea is to incorporate a time-delay embedding by feeding a \textit{window} of past sensor measurements $\left(s_{k-L}, s_{k-L+1},\ldots, s_k\right)$---rather than only the current measurement $s_k$, as in standard autoencoder formulations---into a recurrent network.

Let the sensor history over a window of length $L$ be
\[
\mathbf{s}_{k-L:k} = (s_{k-L}, s_{k-L+1}, \ldots, s_k),
\]
and let $E_T$ denote a recurrent temporal encoder (typically an LSTM) that maps this window to a latent state $h_k \in \mathbb{R}^r$:
\begin{equation}
h_k = E_T(\mathbf{s}_{k-L:k}).
\label{eq:shred_temporal}
\end{equation}
A shallow decoder $D_U$ then maps the latent state $h_k$ to a reconstruction of the full state:
\begin{equation}
\widehat{u}_k = D_U(h_k).
\label{eq:shred_decode}
\end{equation}
Here, the subscript $T$ in $E_T$ emphasizes that the encoder processes a temporal window of sensor measurements, while the subscript $U$ in $D_U$ indicates that the decoder reconstructs the full state variable $u$, rather than the sensor data $s$.

In SHRED, the recurrence provides robustness to sparse/poorly placed sensors by exploiting temporal information: if the instantaneous map $s_k\mapsto u_k$ is ill-conditioned, the trajectory $\mathbf{s}_{k-L:k}$ can still contain enough information to infer the current state when the system is observable through the chosen sensors.  From a nonlinear-wave viewpoint, this is extremely natural: dispersive PDEs transport information, so the time history at a probe often encodes incoming wave content even when the instantaneous value is ambiguous.

A common efficiency upgrade is to reconstruct not the full state but a low-rank representation.  Suppose POD yields a basis $\Psi\in\mathbb{R}^{N\times r}$ such that $u_k\approx \Psi a_k$ for coefficients $a_k\in\mathbb{R}^{r}$.  One can train SHRED to predict the coefficients:
\begin{equation}
\widehat{a}_k = D_U(h_k), \qquad \widehat{u}_k = \Psi \widehat{a}_k.
\label{eq:shred_pod}
\end{equation}
This ``decoder-only'' structure, which functions as a recurrent compression of sensor time series followed by a shallow map into coefficients/state, is attractive for nonlinear waves because it keeps the model small and makes training feasible even when $N$ is very large (high-resolution fields, multi-channel measurements, videos).

SHRED-ROM extends the SHRED sensing architecture to a reduced-order modeling setting where the goal is to reconstruct and emulate high-dimensional spatio-temporal trajectories across multiple scenarios (e.g.,\ varying physical parameters, geometries, or forcing), often without providing the parameter values explicitly to the network \cite{tomasetto2025shredrom}.  The central reconstruction map can be written in a compact parametric form.  Denote by $u_k^\mu$ the state at time $t_k$ for scenario/parameter $\mu$, and by $s_k^\mu$ the corresponding sparse sensors.  SHRED-ROM combines a recurrent temporal encoder $E_T$ with a shallow decoder $D_U$ and optionally a basis expansion $\Psi$ so that
\begin{equation}
\widehat{u}_k^\mu
=
\Psi\,\widehat{a}_k^\mu
=
\Psi\, D_U\!\Big( E_T(s_{k-L}^\mu,\ldots,s_k^\mu)\Big).
\label{eq:shredrom_core}
\end{equation}
The use of $\Psi$ can be data-driven (POD) or physics-driven (handcrafted basis expansions), and is motivated by compressive training: instead of learning a heavy end-to-end map into $\mathbb{R}^N$, the network learns to predict only the reduced coefficients \cite{tomasetto2025shredrom}. In addition, because the network ingests trajectories of sensors rather than instantaneous measurements, it can accommodate fixed or mobile sensors and may be less sensitive to sensor placement than snapshot-based approaches, provided the sensing configuration yields sufficient observability over time \cite{tomasetto2025shredrom,williams2024shred}.

From the nonlinear-wave perspective, \eqref{eq:shredrom_core} is compelling for three concrete reasons. First, it targets the \emph{actual experimental bottleneck}: sparse measurements.  Many nonlinear-wave platforms (water-wave tanks, optical fibres, plasmas, geophysical waves, BECs) provide only partial observations.  A sensor-driven ROM that outputs a coherent field estimate enables downstream tasks that are otherwise inaccessible: estimating conserved quantities from incomplete data, classifying regimes (e.g.,\ soliton gas vs wave turbulence), and initializing short-term forecasts.

Second, the architecture naturally supports \emph{multi-scenario generalization}.  In wave problems, varying dispersion coefficients, nonlinearity strengths, boundary conditions, or forcing can produce qualitatively different wavefields.  Classical POD bases can struggle because they are scenario-specific; SHRED-ROM is designed to learn across scenarios and reconstruct new ones without explicit parameter input, leveraging the sensor trajectory as an implicit signature of the scenario \cite{tomasetto2025shredrom}.

Third, SHRED-ROM cleanly separates (i) compressive representation (via $\Psi$) and (ii) inference from sensor trajectories (via $E_T,D_U$).  This separation makes it straightforward to incorporate domain structure when available.  For example, one may choose $\Psi$ to respect symmetries (Fourier modes for periodic domains, localized wavelets for solitary waves) or to encode known invariants (e.g.,\ mean removal, mass/energy normalization).  Likewise, training losses can be augmented with physics constraints without changing the core architecture: if one reconstructs $\widehat{u}^\mu(x,t)$, one can penalize residuals of known PDE operators, mismatch of conserved quantities, or violations of symmetry constraints.  For nonlinear waves, this is a practical path to pushing beyond ``rediscovery from synthetic data'' toward robust field inference in partially observed regimes.

Of course, limitations remain.  Any sensor-to-field map depends on the training distribution: if the wave dynamics encountered online differ substantially (new coherent structures, different noise statistics, unseen boundary effects), performance can degrade.  This is not unique to SHRED-ROM; it is a fundamental challenge for learned observers.  Still, the combination of (a) explicitly low-dimensional outputs (coefficients $a_k$), (b) time-history conditioning, and (c) the ability to incorporate physically motivated bases and losses makes SHRED-ROM a strong candidate methodology for nonlinear-wave problems where sparse sensing is the primary constraint.
}}

\section{Learning structure from data}
\label{sec5}

\subsection{Learning Hamiltonians from data}
\label{sec:hnn}

Data‑driven modeling of dynamical systems has progressed rapidly over the past decade, yielding methods that infer the governing equations directly from time‑series measurements. As discussed above, SINDy recovers parsimonious ODEs or PDEs by selecting a few basis functions that best fit observed time-derivatives~\cite{brunton2016discovering,rudy2017data}. Symbolic‑regression approaches such as Eureqa~\cite{schmidt2009distilling} likewise search the space of closed‑form expressions. Deep‑learning techniques broadened the toolkit: PDE‑Net learns convolutional filters that approximate unknown spatial operators~\cite{long2018pde}; Koopman autoencoders embed nonlinear flows in linearly evolving latent coordinates~\cite{lusch2018deep}; Neural ODEs estimate continuous-time dynamics with adjoint backpropagation~\cite{chen2018neural}; and PINNs encode differential-equation residuals as soft constraints during training~\cite{raissi2019physics}. While these approaches can capture complex dynamics when ample data are available, they generally do not enforce energy conservation or symplectic structure, which are crucial for accurately modeling Hamiltonian systems. This limitation has motivated the incorporation of Hamiltonian structure as a physical prior, enhancing model generalization, particularly in low-data regimes.

Consider a \textit{Hamiltonian system} of $d$ degrees of freedom,
\begin{equation}
     \frac{d\mathbf{x}}{dt} = \mathbf{f}(\mathbf{x}), \quad \mathbf{f}(\mathbf{x}) = J(\mathbf{x})\nabla H(\mathbf{x}), \quad \forall \mathbf{x}\in D \subset \mathbb{R}^{2d},
     \label{eq:d-hamiltonian}
\end{equation}
where $H: D \to \mathbb{R}$ is the Hamiltonian, and $J(\mathbf{x}) \in \mathbb{R}^{2d \times 2d}$ is antisymmetric. In particular, under canonical coordinates $\mathbf{x} = (\mathbf{q}, \mathbf{p})$, with positions $\mathbf{q}$ and momenta $\mathbf{p}$, the dynamics simplifies to:
\begin{align}
\label{eq:d-hamiltonian-canonical}
    J(\mathbf{x})\equiv
    \begin{bmatrix}
        0 & I_d\\
        -I_d & 0
    \end{bmatrix},\quad \text{and}~
    \left\{
    \begin{aligned}
        &\frac{\mathrm{d}\bq}{\mathrm{d}t} = \nabla_{\bp} H,\\
        &\frac{\mathrm{d}\bp}{\mathrm{d}t} = -\nabla_{\bq} H.
    \end{aligned}\right.
\end{align}

Instead of directly approximating the vector field $\mathbf{f}(\bfx)$ in Eq.~\eqref{eq:d-hamiltonian}, Hamiltonian Neural Networks (HNNs)~\cite{greydanus2019hamiltonianneuralnetworks,bertalan2019learning}  leverage prior knowledge of Hamiltonian structure and aim to learn the Hamiltonian $H(\mathbf{x})$ itself. 
The corresponding vector field $\mathbf{f}(\mathbf{x}) = J(\mathbf{x}) \nabla H(\mathbf{x})$ can then be obtained via AD. More specifically, consider a dataset of uniformly sampled trajectories in canonical coordinates:
\begin{align}
\label{eq:data_D}
    \mathcal{D} = \left\{\bfx^{(i,j)} = \left(\bq^{(i,j)}, \bp^{(i,j)}\right) : 1\le i \le I, ~1\le j\le J\right\},
\end{align}
with a small sampling interval $\Delta t$. The time derivatives can be approximated via finite differences:
\begin{align}
\label{eq:derivate_data}
    \dot{\mathcal{D}} = \left\{\dot{\bfx}^{(i,j)} = \left(\dot{\bq}^{(i,j)}, \dot{\bp}^{(i,j)}\right) : 1\le i \le I, ~1\le j\le J\right\}.
\end{align}
Greydanus et al.~\cite{greydanus2019hamiltonianneuralnetworks} parameterize the Hamiltonian as a neural network $H_{\bm{\theta}}(\mathbf{q}, \mathbf{p})$, trained to minimize the mean squared error between the symplectic gradient $(\nabla_{\bp} H_{\bmtheta}, -\nabla_{\bq} H_{\bmtheta})$ and the time derivatives $(\dot{\bq}, \dot{\bp})$ over the dataset $\mathcal{D}$:
\begin{align}
\label{eq:l2_hnn}
        \mathcal{L}_{\text{HNN}}= \frac{1}{IJ}\sum_{i,j}\Big\| \nabla_{\bp} H_{\bmtheta}\left(\bq^{(i,j)}, \bp^{(i,j)}\right) - \dot{\bq}^{(i,j)} \Big\|^2 +\Big\|  \nabla_{\bq}H_{\bmtheta}\left(\bq^{(i,j)}, \bp^{(i,j)}\right) + \dot{\bp}^{(i,j)}\Big\|^2.
\end{align}

However, accurate estimation of the time derivatives in~\eqref{eq:derivate_data} requires trajectories sampled at small time intervals $\Delta t$. In the presence of observation noise, finite difference approximations of time derivatives become extremely sensitive, rendering pointwise $l^2$ loss~\eqref{eq:l2_hnn} between the vector field and the symplectic gradient impractical in realistic settings. Consequently, many follow-up works instead use an ODE solver (based on the symplectic gradient of the parameterized Hamiltonian) to generate predicted trajectories, and define the loss based on the mismatch between predicted and observed (ground-truth) trajectories. To train the network, gradients can be computed either by backpropagating through the numerical integrator or via constant-memory adjoint methods as in Neural ODEs~\cite{chen2018neural}. Early approaches employed generic integrators such as the Runge–Kutta method (RK4)~\cite{Zhong2020Symplectic}, while subsequent works explored symplectic integrators to better preserve the phase space structure. These include leapfrog~\cite{Chen2020Symplectic}, symplectic Euler and implicit midpoint rule~\cite{david2023symplectic}, variational integrators~\cite{saemundsson2020variational}, and higher-order symplectic schemes~\cite{dipietro2020sparse,xiong2021nonseparable}.

Beyond incremental refinements to HNNs, a parallel line of work seeks to bypass explicit Hamiltonian discovery and numerical time stepping altogether. Instead, it aims to approximate the time-$t$ flow map $\bfx \mapsto \phi_t(\bfx)$, which transports an initial state $\bfx$ to its position after time $t$. For any Hamiltonian system described by~\eqref{eq:d-hamiltonian-canonical}, this flow is symplectic---meaning it preserves the canonical two-form---and therefore obeys
\begin{align}
    \label{eq:symplectic}
    \left(\nabla_{\bfx}\phi_t\right)^\top J \nabla_{\bfx}\phi_t = J
\end{align}
Leveraging this geometric constraint, Jin et al.~\cite{jin2020sympnets} constructed ``SympNets,'' deep networks whose layers are composed of analytically symplectic building blocks, and proved that such architectures can uniformly approximate any smooth symplectic map. Chen and Tao~\cite{chen2021data} pursued a complementary route: they parameterized a generating function whose gradient defines $\phi_t$ implicitly, ensuring that~\eqref{eq:symplectic} holds by construction. Empirically, both approaches achieve long-horizon predictions whose global error grows only linearly with $t$, in stark contrast to the exponential error accumulation observed when one first learns the Hamiltonian and then integrates it with ODE solvers.

Other studies extend Hamiltonian learning by addressing generalized systems under noncanonical coordinates~\eqref{eq:d-hamiltonian}~\cite{course2020weak,jin2022learning}; employing (variational) autoencoders that embed high‑dimensional observations into a latent symplectic manifold governed by a learned Hamiltonian~\cite{Toth2020Hamiltonian,Chen2020Symplectic,saemundsson2020variational}; integrating symplectic constraints into message‑passing layers through Hamiltonian GNNs~\cite{sanchez2019hamiltonian}; and developing adaptive HNNs that, once trained on trajectories from only a few bifurcation‑parameter values, accurately predict dynamics at unseen parameters~\cite{han2021adaptable}.

\subsection{Methods using symplectic transformations and the discovery of action-angle
variables}

A cornerstone of integrable Hamiltonian dynamics is the existence of action–angle coordinates $(\mathbf{I}, \bm{\varphi})$, in which the Hamiltonian depends only on the conserved actions $\mathbf{I}$, and the angle variables evolve linearly in time,
\begin{align}
    \dot{\bm{\varphi}} = \omega (\mathbf{I}), \quad \text{and}\quad \dot{\mathbf{I}} = 0,
\end{align}
as guaranteed by the Liouville–Arnold theorem~\cite{arnold1989mathematical}.

However, constructing explicit action–angle transformations from canonical coordinates $(\bq, \bp)$ is notoriously challenging for all but a few classical models. Recognizing this gap, Bondesan and Lamacraft~\cite{bondesan2019learning} proposed a neural-network-based method to learn such transformations directly from trajectory data by enforcing symplectic structure in the mapping.

Their framework is built around a \textit{symplectic normalizing flow}, parameterized as a sequence of symplectic invertible layers that ensure the learned map
\begin{align}
    T:(\bq, \bp)\mapsto \left(\hat{\bq}, \hat{\bp}\right) \equiv (\mathbf{I}, \bm{\varphi})
\end{align}
is a canonical symplectic embedding. Each layer preserves the Hamiltonian structure, enabling efficient learning of integrable transformations.

They demonstrated this approach on three paradigmatic integrable systems: the Kepler problem, Neumann model, and Calogero–Moser system. The training objective encourages trajectories in $(\mathbf{I}, \bm{\varphi})$-space to map to simple circular motion (constant actions, linear phase), reflecting the canonical action–angle flow. Once trained, the network maps observed $(\bq, \bp)$-trajectories into linear dynamics on tori, effectively identifying the action–angle coordinates.

Building on this idea, Daigavane et al.~\cite{daigavane2022learning} propose a related framework that leverages symplectic neural networks~\cite{jin2020sympnets} to learn action–angle representations from data. Their approach focuses on the complete dynamics of integrable systems, building a simulator in action–angle space that captures both the conserved quantities and the linear evolution of angle variables.

\subsection{Learning conserved quantities using deep learning}

In recent years, data‑driven approaches for discovering conservation laws and assessing integrability have advanced rapidly, fuelled by ML techniques. Broadly, existing algorithms fall into two classes: those that incorporate explicit knowledge of the governing DEs and those that operate solely based on observed trajectories.

For DE‑informed methods, recent contributions include the approaches described in~\cite{Liu_2022, liu2024interpretable, zhu2023machine}. Liu et al.~\cite{Liu_2022} proposed a regularized loss function to train a family of conservation laws simultaneously, with a penalty term that encourages pointwise orthogonality of their gradients to promote functional independence. Building on this idea, Liu et al.~\cite{liu2024interpretable} further combined sparse regression with a prescribed dictionary of basis functions to enhance interpretability. The neural deflation framework~\cite{zhu2023machine} adopts an iterative perspective: each newly identified integral is learned with a deflated loss that enforces functional independence with respect to those already discovered, yielding complete families of functionally independent conservation laws across a range of test problems. It is worth noting that unlike the orthogonality-based penalty in~\cite{Liu_2022}, which is sufficient but not necessary for functional independence, the deflation approach enforces independence directly and is provably consistent in the infinite-sample limit: a function is a new conservation law \textit{if and only if} the deflated loss achieves zero, a guarantee not provided by the earlier method. 

Trajectory‑based methods address the more practical scenario in which the governing equations are unavailable. Siamese neural networks~\cite{PhysRevResearch.2.033499} extract a single invariant by comparing pairs of states, whereas the approach of Ha and Jeong~\cite{ha2021discovering} leverages grouped data sampled from level sets to learn multiple  conservation laws. Model‑agnostic techniques such as the method of Arora et al.~\cite{arora2023model} incorporate prior information about the number of invariants, and manifold‑learning formulations~\cite{teg1, lu2023discovering} aim to uncover the geometry of invariant manifolds directly in state space.

While each of these methodologies has its own strengths and limitations, we now focus on the neural deflation method~\cite{zhu2023machine}, which has consistently demonstrated the ability to recover complete sets of functionally independent, Poisson-commuting conservation laws in Hamiltonian systems.

Consider the $d$-dimensional Hamiltonian system~\eqref{eq:d-hamiltonian}. Let $F$ and $G$ be two smooth functions defined on the phase space $D$. Their \textit{Poisson bracket}, $\{F, G\}:D\to\R$, is given by
\begin{equation}
     \{F, G \}(\mathbf{x}) \coloneqq \nabla F(\mathbf{x})^T J(\mathbf{x}) \nabla G(\mathbf{x}), \quad \forall \mathbf{x}\in D.
     \label{eq: poisson bracket}
\end{equation}
A continuously differentiable function $I: D \to \mathbb{R}$ is called a \textit{conservation law} of system~\eqref{eq:d-hamiltonian} if it remains constant along system trajectories. That is, for any solution $\mathbf{x}(t)$ to Eq.~\eqref{eq:d-hamiltonian}, we have
\begin{align}
    I(\mathbf{x}(t))\equiv I(\mathbf{x}(0)), \quad \forall t\geq 0.
\end{align}
It is straightforward to verify that $I$ is a conservation law if and only if its Poisson bracket with the Hamiltonian $H$ vanishes on $D$, i.e.,
\begin{equation}\label{eq:CQPBracekt}
     \{I, H \}(\mathbf{x}) = \nabla I(\mathbf{x}) \cdot \mathbf{f} (\mathbf{x}) = 0, \quad \forall \mathbf{x} \in D
\end{equation}

A collection $\{I_k:D\to\R \}_{k=1}^K$ of $K$ conservation laws of system~\eqref{eq:d-hamiltonian} is called \textit{functionally independent} if their gradients $\{ \nabla I_k(\mathbf{x}) \}_{k=1}^{K} $ are linearly
independent vectors in $\mathbb{R}^{2d}$ for almost every $\mathbf{x}\in D$. Intuitively, this means that none of the conserved quantities can be expressed as a (nonlinear) function of the others. Moreover, these integrals are said to be \textit{in involution} or \textit{Poisson commuting} if their pairwise Poisson brackets vanish, i.e., $\ \{ I_j,I_k \} = 0, \forall j \ne k$. For a Hamiltonian system with $d$ degrees of freedom, there can be at most $d$ functionally independent conservation laws in involution. When such a collection exists, the system is said to be \textit{completely integrable} in the sense of Liouville~\cite{arnold1989mathematical}.

The neural deflation framework~\cite{zhu2023machine} offers a principled, data-driven strategy for identifying a maximal set of functionally independent, Poisson-commuting conservation laws when the underlying dynamics is explicitly known. The central idea is to learn each conserved quantity in sequence, modifying the loss function at every step to enforce independence and Poisson-commutativity with those already identified. This strategy draws motivation from the idea of ``deflation'' in numerical PDEs, where previously found solutions are actively avoided to uncover new ones~\cite{patrick}.

The process begins by randomly sampling a training set \(\mathcal{T}\) and a validation set \(\mathcal{V}\) from the phase space \(D \subset \mathbb{R}^{2d}\). Each candidate conservation law is represented by a neural network $I_k(\mathbf{x}; \bm{\theta}_k)$ with parameters $\bm{\theta}_k$. The first such network, $I_1$, is trained by minimizing the following loss:
\begin{align}
\label{eq:l1}
    \mathcal{L}_1(\bm{\theta}_1; \mathcal{T}) \coloneqq \frac{1}{|\mathcal{T}|} \sum_{\mathbf{x}\in \mathcal{T}} \left| \widehat{\mathbf{f}}(\mathbf{x}) \cdot \widehat{\nabla I_1 }(\mathbf{x};\bm{\theta}_1) \right|^{2},    
\end{align}
where the hat notations, $\widehat{\mathbf{f}}$ and $\widehat{\nabla I_1}$, here denote the $l^2$-normalized versions of $\mathbf{f}$ and $\nabla I_1$, respectively. This normalization ensures the training process does not converge to trivial constant solutions and makes the loss independent of scaling.

To learn additional conserved quantities, the method proceeds inductively. Suppose $K-1$ such laws $\{I_k\}_{k=1}^{K-1}$ have been obtained. Then the $K$-th candidate $I_K(\mathbf{x}; \bm{\theta}_K)$ is learned by minimizing a new loss that combines three objectives: (i) it must be conserved under the dynamics, (ii) it must Poisson-commute with each previously learned law, and (iii) it must be functionally independent from the identified laws. The combined loss is given by:
\begin{align}
\label{eq:lK}
\mathcal{L}_K(\bm{\theta}_K; \mathcal{T}) \coloneqq \frac{1}{|\mathcal{T}|}\sum_{\mathbf{x}\in\mathcal{T}}\frac{\displaystyle
     \overbrace{\ell_{\text{conserv}}[\bm{\theta}_K; \mathbf{x}]}^{\text{conservation loss}} + \overbrace{\sum^{K-1}_{k=1} \ell_{\text{inv}}[\bm{\theta}^{*}_k, \bm{\theta}_K; \mathbf{x}]
     }^{\text{involution loss}}}{\displaystyle \underbrace{K \left| \ell_{\text{ind}}[\bm{\theta}_K|\bm{\theta}^{*}_1, \dots, \bm{\theta}^{*}_{K-1}; \mathbf{x}] \right|^{\alpha}}_{\text{independent loss}}},
\end{align}
where each term plays a specific role:
\begin{itemize}
    \item The \textit{conservation loss} enforces the defining property of conserved quantities:
    \begin{align}
        \ell_{\text{conserv}}[\bm{\theta}_K; \mathbf{x}] \coloneqq \left| \widehat{\mathbf{f}}(\mathbf{x}) \cdot \widehat{\nabla I_K }(\mathbf{x};\bm{\theta}_K) \right|^{2}.
    \end{align}
    \item  The \textit{involution loss} penalizes nonzero Poisson brackets with earlier conservation laws:
    \begin{align}
        \ell_{\text{inv}}[\bm{\theta}^{*}_k, \bm{\theta}_K; \mathbf{x}] \coloneqq \left| \{ I_{k} (\cdot ; \bm{\theta}^{*}_k),  I_{K} (\cdot ; \bm{\theta}_K) \}(\mathbf{x}) \right|^{2}
    \end{align}
    \item The \textit{independent loss} in the denominator is given by:
    \begin{align}
    \label{eq:ind_loss}
        \ell_{\text{ind}}[\bm{\theta}_K| \bm{\theta}^{*}_1, \cdots,\bm{\theta}^{*}_{K-1};\mathbf{x}] \coloneqq \left\|\text{Proj}_{\text{span}\{\widehat{\nabla I_k}(\mathbf{x}; \bm{\theta}^{*}_k)  \}_{k\in[K-1]}^{\perp}} \widehat{\nabla I_K}(\mathbf{x}; \bm{\theta}_K)\right\|^{2},
    \end{align}
    where \(\text{Proj}_{\text{span}\{\widehat{\nabla I_k}(\mathbf{x}; \bm{\theta}^{*}_k) \}_{k\in[K-1]}^{\perp}} \widehat{\nabla I_K}(\mathbf{x}; \bm{\theta}_K)\) denotes the projection of \(\widehat{\nabla I_K}(\mathbf{x}; \bm{\theta}_K)\) onto the orthogonal complement of the subspace spanned by \(\{ \widehat{\nabla I_k}(\mathbf{x}; \bm{\theta}^{*}_k)\}_{k\in[K-1]}\) in $\R^{2d}$. This term introduces a singularity that penalizes any lack of independence between \(\widehat{\nabla I_K}(\mathbf{x}; \bm{\theta}_K)\) and the previously learned \(\{ \widehat{\nabla I_k}(\mathbf{x}; \bm{\theta}^{*}_k)\}_{k\in[K-1]}\). The hyperparameter $\alpha > 0$ controls how strongly this constraint is enforced.
\end{itemize}

This formulation can be shown to be consistent in the limit of infinite data. Specifically, if the earlier conservation laws exactly capture a true set of independent, commuting integrals, and if empirical averages are replaced by expectations under an absolutely continuous probability distribution, then the minimum of $\mathcal{L}_K$ is zero if and only if $I_K$ extends the set to $K$ such laws. A rigorous proof is given in~\cite{zhu2023machine}.

The algorithm continues adding new conservation laws until training one more leads to a significant rise in the loss on the validation set, indicating that no further independent commuting conserved quantities can be found. At that point, the process terminates with a maximal set of $d_0 = K - 1$ such laws.

\begin{figure}
  \centering
  \begin{minipage}{0.48\textwidth}
    \centering
    \includegraphics[width=\textwidth]{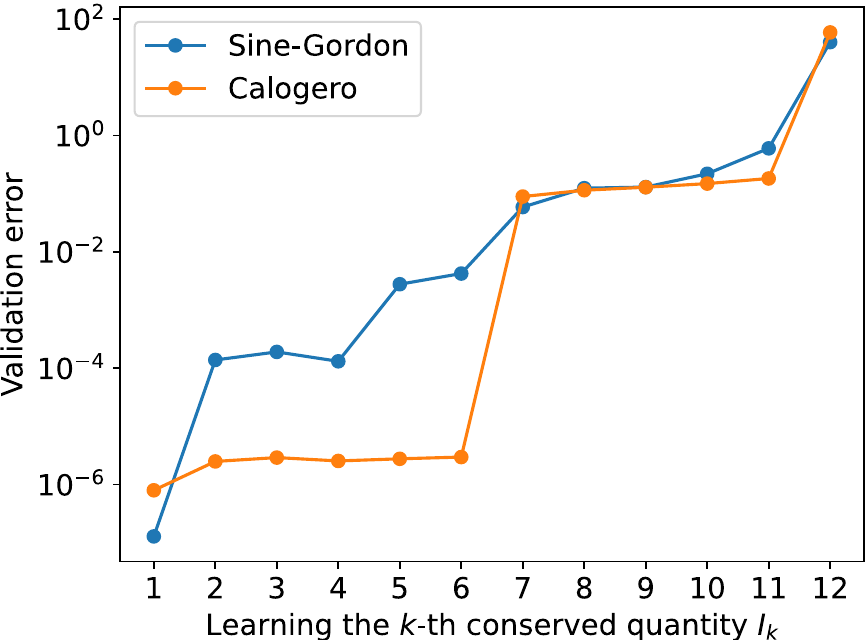}
    \caption*{(a) Number of lattice sites $N=6$}
    \label{fig:sg_calogero_6}
  \end{minipage}
  \hfill
  \begin{minipage}{0.48\textwidth}
    \centering
    \includegraphics[width=\textwidth]{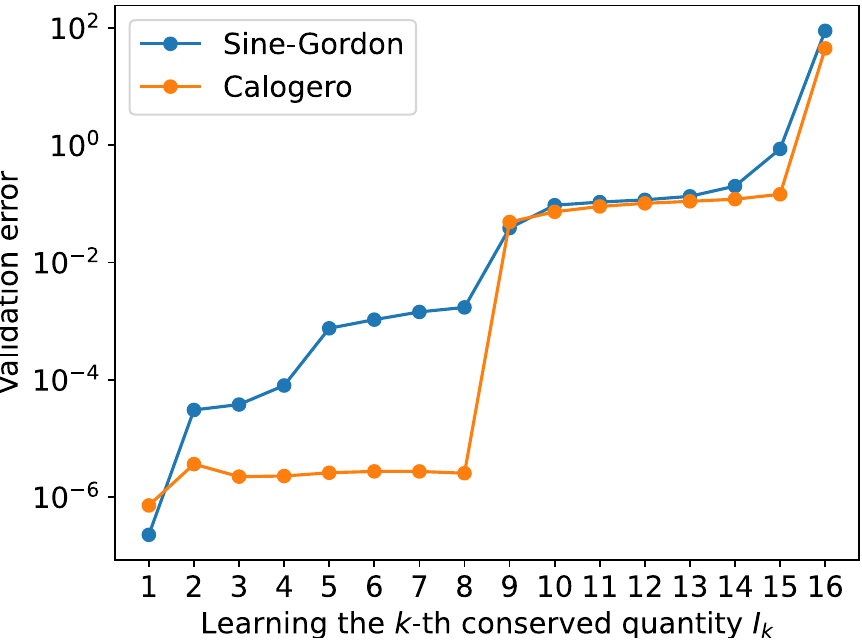}
    \caption*{(b) Number of lattice sites $N=8$}
    \label{fig:sg_calogero_8}
  \end{minipage}
  \caption{Validation losses for the learned conservation laws in the non-integrable discrete Sine--Gordon system and the integrable Calogero--Moser system, each with $d=N$ degrees of freedom. Reproduced from~\cite{zhu2023machine}. CC BY 4.0.}
  \label{fig:results_sg_calogero}
\end{figure}

Figure~\ref{fig:results_sg_calogero} illustrates the neural deflation method applied to two nonlinear dynamical lattices of $N$ identical nodes ($d=N$ degrees of freedom): the \textit{non-integrable} DsG system and the \textit{integrable} CM system~\cite{calogero,MOSER1975197}. For the CM system, a sharp increase in validation loss at $k = d + 1 = N + 1$ indicates that the algorithm accurately detects integrability and recovers a \textit{complete} set of independent conserved quantities in involution. In contrast, for the DsG system, the loss always jumps at $k = 2$, consistent with the presence of only \textit{one} conserved quantity (the Hamiltonian), independent of lattice size. Similar patterns arise in other systems, such as the Toda lattice~\cite{toda} and Fermi--Pasta--Ulam--Tsingou chains~\cite{fermi1955studies,FPUreview}.

These results demonstrate that the neural deflation method not only provides a principled and effective framework for identifying complete sets of functionally independent, Poisson-commuting conservation laws when the governing equations are known, but also serves as a foundation for data-driven integrability detection. By combining it with Hamiltonian neural networks~\cite{greydanus2019hamiltonianneuralnetworks}, recent extensions~\cite{CHEN2025108563} enable direct application to trajectory data, thereby broadening its applicability to settings where only observational data are available.

\subsection{Learning Hamiltonian structure in dissipative systems}
\label{subsec:generic}
As discussed extensively throughout this section, a central objective in the data-driven discovery of dynamical systems is to uncover an underlying \emph{Hamiltonian structure} directly from observations. Yet, many real-world systems are not Hamiltonian due to, for example, the presence of dissipative or time-dependent forces. To this end, there exist several geometric formalisms that extend Hamiltonian ideas to broader classes of systems. For example, the \emph{port-Hamiltonian} framework generalizes Hamiltonian dynamics to open systems with inputs and outputs, making it widely used in control theory \cite{duindam2009modeling}. The \emph{metriplectic} and \emph{GENERIC} formalisms extend Hamiltonian structure to include dissipative processes while maintaining thermodynamic consistency \cite{morrison1984bracket, grmela1997dynamics, ottinger1997dynamics}. Other extensions include Lie–Poisson systems \cite{marsden1984semidirect} and multi-symplectic formulations for PDEs \cite{bridges1997multi}. Of course, a comprehensive treatment of all such frameworks is beyond the scope of this work. For this reason, we focus on GENERIC-type structures, which encompass both reversible (Hamiltonian) and irreversible (dissipative) dynamics in a unified, physically interpretable way.

To begin, we first discuss the general nature of a \emph{metriplectic} system. Metriplecticity provides a geometric framework for dissipative dynamics by combining a Poisson bracket $\{\cdot,\cdot\}$, encoding the reversible part, with a symmetric, positive semi-definite bracket $(\cdot,\cdot)$, encoding the irreversible (dissipative) part. For a state variable $z$ in the phase space and functionals $F$ and $G$, these brackets satisfy $\{F,G\} = -\{G,F\}$, $(F,G) = (G,F) \ge 0$ and $(F,F)\ge 0$. Given an energy functional $E$ and an entropy functional $S$, the metriplectic evolution of any observable $F(z)$ is governed by
\begin{equation}
\frac{dF}{dt} = \{F,E\} + (F,S),
\label{eq:metriplectic}
\end{equation}
with the degeneracy conditions $\{S,E\} = 0$ and $(E,F) = 0$ for all $F$, ensuring that $E$ is conserved and $S$ is non-decreasing.

The \emph{GENERIC} (General Equation for Non-Equilibrium Reversible-Irreversible Coupling) formalism generalizes this structure. Let $z$ denote the state variables, $E(z)$ the total energy, and $S(z)$ the total entropy. The evolution equation takes the form
\begin{equation}
\dot{z} = L(z) \frac{\delta E}{\delta z} + M(z) \frac{\delta S}{\delta z},
\label{eq:generic}
\end{equation}
where $L(z)$ is an antisymmetric operator defining the Poisson bracket, $M(z)$ is a symmetric, positive semi-definite operator defining the dissipative bracket, and the state $z$ lives in some abstract Banach space $\Omega$. The degeneracy conditions are now
\begin{equation}
L(z) \frac{\delta S}{\delta z} = 0, \quad M(z) \frac{\delta E}{\delta z} = 0
\end{equation}
which recover the conservation of energy and the monotonicity of entropy, reducing equations~\eqref{eq:generic} to equation~\eqref{eq:metriplectic} in the metriplectic case. GENERIC provides a unifying geometric framework for coupled reversible-irreversible dynamics, ensuring compatibility with the first and second laws of thermodynamics for both finite- and infinite-dimensional systems.

Work by Zhang et al.~\cite{Zhang2022GFINNs}, introduces \emph{GENERIC Formalism Informed Neural Networks} (GFINNs). GFINNs are physics-informed architectures designed to learn dynamical systems while preserving the exact structure of the GENERIC formalism. Let us consider the case where $L$ and $M$ are known, and the task is to learn the functionals $E$ and $S$. It may be tempting to directly parametrize the gradients of $E$ and $S$, as they appear in Equation~\eqref{eq:generic}. This approach loses the direct recovery of the functionals since not every vector function is the gradient of a scalar function. Thus, GFINNs introduce fully connected feedforward neural networks to model the energy and entropy functional, computing their gradients via AD. 

Having parametrized the functionals directly, care must be taken to satisfy the degeneracy conditions. A key to this is to recognize the multiplicative nature of gradients of $l-$layer neural networks. If one lets $f(x ; \theta): \mathbb{R}^d \rightarrow \mathbb{R}$ denote an $l$-layer neural network and $g: \mathbb{R}^d \rightarrow \mathbb{R}^d$ be any differentiable function, the gradient of $(f \circ g)(x):=f(g(x))$ with respect to $x$ is given by
\begin{equation}\label{eq:MultGrad}
\nabla_x f(g(x) ; \theta)=(J g(x))^{\top}\left(W^l D_{l-1} \cdots W^2 D_1 W^1\right)^{\top}=(J g(x))^{\top} \nabla_x f \circ g(x).
\end{equation}
Here, $D_j$ is a diagonal matrix whose $(i, i)$-entry is $\phi^{\prime}\left(z_i^j(g(x))\right),$ $\phi$ denoting a nonlinear activation function, for $1 \leq j<l$ and $1 \leq i \leq n_j$, and $\mathrm{J} g(x)$ is the Jacobian of $g$ at $x$. The product form on the right-hand side of Equation~\eqref{eq:MultGrad} is what Zhang et al. refer to as the multiplicative structure of the neural network gradient.

Now, without loss of generality, let us only consider the pair $(L,S)$, as what follows will also apply to the pair $(M,E)$. Very loosely speaking and for sake of brevity, the trick GFINNs introduce is to use a tailored projection-like transformation $\mathcal{P}_L$ in the very first layer of the neural network. To define the transformation, one first assumes that the nullspace of $L(z)$ is constant over all possible state variables $z$. Define the matrices $\tilde{Q}_L(\boldsymbol{z})=\left[q_L^1(\boldsymbol{z}), \ldots, q_L^{\tilde{n}_L}(\boldsymbol{z})\right]$, where each $q_L^j$ is a vector function in the largest, hence invariant, subspace of the nullspace of $L$, and $F_L(\boldsymbol{z})=\left[F_L^1(\boldsymbol{z}), \ldots, F_L^{\hat{n}_L}(\boldsymbol{z})\right]^{\top}$ assuming that there exist $F^j$ such that
$$
\operatorname{span}\left\{\nabla F_L^j(\boldsymbol{z}): j=1, \ldots, \hat{n}_L\right\} \bigoplus \operatorname{null}_{\rm invariant} L(\boldsymbol{z})=\operatorname{null} L(\boldsymbol{z}).
$$
The projection operator is then defined as
$$
\mathcal{P}_L(\boldsymbol{z})=\left[\tilde{Q}_L^{\top}(\boldsymbol{z}) \boldsymbol{z} ; F_L(\boldsymbol{z})\right] \in \mathbb{R}^{n_L}
$$
Note that the first $\tilde{n}_L$ components of $\mathcal{P}_L(\boldsymbol{z})$ are the orthogonal projection coefficients of $\boldsymbol{z}$ onto $\operatorname{null}_{\text {invariant }} L(\boldsymbol{z})$ and the remaining components that complete the null space of $L$. 

For neural networks $f(z ; \theta): \mathbb{R}^d \rightarrow \mathbb{R}$, if one defines
\begin{equation}\label{eq:GFINNsGrad}
S_{\mathrm{NN}}\left(\boldsymbol{z} ; \theta_L\right)=f\left(\mathcal{P}_L(\boldsymbol{z}) ; \theta_L\right)
\end{equation}
it then follows from the multiplicative structure of Equation~\eqref{eq:MultGrad} and the construction of the projection transformation that any $S_{\mathrm{NN}}$ of the form~\eqref{eq:GFINNsGrad} satisfies $L(\boldsymbol{z}) \nabla S(\boldsymbol{z})=\mathbf{0}$ for all $\boldsymbol{z}$ in the state space $\Omega$. With this in hand, Zhang et al. show that a universal approximation result that exactly preserves the degeneracy condition is readily available; see theorem 1 in~\cite{Zhang2022GFINNs}. Theorem 2 in~\cite{Zhang2022GFINNs}, and surrounding text discuss the extension of a similar universal approximation result in the context of unknown matrix-valued functions $L(z)$ and $M(z).$

We emphasize that the power of GFINNs, given the surrounding assumptions in the construction of the projection operators $\mathcal{P}_L$ and $\mathcal{P}_M$, is that the full learned dynamics take the form of Equation~\eqref{eq:generic} with all GENERIC properties preserved to machine precision. Therefore, because the architecture encodes the GENERIC constraints intrinsically, GFINNs act as universal approximators for dynamics compatible with the GENERIC structure, while ensuring that predictions remain physically consistent. Zhang et al. show that GFINNs are successful in modeling both deterministic and stochastic systems, such as heat-volume exchange in coupled gas containers, thermo-elastic pendulums, and underdamped Langevin dynamics, consistently outperforming unconstrained neural networks in both predictive accuracy and adherence to thermodynamic laws. 

GFINNs have also been pushed beyond finite-dimensional benchmarks toward the regime of field equations. Weak-form GFINNs (WGFINNs) recast the GENERIC constraints in a variational setting to target PDE residuals directly. Related metriplectic/Onsager-inspired models encode the same reversible-irreversible split for continuum mechanics and fluid systems, and pseudo-Hamiltonian neural networks (PHNNs) adapt that decomposition to nonlinear dispersive-dissipative PDEs, offering a template for wave dynamics. Complementary structure-preserving lines include discretization-aware architectures (e.g. FINN adapted to shallow-water equations for bathymetry inference) and neural-operator approaches tailored to Helmholtz/elastic waves, underscoring that GENERIC-style priors play well with PDEs in strongly nonlinear, wave-dominated settings \cite{Zhang2022GFINNs,Park2024WGFINNs,Yu2021OnsagerNet,henon1964applicability,Eidnes2024PHNN_PDE,Horuz2024FINN_SWE,Zou2024HelmholtzNO}.

Another approach that has not been covered explicitly here is approximations via pseudo-Hamiltonian neural networks (PHNNs). PHNNs offer a compelling approach, originally developed for ODEs but later adapted to the PDE setting. PHNNs partition the learned representation into distinct components (internal conservative dynamics, dissipative effects, and external forces) each modelled by separate neural sub-networks. This modular architecture improves interpretability and robustness, particularly when environmental influences change or are removed. Eidnes and Lye~\cite{Eidnes2024PHNN_PDE} have successfully extended this PHNN framework to PDEs, demonstrating, just as with GFINNs, that PHNNs outperform standard, monolithic neural models, such as vanilla PINNs, while maintaining the physics provided by the GENERIC formalism.
{
\section{Applications of Koopman theory}
\label{sec_appK}

This section develops a wave-centred view of Koopman theory. The emphasis is on Koopman analysis as a language for organizing nonlinear-wave dynamics through spectral coordinates, invariant observables, and phase-space geometry. In favourable settings these objects can be identified exactly; more often they must be approximated from data, with the quality of the result determined by the choice of observables and the extent to which coherent structures and their interactions are faithfully represented. The goal here is to connect these perspectives, moving from exact structure to approximation and then to the kinds of wave phenomena where this viewpoint is most informative.

\subsection{Koopman eigenfunctionals of canonical PDE}

Let $S^t:\mathcal{X}\to\mathcal{X}$ denote the solution semigroup of a PDE on a function space $\mathcal{X}$. A Koopman eigenfunctional is a scalar observable $\varphi:\mathcal{X}\to\mathbb{C}$ such that
\begin{equation}
  \varphi(S^t u)=e^{\lambda t}\varphi(u)
  \label{eq:koop_nlw_eigenfunctional_def}
\end{equation}
for some $\lambda\in\mathbb{C}$. For PDEs this definition becomes especially concrete when the dynamics are conjugate to a linear evolution or when the equation is integrable and admits nonlinear coordinates with linear time dependence. In this sense, the analytic examples below are best viewed as exact benchmarks for any data-driven Koopman approximation.

\paragraph{Conjugacies to diffusion and explicit eigenfunctionals.}
Nakao and Mezi\'c formulate a Koopman eigenfunctional framework for PDEs relaxing to stable stationary states and show that canonical dissipative equations already exhibit explicit Koopman eigenfunctionals associated with decay toward equilibrium \cite{nakao2020spectral}. Their prototypical model case is diffusion,
\begin{equation}
  w_t=\nu w_{xx},
  \label{eq:koop_nlw_diffusion}
\end{equation}
on an interval $[0,L]$ with boundary conditions for which $-\partial_{xx}$ has eigenpairs $(\mu_n,\phi_n)$. If $w$ denotes a state profile and $w_\star(x)$ is the stationary profile selected by the boundary data, then
\begin{equation}
  \varphi_n[w]:=\int_0^L \big(w(x)-w_\star(x)\big)\phi_n(x)\,dx
  \label{eq:koop_nlw_diffusion_eigs}
\end{equation}
satisfies
\begin{equation}
  \varphi_n[S^t w]=e^{-\nu\mu_n t}\varphi_n[w].
  \label{eq:koop_nlw_diffusion_eig}
\end{equation}
Thus each Laplacian mode coefficient, viewed as the linear observable $\varphi_n[w]$, is a Koopman eigenfunctional with eigenvalue $-\nu\mu_n$. On a periodic interval $[0,2\pi]$, taking $\phi_k(x)=e^{-ikx}$ gives the familiar Fourier coefficients
\[
a_k(w):=\frac{1}{2\pi}\int_0^{2\pi} w(x)e^{-ikx}\,dx,
\qquad
a_k(S^t w)=e^{-\nu k^2 t}a_k(w),
\]
so the Koopman spectrum directly reflects the Laplacian spectrum. Since products of Koopman eigenfunctionals remain eigenfunctionals, monomials $\prod_{\ell=1}^p a_{k_\ell}$ have eigenvalue $-\nu\sum_{\ell=1}^p k_\ell^2$.

The viscous Burgers equation
\begin{equation}
  u_t + u\,u_x = \nu u_{xx},
  \label{eq:koop_nlw_burgers}
\end{equation}
inherits this structure through the Cole--Hopf transform. Writing $u=\phi_x$ and fixing the additive gauge in $\phi$, define
\[
w=\exp\!\Big(-\frac{\phi}{2\nu}\Big).
\]
Then $w$ solves \eqref{eq:koop_nlw_diffusion} and
\[
u=-2\nu\,\partial_x\log w.
\]
Therefore any diffusion eigenfunctional $\psi[w]$ pulls back to a Burgers eigenfunctional $\varphi[u]:=\psi(e^{-\phi/(2\nu)})$ with the same eigenvalue. What changes is the reconstruction of the state observable. Since $u$ depends nonlinearly on the ratio $w_x/w$, representing $u$ requires an infinite hierarchy of diffusion eigenfunctionals and their products. Consequently Burgers eigenvalues include the additive family
\begin{equation}
  \lambda=-\nu\sum_{\ell=1}^{p}k_\ell^2,
  \qquad k_\ell\in\mathbb{Z},
  \label{eq:koop_nlw_burgers_sumsq}
\end{equation}
with substantial degeneracy. Page and Kerswell~\cite{page2018burgers} showed that this degeneracy leads to initial-condition-dependent mode amplitudes and linear dependence among Koopman modes for state observables, so naive mode fitting from snapshots can be ill-posed. In a complementary small-data Dirichlet setting, Balabane, Mendez, and Najem~\cite{balabane2021koopman} later constructed the full eigenspaces associated with these repeated eigenvalues and proved convergence and a completeness property for the Burgers Koopman decomposition. This is useful conceptually since the intricate Burgers spectrum is a mathematically forced consequence, and not a numerical pathology, of the nonlinear observable map.

Nakao and Mezi\'c also treat a nonlinear phase-diffusion equation,
\begin{equation}
  y_t = y_{xx} + (y_x)^2,
  \label{eq:koop_nlw_phase_diffusion}
\end{equation}
which is conjugate to diffusion through the Cole--Hopf-type substitution $v=e^{y}$. Indeed,
\[
v_t=v_{xx},
\qquad
y=\log v.
\]
If $y_\star$ is the stable stationary solution fixed by the boundary data and $\phi_n$ is a Laplacian eigenfunction, then the functionals
\begin{equation}
  \varphi_n[y]
  :=
  \int_0^{L}\big(e^{y(x)}-e^{y_\star(x)}\big)\phi_n(x)\,dx
  \label{eq:koop_nlw_phase_eigs}
\end{equation}
satisfy
\[
\varphi_n[S^t y]=e^{-\mu_n t}\varphi_n[y].
\]
For Dirichlet problems one may take $\phi_n(x)=\sin(n\pi x/L)$, giving eigenvalues $\lambda_n=-(n\pi/L)^2$. In this formulation, the slowest decaying eigenfunctional defines isostable level sets, while joint zero level sets of higher modes encode inertial-manifold structure \cite{nakao2020spectral}. For dissipative wave-adjacent PDEs, these formulas provide exact targets for what learned Koopman eigenfunctionals should approximate.

\paragraph{Localized coherent structures and region-dependent expansions.}
For localized travelling waves, the main lesson is different, since Koopman expansions need not be globally valid in space-time. Parker and Page show that isolated fronts and solitons admit distinct Koopman decompositions in different observation regions, so the meaning of a data-driven spectrum depends on where the data are sampled \cite{parker2020isolated}. The mechanism is already visible in the one-soliton solution of KdV,
\begin{equation}
  u_t + 6uu_x + u_{xxx}=0,
  \qquad
  u(x,t)=2\kappa^2\,\mathrm{sech}^2\!\big(\kappa(x-4\kappa^2 t-x_0)\big),
  \label{eq:koop_nlw_kdv_soliton}
\end{equation}
where $\xi:=x-4\kappa^2 t-x_0$ is the comoving coordinate. For $\xi>0$, the identity
\[
\mathrm{sech}^2 z = 4\sum_{n=1}^\infty (-1)^{n-1} n e^{-2nz},
\qquad z>0,
\]
gives
\begin{equation}
  u(x,t)
  =
  8\kappa^2\sum_{n=1}^\infty (-1)^{n-1}n\,e^{-2n\kappa\xi}
  =
  \sum_{n=1}^\infty
  \Big(8\kappa^2(-1)^{n-1}n\,e^{-2n\kappa(x-x_0)}\Big)e^{8n\kappa^3 t}.
  \label{eq:koop_nlw_kdv_downstream}
\end{equation}
For $\xi<0$, using $\mathrm{sech}^2 z=\mathrm{sech}^2(-z)$ yields instead
\begin{equation}
  u(x,t)
  =
  8\kappa^2\sum_{n=1}^\infty (-1)^{n-1}n\,e^{2n\kappa\xi}
  =
  \sum_{n=1}^\infty
  \Big(8\kappa^2(-1)^{n-1}n\,e^{2n\kappa(x-x_0)}\Big)e^{-8n\kappa^3 t}.
  \label{eq:koop_nlw_kdv_upstream}
\end{equation}
Thus the same travelling wave admits one convergent Koopman-type expansion downstream and a different convergent expansion upstream, with opposite time exponents. Each expansion is perfectly valid, but only in its own convergence region. This is exactly why a global DMD fit across a window that mixes both regions can return misleading eigenvalues. Parker and Page show that the same phenomenon occurs for Burgers fronts and provide numerical evidence that it persists for moving SG breathers \cite{parker2020isolated}.

\paragraph{Integrable wave coordinates as Koopman eigenfunctionals.}
A useful Hamiltonian counterpoint is periodic KdV. Parker and Valva show that finite-gap solutions on a periodic interval admit analytic Koopman eigenfunctions through periodic inverse scattering and algebraic-geometric coordinates \cite{parker2023kdv}. On a fixed $g$-dimensional invariant torus with angle variables $\theta\in\mathbb{T}^g$ and frequency vector $\omega\in\mathbb{R}^g$,
\begin{equation}
  \dot\theta=\omega,
  \qquad
  \varphi_m(u):=e^{i m\cdot \theta(u)},
  \qquad
  m\in\mathbb{Z}^g,
  \label{eq:koop_nlw_periodic_kdv_torus}
\end{equation}
so that
\[
\varphi_m(S^t u)=e^{i m\cdot\omega\, t}\varphi_m(u).
\]
Hence the Koopman eigenvalues on that torus are
$
i\,m\cdot\omega, \ m\in\mathbb Z^g,
$
that is, purely imaginary numbers indexed by integer combinations of the fundamental frequencies. For $g\geq 2$, these frequencies are generically incommensurate and the resulting spectrum densely fills the imaginary axis. This provides a clean canonical example in which Koopman eigenfunctionals are exact, but the spectrum is not a small discrete set of isolated decay rates. It also clarifies the sharp contrast with the dissipative examples above since canonical PDE theory already contains both exponentially decaying eigenfunctionals near sinks and purely oscillatory eigenfunctionals on invariant tori.

Taken together, these canonical PDEs explain what Koopman methods should and should not be expected to deliver for nonlinear waves. When the underlying analytic coordinates are aligned with the observables, one can recover exact decay rates, frequencies, and invariant coordinates. When the PDE forces spectral degeneracy, region-dependent expansions, or dense imaginary lattices, those complications are intrinsic to the dynamics and should be treated as such in any data-driven approximation.

\subsection{Approximating Koopman structure from nonlinear-wave data}

Exact eigenfunctionals are exceptional, and in most nonlinear-wave problems one instead approximates the Koopman action from snapshots, sparse probes, or reduced coordinates. The main issue is not finite-dimensional regression per se, but whether the chosen observables respect the structures already seen in the canonical PDE examples such as local validity of expansions, translation/phase symmetry, large spectral degeneracies, and, in broadband regimes, substantial continuous spectral content.

\paragraph{Projection viewpoint and observable dependence.}
DMD and EDMD are best understood as projection methods. If $x_k=g(u_k)$ denotes a measured observable, DMD fits $Y\approx AX$; if $\psi$ is a nonlinear feature map, EDMD fits
\[
\psi(u_{k+1})\approx K^\top \psi(u_k).
\]
Thus DMD is the special case of EDMD in which the observable class is simply the span of the measurements themselves \cite{rowley2009spectral,schmid2010dmd,tu2014dmd,williams2015edmd}. In laboratory-frame wave data, neutral transport, phase drift, internal modulation, and dispersive radiation are all mixed together. A travelling pulse that is nearly rigid in a comoving frame may therefore produce a cloud of near-imaginary DMD eigenvalues even when its physically relevant internal dynamics are genuinely low-dimensional. What DMD/EDMD recover is the Koopman action seen through the chosen observables, not an absolute ``Koopman spectrum'' detached from representation.

\paragraph{Localization, windowing, and partial observations.}
The local-expansion phenomenon already visible in \eqref{eq:koop_nlw_kdv_downstream} has an immediate numerical consequence. If a single least-squares fit mixes upstream and downstream data, or combines pre-collision and post-collision dynamics of interacting coherent structures, there is no reason to expect one global spectrum to be meaningful. Windowed DMD/EDMD, piecewise fits, and comoving-coordinate constructions are the approximation-theoretic counterpart of the fact that Koopman expansions may only be valid locally in space-time \cite{parker2020isolated}. This point becomes even sharper in experiments, where one frequently observes only a few probes of the entire field. In that setting, delay-coordinate and Hankel lifts provide a Koopman-compatible enlargement of the observable class from time series alone, and they are on especially firm footing in ergodic settings \cite{arbabi2017koopman,brunton2017havok}. For nonlinear waves, such delay lifts are often the natural way to encode phase propagation and recurrent modulation when full-field snapshots are unavailable.

\paragraph{High-capacity observable classes: kernels and generators.}
Once observable dependence is accepted, the next question is expressive power. Low-degree global dictionaries are often too rigid for wave fields containing localized cores, weak but dynamically important radiation tails, and multiple active scales. Kernel EDMD addresses this by replacing an explicit dictionary with an RKHS feature space, so that one still approximates the Koopman operator, but does so in an implicitly high-capacity observable class \cite{williams2015kernel}. The main advantage in wave problems is that the kernel can be matched to the structure of the dynamics, for example, Fourier-sensitive kernels for periodic wave trains, localized kernels for fronts and solitons, and symmetry-invariant kernels when the relevant group action is known. When time derivatives are available, one may instead approximate the generator through
$$
D\varphi[u](\mathcal N(u))=\lambda \varphi(u),
$$
which is the continuous-time Koopman eigenproblem for
$
u_t=\mathcal N(u)
$\cite{klus2020data}. For weakly dissipative wave equations, where decay rates and oscillatory frequencies often coexist across widely separated scales, this generator viewpoint can be cleaner than fitting discrete-time multipliers and then passing through a logarithm.

\paragraph{Symmetry-aware approximation.}
Translation symmetry is the dominant obstruction to naive Koopman fitting for wave PDEs. If $(\tau_s u)(x)=u(x+s)$ and the flow is equivariant,
\[
\Phi^t(\tau_s u)=\tau_s\Phi^t(u),
\]
then the induced pullback on observables satisfies
\begin{equation}
  (\mathcal{T}_s g)(u):=g(\tau_s u),
  \qquad
  \mathcal{K}^t\mathcal{T}_s=\mathcal{T}_s\mathcal{K}^t,
  \qquad s\in\mathbb{T}.
  \label{eq:koop_nlw_commutation_translation_adv}
\end{equation}
Accordingly, Koopman-invariant subspaces can be organized in symmetry-adapted bases, and on periodic domains this leads naturally to Fourier or isotypic block structure \cite{salova2019symmetries}. Two recent constructions exploit this directly. Hochrainer and Kar enforce the translation-induced Fourier block structure in tieDMD, with demonstrations on periodic Burgers, KdV, nonlinear phase diffusion, and coupled FitzHugh--Nagumo systems \cite{hochrainer2024tiedmd}. Harder \emph{et al.}\ develop group-convolutional EDMD, showing that under equivariance assumptions the optimal finite-dimensional Koopman approximation inherits the same symmetry and can be represented through group convolutions and generalized Fourier transforms \cite{harder2025gcedmd}. For nonlinear waves this is often the only way to prevent neutral drift from being misread as internal spectral complexity.

\paragraph{Beyond isolated eigenvalues: spectral measures in broadband regimes.}
Not every nonlinear-wave regime is well described by a finite list of eigenpairs. When the continuous component of the spectrum is substantial, as in mixing or weakly chaotic spatiotemporal dynamics, a set of DMD exponents is not in general a stable or complete object. In sampled measure-preserving settings, a more robust alternative is to target the spectral measure of an observable. If $g\in L^2(\mu)$ and data are sampled at interval $\Delta t$, define
\begin{equation}
  m_k:=\left\langle g,(\mathcal{K}^{\Delta t})^k g\right\rangle_{L^2(\mu)},
  \qquad k\in\mathbb{Z},
  \label{eq:koop_nlw_correlation_moments}
\end{equation}
so that
\[
m_k=\int_0^{2\pi}e^{ik\theta}\,d\mu_g(\theta)
\]
for a spectral measure $\mu_g$ on the unit circle \cite{korda2020spectral}. This is well aligned with broadband wave dynamics, where the relevant object may be a spectral density or a mixed point--continuous measure. In this sense, advanced EDMD variants are useful precisely to the extent that they respect the wave-specific structures already exposed by the exact examples involving degeneracy, local expansions, symmetry-induced mixing, and genuinely non-discrete spectral content \cite{korda2020spectral,page2018burgers,parker2020isolated}.

\subsection{Koopman geometry: phase, basins, and ergodic partitions}

Koopman eigenfunctionals do more than support modal decompositions. When defined on a basin or invariant set, they can furnish phase/amplitude or partition coordinates for the dynamics. For nonlinear waves, this geometric viewpoint is often the more revealing one, because the relevant state space is organized by coherent structures, their stable and unstable manifolds, and, in conservative or weakly dissipative regimes, by invariant tori and ergodic components.

\paragraph{Phase and amplitude coordinates near coherent states.}
For a stable equilibrium---or, after symmetry reduction, for a travelling or rotating coherent structure---eigenfunctions with $\Re(\lambda)<0$ define amplitude coordinates measuring asymptotic distance to the attractor. For a stable periodic orbit or wave train with fundamental frequency $\omega$, the asymptotic phase $\theta(u)$ satisfies
\[
\theta(\Phi^t(u))=\theta(u)+\omega t \quad (\mathrm{mod}\ 2\pi),
\]
so $e^{i\theta(u)}$ is a Koopman eigenfunction with eigenvalue $i\omega$ and its level sets are the isochrons \cite{mauroy2013isostables}. For stable equilibria, the slowest-decaying eigenfunction defines isostables. A constructive route to such coordinates is provided by Laplace and Fourier averages:
\begin{equation}
  \varphi_\lambda(u)
  =
  \lim_{T\to\infty}\frac{1}{T}\int_0^T e^{-\lambda t} g(\Phi^t(u))\,dt,
  \label{eq:koop_nlw_laplace_average_geom}
\end{equation}
with the harmonic version obtained by taking $\lambda=i\omega$ \cite{budisic2012applied,mauroy2013isostables}. In wave problems, this is valuable because it produces phase and amplitude coordinates directly from trajectories, without first positing a collective-coordinate model or a reduced manifold ansatz.

\paragraph{Stable manifolds, separatrices, and basin boundaries.}
The same eigenfunction viewpoint gives geometric access to threshold dynamics. If $u_\dagger$ is a saddle-type coherent state and $\{\varphi_j\}$ are Koopman eigenfunctions associated with unstable directions, then boundedness forward in time forces those coordinates to vanish on the stable manifold. Formally,
\begin{equation}
  W^s(u_\dagger)
  \subseteq
  \bigcap_{\Re(\lambda_j)>0}\{u:\varphi_j(u)=0\},
  \label{eq:koop_nlw_stable_manifold_zero}
\end{equation}
with local equality under standard nondegeneracy assumptions \cite{mauroy2016global,mezic2020statespace_geometry}. For nonlinear waves involving scattering problems, depinning transitions, capture-versus-transmission scenarios, and pattern-selection problems, basin boundaries are often organized by unstable coherent structures. The Koopman formulation says that these boundaries may be sought as level-set geometry of eigenfunctions instead of through direct state-space shooting or continuation.

\paragraph{Zero-eigenvalue coordinates: invariants and slow drift.}
Conserved quantities are precisely Koopman eigenfunctionals at $\lambda=0$. If $I[u(t)]$ is invariant along trajectories, then
\begin{equation}
  I(\Phi^t(u))=I(u)
  \quad\Longrightarrow\quad
  \mathcal{K}^t I = I,
  \qquad
  \mathcal{L} I = 0.
  \label{eq:koop_nlw_zero_eig_geometry}
\end{equation}
Joint level sets of such functionals define invariant leaves of the dynamics; in Hamiltonian wave PDEs, these may be the level sets of mass, momentum, Hamiltonian, wave action, or more refined integrals when available \cite{mezic2013analysis}. Even when exact conservation is lost, as in weakly damped or weakly forced wave systems, slowly varying quantities appear naturally as near-zero Koopman coordinates. This is the geometric content of adiabatic modulation and near-integrability where the dynamics drift along almost-invariant leaves. In dissipative PDEs, the explicit eigenfunctionals of Nakao and Mezi\'c furnish the same idea near attracting states, where leading decaying eigenfunctionals foliate the basin by isostable-type level sets \cite{nakao2020spectral}.

\paragraph{Ergodic partitions from time averages.}
A complementary form of Koopman geometry arises from long-time averages. For any observable $f$ for which the limit exists, define
\begin{equation}
  f^\ast(u)
  :=
  \lim_{T\to\infty}\frac{1}{T}\int_0^T f(\Phi^t(u))\,dt.
  \label{eq:koop_nlw_time_average}
\end{equation}
Then $f^\ast(\Phi^t(u))=f^\ast(u)$, so $f^\ast$ is a $\lambda=0$ Koopman eigenfunctional. A sufficiently rich family of such averages yields a map
\[
u\mapsto \big(f_1^\ast(u),\ldots,f_m^\ast(u)\big),
\]
whose joint level sets approximate the ergodic partition of the state space \cite{mezic1999ergodic_partition,levnajic2010mesochronic,susuki2020invariant_sets}. This is the basis of mesochronic-plot constructions, in which one visualizes low-dimensional slices of the ergodic quotient using finitely many averages \cite{levnajic2010mesochronic}. In dissipative settings these averages can separate basins of attraction; in conservative or weakly dissipative settings they distinguish invariant components that may be geometrically intertwined in raw state coordinates. For nonlinear waves this suggests a useful state-space picture beyond isolated modes. Different phase-locked wave trains, quasiperiodic tori, chaotic seas, or coexisting attractors may appear as distinct regions in time-averaged coordinates even when they are hard to separate pointwise in the original field representation.

\paragraph{Harmonic averages and quasiperiodic wave geometry.}
Time averages at nonzero frequencies refine this partition by targeting oscillatory rather than invariant content. Define
\begin{equation}
  f^\ast_\omega(u)
  :=
  \lim_{T\to\infty}\frac{1}{T}\int_0^T e^{-i\omega t} f(\Phi^t(u))\,dt.
  \label{eq:koop_nlw_harmonic_average}
\end{equation}
Whenever $f^\ast_\omega$ is nonzero, it transforms as a Koopman eigenfunction with eigenvalue $i\omega$, and its level sets reveal periodic or quasiperiodic structures at that frequency \cite{levnajic2015fourier_mesochronic}. This is precisely the idea behind Fourier mesochronic plots. For nonlinear waves it gives a geometric counterpart to the spectral viewpoint of periodic KdV and other near-integrable systems: one is not only extracting frequencies, but also identifying where in state space those frequencies live. In problems with coexisting regular and irregular dynamics, harmonic averages can therefore separate quasiperiodic islands from surrounding broadband or chaotic regions in a way that is especially natural for wave trains, breathers, and modulated coherent structures.

Koopman geometry is thus the phase-space content of the eigenfunctional viewpoint. Modal decompositions simply describe how observables evolve. Koopman eigenfunction level sets describe where trajectories live and how the state space is partitioned into dynamically distinct regions \cite{mezic2020statespace_geometry}.

\subsection{Representative nonlinear-wave applications}
A non-exhaustive but wave-centred set of applications shows where Koopman and DMD ideas become genuinely informative in practice, even when exact eigenfunctionals are unavailable. What makes these examples interesting is that each highlights a different obstruction already visible in the theory above including poor global observables, symmetry contamination, local rather than global spectral descriptions, or the need to separate coherent wave content from broadband backgrounds.

\emph{Excitable and reaction--diffusion waves.} Kernel-based Koopman methods have been demonstrated on canonical excitable media such as FitzHugh--Nagumo \cite{williams2015kernel}. These systems are especially revealing because they combine travelling pulses, recovery dynamics, and dissipation in a way that defeats small global dictionaries. The point is that (along with the field being high-dimensional) propagation and relaxation live on different geometric and temporal scales. This makes reaction--diffusion waves a natural testbed for whether a lifted observable class can recover dynamics in coordinates that remain spectrally meaningful.

\emph{Chaotic and pattern-forming wave PDEs.} Symmetry-aware Koopman learning has been tested on Kuramoto--Sivashinsky and related pattern-forming PDEs \cite{harder2025gcedmd}. These are important wave applications because they sit at the boundary between coherent-structure dynamics and broadband spatiotemporal chaos. Translations, phase drift, and intermittent pattern reorganization generate large neutral or near-neutral directions, so a naive spectral fit tends to confuse transport with internal dynamics. In this regime, equivariance is a mathematical necessity if one wants the recovered spectrum to say anything about modulation, instability, or coherent pattern turnover.

\emph{Strongly nonlinear internal waves in the laboratory.} Piecewise DMD and local Koopman constructions have been used to analyse internal solitary waves interacting with topography \cite{zhang2020pdmd_isw}. This is an especially compelling setting because the underlying wave physics is visibly nonstationary: one observes propagation, shoaling, deformation, fission, and sometimes breaking within a single experiment. There is therefore no reason to expect one globally valid Koopman model. What makes these studies valuable is that they operationalize the local-expansion viewpoint suggested by exact front and soliton theory, using segmentation to track regime-dependent spectral content instead of forcing one spectrum across incompatible wave phases.

\emph{Detonation and sharp-front wave phenomena.} Data-driven reduced-order modeling of detonation fronts provides another wave-relevant regime in which localized structure and symmetry dominate the dynamics \cite{mendible2022detonation}. Here the state is organized by translating and curving fronts, their mutual interaction, and the onset of instabilities. Koopman or DMD-type reductions are interesting in this context because they test whether front location, curvature, separation, and interaction phase can be encoded into observables that retain predictive content even when the front geometry changes rapidly. This makes detonation-type problems a good stress test for any wave-adapted observable design.

\emph{Wave--turbulence separation in environmental signals.} DMD has been proposed as an equation-free tool to separate coherent wave components from turbulent fluctuations in oceanographic data when conventional spectral filtering is ineffective \cite{chavezDorado2024waveturb}. The appeal here is that the physically meaningful distinction is that coherent waves and turbulent backgrounds may overlap in the Fourier domain while remaining dynamically quite different. Koopman-style decompositions are interesting precisely because they target dynamical coherence rather than only spectral energy, making them potentially useful when one wants to identify transport-relevant wave structure in noisy field measurements.

\emph{Pilot-wave hydrodynamics and wave-driven chaos.} Pilot-wave systems provide a particularly vivid application because the wave field itself participates in the effective particle dynamics and undergoes qualitative changes across bifurcations and routes to chaos \cite{kutz2023pilotwave}. This makes them a natural laboratory for Koopman diagnostics since a nonlinear wavefield whose recurrent structures, dominant frequencies, and instability mechanisms shift with forcing. What makes this class attractive is that it sits between continuum wave dynamics, low-dimensional bifurcation theory, and experimental nonlinear physics, so Koopman analysis can be used both to detect emergent organization and to monitor its breakdown.

Taken together, these applications show that Koopman and DMD methods are most persuasive in nonlinear-wave settings when they are used as \emph{physics-aware observable methodologies}. Reaction--diffusion waves emphasize the need for expressive observables, pattern-forming PDEs emphasize symmetry adaptation, internal and sharp-front waves emphasize locality and regime dependence, and environmental or pilot-wave examples emphasize diagnostic separation of coherent structure from broadband complexity. In that sense, we believe the most successful applications are those that take seriously the lessons of the analytic theory, namely that eigenvalue degeneracy may be intrinsic, expansions may only be locally valid, symmetries must be handled explicitly, and the relevant spectral object may be a geometric coordinate or structured partition of state space.

\section{Learning Lax integrability of DEs}
\label{sec6}

\subsection{Summary of recent community efforts}

Given a dynamical system with state $x$ and equations of motion $\dot x = f(x)$, the goal here is to \emph{identify integrability features} directly from either numerical or symbolic data. One approach is to try to learn a \emph{Lax pair} $(L,P)$ in mechanics or a \emph{Lax connection} $(U,V)$ in field theory. Recall from Section~\ref{section:LLPN}, for a Lax pair one requires
$
\frac{dL}{dt} = [L,P]
$
so that spectral invariants $\mathrm{tr}\,L^k$ are conserved; for field theories recall the \emph{zero-curvature} (flatness) condition
$$
U_t - V_x + [U,V] = 0
$$
holds for the relevant and compatible dynamics. 

In addition to the zero-curvature formulation, integrability can also be encoded at the level of the underlying Poisson structure. In the so–called $r$-matrix formulation one prescribes a Lie–Poisson bracket for the Lax operator,
$$
\{L_1,L_2\} = [r_{12},L_1]-[r_{21},L_2],
$$
where the subscripts denote tensor placement: $L_1 = L \otimes I$ acts on the first slot, $L_2 = I \otimes L$ on the second, and $r_{12}$ (respectively $r_{21}$) acts on both slots of $V \otimes V$ (with the tensor factors ordered as $1,2$ or $2,1$). In components, this compact tensorial notation encodes all Poisson brackets between the entries of $L$. The condition guarantees that the spectral invariants of $L$ are in involution and hence generate commuting flows. Importantly, once a candidate $L$ is identified, the associated $r$-matrix can in principle also be learned directly from data, providing access not just to the dynamics but also to the Hamiltonian structure of the system.

Pioneering work by Krippendorf et al.~\cite{krippendorf2021integrability}, considers the deep learning of Lax pairs. This approach is unsupervised, physics-constrained (through the equation of motion), and returns \emph{analytic} candidates for Lax operators (and $r$-matrices) that can be verified a posteriori. Depending on the mechanical context, a type of soft constraint on the Lax/flatness residual is enforced
$$
\mathcal{L}_{\mathrm{Lax}}=\bigl\lVert \partial_t L - [L,P]\bigr\rVert^2
\quad\text{or}\quad
\mathcal{L}_{\mathrm{flat}}=\bigl\lVert U_t - V_x + [U,V]\bigr\rVert^2.
$$

Time derivatives are easily accessed by the chain rule of the equation of motion, e.g.,
$$
\dot L=\frac{\partial L}{\partial x}\,\dot x=\frac{\partial L}{\partial x}\,f(x).
$$

To align with the expected behaviour of Lax operators, Krippendorf, et al.~\cite{krippendorf2021integrability} write each learned spectral operator entry as $z_k(x)$, enforcing that $\dot z_k$ is linear in a small set of elementary time derivatives and proportional to at least one such element, which yields auxiliary losses
$$
\mathcal{L}_t=\sum_k \min_{\{c_{kj}\}}\bigl\lVert \dot z_k-\sum_j c_{kj}\,\xi_j\bigr\rVert^2,
\qquad
\mathcal{L}_M=\sum_k \min_{\{d_{kj}\}}\bigl\lVert [L,P]_k-\sum_j d_{kj}\,\xi_j \bigr\rVert^2,
$$
with $\xi_j$ chosen from the components of $f(x)$ and their linear combinations. To avoid degenerate solutions $L\equiv0$, a type of \emph{mode-collapse} penalty (see terms with subscript MC in 
what follows) encourages nontrivial entries, and the total objective for pairs or connections is
$$
\mathcal{L}_{\mathrm{pair}}=a_1\mathcal{L}_{\mathrm{Lax}}+a_2\mathcal{L}_t+a_3\mathcal{L}_M+a_4\mathcal{L}_{\mathrm{MC}},
\qquad
\mathcal{L}_{\mathrm{conn}}=a_1\mathcal{L}_{\mathrm{flat}}+a_2\mathcal{L}_t+a_3\mathcal{L}_M+a_4\mathcal{L}_{\mathrm{MC}}.
$$
With this loss function in hand, training is performed using typical deep learning machinery. Once $L$ is known, the classical $r$-matrix can be fit by minimizing
$$
\mathcal{L}_r=\bigl\lVert \{L_1,L_2\}-[r_{12},L_1]+[r_{21},L_2]\bigr\rVert^2,
$$
where, again, tensor indices denote action on separate copies of phase space.

We summarize the results of Krippendorf, et al. here:
\begin{itemize}
    \item \textbf{Harmonic oscillator:} The method recovers a valid $2\times2$ Lax pair with $\mathrm{tr}\,L^2 \propto H$. 
\item \textbf{KdV:} A $2\times2$ operator connection $(A_x,A_t)$ with quadratic/derivative terms matches the known KdV Lax form up to benign gauge shifts.
Here the authors needed to enforce constraints in the
connection (e.g. the existence of off-diagonal elements)
and then got approximations up to small constant terms  of the KdV model.
\item\textbf{Heisenberg ferromagnet:} Using an $\mathrm{SU}(2)$-equivariant ansatz and a suitably
restricted form of $A_x$, the algorithm reproduces the standard Lax connection. 
\item \textbf{SG and principal chiral model:} The zero-curvature condition is satisfied if and only if the respective equations of motion hold, again
upon suitable choice of both $A_t$ and $A_x$
(e.g. with the latter depending only on the
first spatial derivative of the field).
\end{itemize}

The work of~\cite{krippendorf2021integrability} also finds that for perturbed systems (e.g. a 2D coupled oscillator or a perturbed Heisenberg model), training losses \emph{adapt} and decrease over time for integrable perturbations (the network can retune the Lax structure), but plateau at larger values for non-integrable ones, furnishing a practical diagnostic.

The more recent work of de Koster and Wahls~\cite{de2024data} takes a different approach to learning integrability of DEs, using a more traditional data-driven approach. Since many nonlinear wave PDEs can be cast in Lax form and solved by the inverse scattering method, one would hope to be able to leverage the spectral operator $L$, which induces a nonlinear Fourier transform (NFT), in practical applications. When only data are available and the governing equation is unknown (or only approximately Lax-integrable), the question is whether one can \emph{identify} a Lax structure from data, sufficient to enable analysis via NFT.

The paradigm on which~\cite{de2024data} focuses concerns
the celebrated Ablowitz--Kaup--Newell--Segur (AKNS) class, whose spectral operator has the canonical $2\times 2$ form
$$
L =
\begin{pmatrix}
i\partial_x & -\,i\,q(u) \\
i\,r(q) & -i\partial_x
\end{pmatrix},
\qquad
P =
\begin{pmatrix}
P_{11}(u) & P_{12}(u) \\
P_{21}(u) & -P_{11}(u)
\end{pmatrix},
$$
with potentials $q(u)$ and $r(q)$ that distinguish members of the AKNS hierarchy (e.g. NLSE, mKdV, SG) and relatives via simple choices such as $r=\pm q$, $r=\pm q^\ast$, or $r=-1$. Together with the linearized dispersion relation, the choice of $(q,r)$ \emph{completely determines} an integrable PDE within the AKNS family, hence also $A(u)$.

AKNS systems admit an infinite sequence of conserved integrals $C_k[q,r]$ obtained recursively from the spectral problem. The first few are explicit functionals of $q$, $r$ and their derivatives; for example
$$
C_1=\int q\,r\,dx,\quad
C_2=\int (r\,q_x - r_x\,q)\,dx,\quad
C_3=\int \bigl(q^2 r^2 + q_x r_x\bigr)\,dx,
$$
with higher $C_k$ mixing higher derivatives and nonlinear combinations. These $C_k$ serve as \emph{invariants} that should remain (approximately) constant along the evolution for a Lax-integrable model. This is the key feature of AKNS that de Koster and Wahls exploit in~\cite{de2024data}. Naturally, of course,
this raises the question about how these $C_i$ quantities 
will be known without expert insight into the model.
In the work of~\cite{de2024data}, they are assumed to be
a priori ``given'' and their constancy practically 
formulates the loss function of choice.

Given just two snapshots $u(\cdot,t_0)$ and $u(\cdot,t_1)$ (more can be used), the method of~\cite{de2024data} searches over parameters $c_d$ in a parametrized library for $q(u)$:
$$
\begin{aligned}
q(u) & =\sum_{d=1}^D c_d g_d(u), \text { with } D=5, \\
G & =\left\{g_1=u, g_2=u_x, g_3=u_{x x}, g_4=u^2, g_5=u u_x\right\},
\end{aligned}
$$
which also implies a library once a functional form of $r(q)$ is specified. For each candidate, the method computes the first few conserved quantities $C_k[q,r]$ at $t_0$ and $t_1$, measures their variation $\Delta C_k = C_k(t_1)-C_k(t_0)$, and selects the $(q,r)$ that \emph{minimizes} a weighted norm of $\{\Delta C_k\}_k$. More precisely, de Koster and Wahls solve the following optimization problem to identify $r$ and $q$ within the AKNS family
\begin{equation}\label{eq:dKWObj}
\left(r^{(\mathrm{ID})}, c^{(\mathrm{ID})}\right)=\underset{r \in\left\{-1, \pm q, \pm q^*\right\}}{\operatorname{argmin}} \underset{c \in \mathcal{C}}{\operatorname{argmin}} 
\left[ w(c) E(c, r ; u) \right], \quad \mathcal{C} \subseteq \mathbb{R}^d
\end{equation}
with $ w(c)$ denoting the weight of coefficient $c$ and where the function
$$
E(c, r ; u)=\sum_{k \in\{1,3,5\}}\left(\sum_{n=1}^N \frac{\sigma\left[C_k\left(t ; c, r, u^{(n)}\right)\right]}{\mu\left[C_k\left(t ; c, r, u^{(n)}\right)\right]}\right)
$$
with the mean and standard deviation defined by
$$
\mu\left[C_k(t)\right]  =\frac{1}{M} \sum_{m=1}^M C_k\left(t_m\right),\qquad \sigma\left[C_k(t)\right]=\sqrt{\frac{1}{M} \sum_{m=1}^M\left(C_k\left(t_m\right)-\mu\left[C_k(t)\right]\right)^2}.
$$
Simply put, de Koster and Wahls optimize over the parameter $c,$ and then perform a brute force search over the 5 options of $r$ implied by Equation~\eqref{eq:dKWObj}.

The paper demonstrates identification on (nearly) Lax-integrable data from mKdV, NLSE, SG, and a transformed KdV example, and explores a viscous Burgers case with complex viscosity as a stress test. In these examples, the correct AKNS spectral operator is recovered from noisy data (on the order of $\lesssim 1\%$ of the signal amplitude in most cases).

It is important to note that the method of~\cite{de2024data} is \emph{class}-specific to AKNS systems and targets the \emph{spectral operator} rather than a free-form PDE. This specialization yields robustness and interpretability: once $L$ (and the dispersion) are determined, one obtains an integrable surrogate PDE and inherits the entire AKNS solution machinery. Compared with general Lax-pair learning via neural {\"A}nsatze optimization, this approach trades breadth for reliability and data efficiency.
However, we also remind the reader of the specificity 
requirement of being aware of the relevant conservation laws.
Clearly, this work leaves room for adaptability to wider classes of integrable and nearly integrable dynamics.

Kantamneni et al.~\cite{kantamneni2024optpde,Kantamneni2025OptPDE}, use yet another approach to learning integrability. Their insight is based on the recognition that integrable PDEs are characterized by having infinitely many conserved quantities. Despite the fact that integrable PDEs are exceptionally rare and often discovered by serendipity, the authors of~\cite{kantamneni2024optpde,Kantamneni2025OptPDE} propose to learn new candidate integrable PDEs by fitting coefficients in a parametrized PDE form to maximize the number of conserved quantities discovered, denoted \(n_{\rm CQ}\).

Consider a parametric PDE of the form
$$
u_t = \mathcal{N}[u;\, \alpha],
$$
with \(\alpha\) denoting a finite set of coefficients (e.g. in differential operators or nonlinear terms). The goal is to adjust \(\alpha\) so that the resulting PDE admits as many independent conserved quantities \(C_k[u]\) as possible. The conserved quantities \(C_k[u]\) are defined by satisfying
$$
\frac{d}{dt} C_k[u(t)] = 0
$$
under the PDE evolution. In practice, the authors search over $\alpha$ using backpropagation. They simultaneously identify candidate invariants \(C_k\) (via weak form or PDE-specific templates), evaluate their rate of change \(\dot C_k\) numerically,
and maximize the count of invariants with \(\|\dot C_k\|\) below a prescribed tolerance.

This framework, called OptPDE, successfully rediscovers known integrable PDEs (such as KdV-like models) within its coefficient search space. Interestingly, it also generates \emph{novel families} of PDEs that admit at least one non-trivial conserved quantity. An example highlighted is
\begin{equation}\label{eq:OptPDE}
u_t = \bigl(u_x + a^2\,u_{xxx}\bigr)^3,
\end{equation}
for which analytical exploration reveals interesting conserved properties, though integrability is not guaranteed. Kantamneni et al. study, by hand, the case $a=0$ of Equation~\eqref{eq:OptPDE}. They show that this PDE indeed possesses an infinite number of conserved quantities.

Of course, it is important to point out that maximizing \(n_{\rm CQ}\) is \emph{not sufficient} to prove integrability. This merely identifies promising candidates, and the method depends critically on the choice of invariant templates and detection criteria. Nevertheless, by proposing novel PDEs with nontrivial invariants, OptPDE enables a human-AI collaboration loop where a machine hypothesizes and domain scientists verify. One can argue 
based on the proposed examples that this is a methodology
that may yield intriguing results regarding identifying
PDEs with conservation laws, which occasionally may have
a bearing on integrability studies, although it is not directly
(a priori) geared in that direction.

\subsection{Learning ODE integrability using sparse identification of Lax operators (SILO)}

SILO~\cite{adriazola2025computer} takes a more basic approach to learning integrability. Instead of using deep learning, using data-driven approaches, or trying to maximize the total number of conserved quantities, SILO is a symbolic operator learning framework. The main idea behind SILO, motivated by the SINDy methodology, is to identify sparse and interpretable representations of a Lax pair compatible with a given Hamiltonian system. To describe this approach, we first illustrate its application in identifying Lax pairs for Hamiltonian ODE systems.

Let us begin with the simple harmonic oscillator $\dot{q}=p,\ \dot{p}=-q$. We may hypothesize that the Lax pair is of the form
\begin{equation}\label{eq:SHOHyp}
\tilde{L}(q,p) = \sum_{k=1}^{N_{\xi}} \xi_k \Theta_k^{(L)} (q,p),\qquad
\tilde{P}(q,p) = \sum_{k=1}^{N_{\zeta}} \zeta_k \Theta_k^{(P)}(q,p),
\end{equation}
where $\Theta_k^{(L)}$ and $\Theta_k^{(P)}$ are matrices that are at most linear in the canonical variables $q$ and $p$. A simple count shows that the vector $\eta=[\xi\ \zeta]$ belongs to $\mathbb{R}^{24}$. Now, to build the loss function, observe by the chain rule that
$$
\frac{dL}{dt} = \frac{\partial L}{\partial q} \dot{q} + \frac{\partial L}{\partial p} \dot{p}= \frac{\partial L}{\partial q} \frac{\partial H}{\partial p} - \frac{\partial L}{\partial p} \frac{\partial H}{\partial q}= \{L, H\}.
$$
Given the symbolically tractable hypothesis (equation~\eqref{eq:SHOHyp}), analytically calculating the Poisson bracket $\{L,H\}$ is straightforward. Since $\frac{d L}{d t} =[L,P],$ it follows that $\{L,H\}-[L,P]$ should vanish for any point in phase space $(q,p)\in\mathbb{R}^2$. Lastly, we comment that the trivial solution $L\equiv0$ is viable, so care should be taken to penalize away from $\xi\equiv0$.

SILO is thus formulated as the following sparse, empirical risk minimization problem
\begin{equation}\label{eq:SHOProb}
\min _{\eta\in\mathbb{R}^{N_{\eta}}}\mathcal{J}[\eta]=\min _{\eta\in\mathbb{R}^{N_{\eta}}} \mathbb{E}_{(q,p)\sim\rho}\left\{\sum_{i, j} \frac{\left(\{\tilde{L}, H\}-[\tilde{L}, \tilde{P}]\right)_{i, j}^2}{\{\tilde{L}, H\}^2_{i,j}}\right\}+r\mathcal{R}(\eta),
\end{equation}
where $r\in [0, 1)$ controls the amount of desired sparsification in the search.  Since the numerator, rewarding the discovery of a Lax pair compatible with Hamiltonian $H$, is divided by the Poisson bracket acting on $\tilde{L}$, this formulation discourages trivial solutions where $\tilde{L}\equiv0$. Since it is infeasible to access the continuous nature of the phase space, we restrict to a finite number of samples from a uniform distribution $\rho$ supported on a subset of the phase space $\mathbb{R}^2$. 

By using sequential thresholding, SILO correctly identifies the following two families of Lax pairs
$$
\tilde{L}_1= 	\begin{pmatrix}
	\eta_1 q  &\ \ \eta_4 p \\
	\eta_6 p  &\eta_7 q
\end{pmatrix},\qquad
\tilde{L}_2= 	\begin{pmatrix}
	\eta_2 p  &\ \ \eta_3 q \\
	\eta_5 q  &\eta_8 p
\end{pmatrix},\qquad \tilde{P}= 	\begin{pmatrix}
	0  & \eta_{22}\\
	\eta_{23} & 0
\end{pmatrix},
$$
with all other 18 entries of $\eta$ equal to zero in both families. The recovered Lax pairs reproduce the equations of motion with seven-digit precision with an unsparsified loss evaluation on the order of $10^{-15}$. We comment that this precision is difficult to achieve using neural networks. See~\cite{adriazola2025computer} for more details on the numerical implementation.

Using SILO, high-precision identification of the integrability of the famous Henon--Heiles (HH) system is made possible. This is a system with two degrees of freedom defined by the Hamiltonian~\cite{Hernandez2024TIGNN,fordy1991henon}
\begin{equation}\label{eq:HHHam}
\begin{aligned}
H=\frac{1}{2}\left(p_x^2+p_y^2\right) & +\frac{1}{2}\left(A x^2+B y^2\right) +x^2 y+\varepsilon y^3,
\end{aligned}
\end{equation}
where the parameters $A,\ B,$ and $\varepsilon$ are arbitrary. This system is known to be integrable for three different cases of these parameters~\cite{ravoson1993separability}, and we will focus on one such case: $A=B=1$ and $\varepsilon=1/3$.

To adapt the optimization of~\eqref{eq:SHOProb} to the HH setting, we introduce another simplified symbolic hypothesis on the Lax pair; please see~\cite{adriazola2025computer} for details. Furthermore, by using
$$
\{L,H\}=\frac{\partial L}{\partial x} \frac{\partial H}{\partial p_x}+\frac{\partial L}{\partial y} \frac{\partial H}{\partial p_y}-\frac{\partial L}{\partial p_x} \frac{\partial H}{\partial x}-\frac{\partial L}{\partial p_y} \frac{\partial H}{\partial y} \\
$$
and generalizing the sampling of the phase space to four dimensions, that is, $(x_k,p_{x_k},y_k,p_{y_k})\sim U([-1,1]^4)$, SILO is ready to be deployed. Indeed, SILO once again finds a loss on the order of $10^{-15}$, in the integrable case of $A=B=1$ and $\varepsilon=1/3$. 

From this point, we further investigate how SILO responds to training on parameter sets where integrability is unknown.  We show, in figure~\ref{fig:HHdetection}, that SILO experiences a significant increase in the computed loss--by several orders of magnitude--across the parameter space. Thus, restricted to the operator hypothesis used, SILO determines where the integrability lies in the parameter space $(A,\varepsilon)$ for fixed $B$. 

\begin{figure}[htbp]
\centering
\begin{minipage}{0.48\textwidth}
  \centering
  \includegraphics[width=\textwidth]{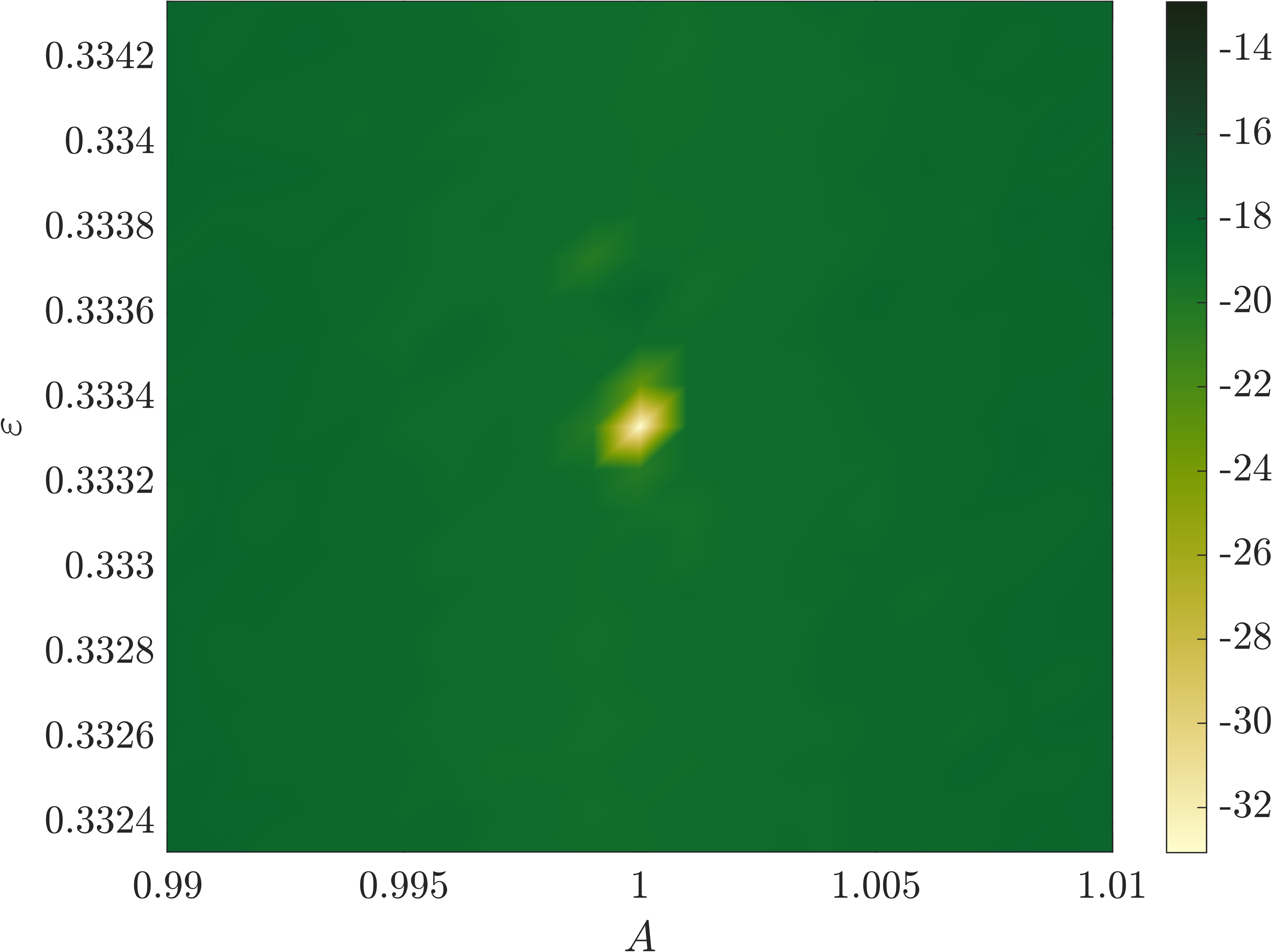}
\end{minipage}
\hfill
\begin{minipage}{0.48\textwidth}
  \centering
  \includegraphics[width=\textwidth]{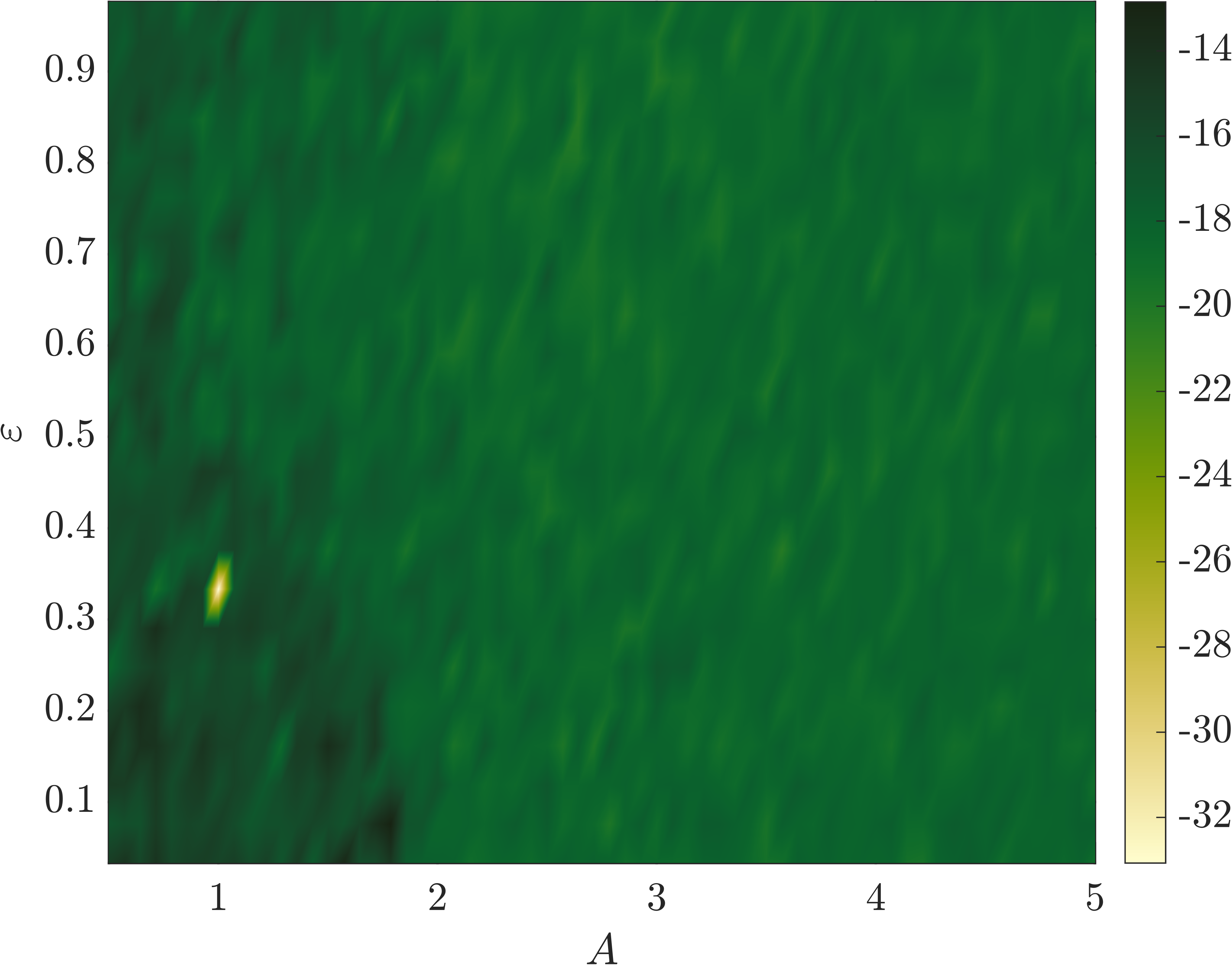}
\end{minipage}
\caption{A broad parameter search (with $B$ fixed to 1) for integrability detection in the HH system given by Hamiltonian~\eqref{eq:HHHam} from the work of~\cite{adriazola2025computer}. The optimization of the loss function, shown on a logarithmic scale, identifies a distinct position at $(A, \varepsilon) = (1, 1/3)$ where integrability is meaningfully detected, differing by several orders of magnitude from background loss values. Cubic spline interpolation of the landscape is used for visualization. Reproduced from~\cite{adriazola2025computer}. CC BY 4.0.}
\label{fig:HHdetection}
\end{figure}

The last ODE example we study using SILO is an example from rigid body dynamics. The Euler top is a three-dimensional rigid body whose angular velocities $\Omega_i$ are governed by 
\begin{equation}\label{eq:EulerTop}
    I_i \dot{\Omega}_i = \frac{1}{2}\sum_{j,k=1}^3 \epsilon_{ijk} (I_j - I_k)\, \Omega_j \Omega_k
\end{equation}
where each principal moment of inertia $I_j$ is a constant, positive real number and $\epsilon_{ijk}$ is the Levi-Civita symbol. Lax integrability is known because~\eqref{eq:EulerTop} is compatible with the matrices
$$
\begin{aligned}
&L=\left(\begin{array}{ccc}
0 & -M_3 & M_2 \\
M_3 & 0 & -M_1 \\
-M_2 & M_1 & 0
\end{array}\right), \ P=\left(\begin{array}{ccc}
0 & -\Omega_3 & \Omega_2 \\
\Omega_3 & 0 & -\Omega_1 \\
-\Omega_2 & \Omega_1 & 0
\end{array}\right),\\
\end{aligned}
$$
satisfying Lax's equation $dL/dt=[L,P]$, where $
M_i=I_i \Omega_i.$

Straightforward algebra shows this particular $L$ for the Euler top does not generate the well-known two algebraically independent conserved quantities \cite{BabelonBernardTalon2003}. Thus, 
this represents a so-called fake Lax pair~\cite{ReymanSemenov1994}. The trick, first found by Manakov \cite{Manakov1976}, for converting this fake Lax pair into one that leads to Liouville integrability is to introduce a diagonal matrix
$$
J=\frac{1}{2}\left(\begin{array}{ccc}
I_2+I_3-I_1 & 0 & 0 \\
0 & I_1+I_3-I_2 & 0 \\
0 & 0 & I_1+I_2-I_3
\end{array}\right)
$$
that leaves Lax's equation invariant under 
$$
\frac{d}{d t}\left(L+\lambda J^2\right)=\frac{dL}{dt}=\left[L+\lambda J^2,P+\lambda J\right]=[L, P],
$$
where $\lambda$ is arbitrary and often referred to in the literature as the spectral parameter. This provides the remedy matrix $\hat{L}=L+\lambda J^2$ which can be shown to reproduce both algebraically independent conserved quantities via $\frac{d}{dt}\mathrm{tr}\hat{L}^{n}=0,\ n=2,3,$ as long as the spectral parameter $\lambda\neq0.$

It is useful to note that the Euler top's Hamiltonian
$H=\frac{1}{2}\sum_{i=1}^3 I_i\Omega_i^2=\frac{1}{2}\sum_{i=1}^3\frac{M_i^2}{I_i}$
can be used to more compactly write the three equations of motion as $\dot{\Omega}_i=\{H,\Omega_i\},$ where curly brackets indicate once again the Poisson bracket. Knowledge of this Hamiltonian allows us to compactly represent the total derivative, in time, of any operator $L$ satisfying Lax's equation. Using  the chain rule, we observe that
$$
\begin{aligned}
\frac{d L}{d t} & =\sum_i \frac{\partial L}{\partial \Omega_i} \dot{\Omega}_i+\frac{\partial L}{\partial I_i} \dot{I}_i 
 =\sum_i \frac{\partial L}{\partial \Omega_i}\left\{H, \Omega_i\right\} =[L,P]
\end{aligned}
$$

Once again, assuming a hypothesis pair $\left(\tilde{L},\tilde{P}\right),$
and recognizing that Lax's equation holds pointwise in phase space, the loss function enabling statistical learning is given by
$$
\mathcal{J}_{\rm loss}(\tilde{L},\tilde{P})=\mathbb{E}_{\Omega\sim\rho}\sum_{j, k} \left(\sum_i \frac{\partial \tilde{L}}{\partial \Omega_i} \{H, \Omega_i\}-[\tilde{L}, \tilde{P}]\right)_{j, k}^2
$$
where the expected value is taken over a uniform distribution $\rho$ defined on the unit cube centred at the origin. Further assuming a parametrization vector $\eta$ in our Lax pair hypothesis, a sparse identification of Lax pairs is found by solving the following optimization problem
$$
\min _{\eta\in\mathbb{R}^N} (1-r)\mathcal{J}_{\rm loss}[\eta]+r\mathcal{R}[\eta]
$$
where $\mathcal{R}[\eta]$ and $r>0$ in tandem promote sparse discoveries.

We have foregone a penalization to steer away from the trivial Lax pair. This turns out to be a sensitive issue that we now illustrate. Suppose that we use the simple linear hypotheses
\begin{equation}\label{eq:SimpEulHyp}
\tilde{L}_{i,j}=\sum_k\xi_{i,j,k}\Omega_k,\qquad \tilde{P}_{i,j}=\sum_k\zeta_{i,j,k}\Omega_k,
\end{equation}
and that we minimize $\mathcal{J}_{\rm loss}$ over the 54 implied parameters $\xi_{i,j,k}$ and $\zeta_{i,j,k}$, which are vectorized into $\eta\in\mathbb{R}^{54}$. Of course, the first difficulty we encounter is that our numerical results consistently return the (valid) trivial solution $\xi_{i,j,k}=\zeta_{i,j,k}=0.$ We have many choices to steer away from the trivial solution, including the strategy used in the context of the simple harmonic oscillator and HH system earlier in this section. Using that strategy, we found degenerate Lax pairs, such as for example
\begin{equation}\label{eq:fakeEulLax}
L_{\rm degen} = \begin{pmatrix}
0 & 0 & 0 \\
0 & \Omega_1 & \Omega_2 \\
0 & \dfrac{I_2 (I_3 - I_2)}{I_1 (I_3 - I_1)}\, \Omega_2 & -\Omega_1
\end{pmatrix}, \qquad
P_{\rm degen} = \begin{pmatrix}
0 & 0 & 0 \\
0 & 0 & \dfrac{I_3 - I_1}{2 I_2}\, \Omega_3 \\
0 & \dfrac{I_2 - I_3}{2 I_1}\, \Omega_3 & 0
\end{pmatrix}
\end{equation}
which only reproduce the dynamics of two of the three expected Euler velocities.

The solution we found to learning Lax pairs for the Euler top in a principled way has two necessary ingredients. First, we found it necessary to enforce skew-symmetry in~\eqref{eq:SimpEulHyp} by demanding that for each $i,j,$ and $k$
$$
\frac{\partial\tilde{L}_{i,j}}{\partial\Omega_k}=-\frac{\partial\tilde{L}_{j,i}}{\partial\Omega_k},\qquad \frac{\partial\tilde{P}_{i,j}}{\partial\Omega_k}=-\frac{\partial\tilde{P}_{j,i}}{\partial\Omega_k}.
$$
Note that this compact notation only expresses the skew-symmetry condition when the hypothesis is linear. Also note that this choice considerably reduces the problem from a 54 parameter search to an 18 parameter one. The second ingredient is to introduce a soft penalization with the intention that each variable $\Omega_k$ remains represented so that all dynamics are reproduced by the learned Lax pair. We express this through 
\begin{equation}\label{eq:DegenEulObj}
\mathcal{J}_{\rm degen}
=
\prod_{k=1}^{3}
\left\| \frac{\partial \tilde{L}}{\partial \Omega_k} \right\|_F^2
\left\| \frac{\partial \tilde{P}}{\partial \Omega_k} \right\|_F^2
\end{equation}
where $\left\| A \right\|_F^2 := \operatorname{tr}(A^\top A)$ denotes the squared Frobenius norm. The problem is now to solve
$$
\min _{\eta\in\mathbb{R}^N}\mathcal{J}_{\rm Euler}[\eta]=\min _{\eta\in\mathbb{R}^N} (1-r)\left(\delta \mathcal{J}^{-1}_{\rm degen}[\eta]+(1-\delta)\mathcal{J}_{\rm loss}[\eta]\right)+r\mathcal{R}[\eta]
$$
where $\delta\in(0,1)$ balances avoidance of degeneracy with model discovery through the loss $\mathcal{J}_{\rm loss}$. We emphasize that if any factor in~\eqref{eq:DegenEulObj} is not considered in the design of the optimization problem, then numerical results will consistently degenerate to fake Lax pairs such as the one given by~\eqref{eq:fakeEulLax}.

Once the skew-symmetric, non-degenerate Lax pair $(L_*,P_*)$ is found numerically, we then search for the Manakov shift by solving the optimization problem
\begin{equation}\label{eq:ManakovSearch}
    \min_{\mathcal{M}_1,\mathcal{M}_2\in\mathbb{R}^3_{\rm diag}}\mathbb{E}_{\Omega\sim\rho}\mathbb{E}_{\lambda\sim 
    \Lambda}\mathcal{J}_{\rm Manakov}(\mathcal{M}_1,\mathcal{M}_2)
\end{equation}
where the Manakov objective is defined by
$$
\mathcal{J}_{\rm Manakov}=\sum_{j,k}\left([L_*+\lambda\mathcal{M}_1,P_*+\lambda\mathcal{M}_2]-[L_*,P_*]\right)^2_{j,k},
$$
and $\Lambda=\mathrm{Uniform}((-1,1]).$
The solution of this problem thus furnishes the Manakov shift for arbitrary spectral parameter $\lambda\in(0,1].$ We again report that SILO learns the expected Lax pair, with the correct conservation laws through learned Manakov shifts, to about seven digits of precision.

We see here that a drawback of SILO is the amount of user-specified information needed to build robust libraries and loss functions that avoid the degeneracy that plagues Lax's equation. Yet, if one is able to contend with these mathematical obstacles to engineer the correct SILO framework for a given Hamiltonian problem, then one can expect sharper resolution of Lax pairs and integrability of the dynamics.

\subsection{Learning KdV integrability using SILO}
We now show how a PDE system's integrability can be learned using SILO, the KdV equation
\begin{equation}
	\partial_tu-6 u\partial_xu+\partial_x^3u=0.
\end{equation}
The Hamiltonian form of the KdV equation is given by~\cite{zakharov1971korteweg}
$$
u_t=Q \frac{\delta H}{\delta u}
$$
where
\begin{equation}\label{eq:KdVHam}
	H =\int_{-\infty}^{\infty}\left(u^3-\frac{1}{2} u u_{x x}\right) d x:=\int_{-\infty}^{\infty} h\left(u, u_{x x}\right) d x.
\end{equation}
The KdV equation is bi-Hamiltonian, that is, two choices of $Q$ 
(with corresponding choices of $H$) 
yield the equation of motion. For the choice of
$H$ used here, the relevant operator is $Q=\partial_x$. 

The following Lax pair is well-known~\cite{gardner,ablowitz1981solitons}:
\begin{equation}\label{eq:KdVLax}
	L=-\partial_x^2+u, \qquad P=4 \partial_x^3-6 u \partial_x-3 u_x.
\end{equation}
The KdV equation can thus be viewed as the compatibility between these differential operators; that is, the equation
$$
\partial_t L=[L, P]
$$
reproduces Equation~\eqref{eq:KdV}.

To yet again build a loss function for SILO, we use the chain rule. Observe that
$$
	\partial_t L  =\frac{\partial L}{\partial u} \frac{\partial u}{\partial t} 
	= \frac{\partial L}{\partial u} Q\frac{\delta H}{\delta u}=[L, P].
$$
Thus, the expression $\frac{\partial \tilde{L}}{\partial u} Q\frac{\delta H}{\delta u}$ plays the role of the Poisson bracket in the design of the loss functions from previous sections. By direct analogy with Problem~\eqref{eq:SHOProb}, our optimization problem for this Hamiltonian setting is given by
\begin{equation}\label{eq:KdVProb}
	\min _{\eta\in\mathbb{R}^{N_{\eta}}}J[\eta]=\min _{\eta\in\mathbb{R}^{N_{\eta}}}
    (1-r)\mathbb{E}_{u\sim\rho} \frac{\int\left|\frac{\partial \tilde{L}}{\partial u} Q \frac{\delta H}{\delta u}u-[\tilde{L}, \tilde{P}]u\right|^2dx}{\int\left|\frac{\partial \tilde{L}}{\partial u} Q \frac{\delta H}{\delta u}u\right|^2dx}+r\mathcal{R}^*(\eta)
\end{equation}
where $\rho$ is a subset of the KdV phase space and $\mathcal{R}^*$ is a sparsification function.

In this PDE setting, there are only two additional modifications we need to make to the numerical framework from previous sections. First, we must carefully consider how we sample the phase space and how we construct the operator hypotheses. This amounts to sampling from the function space $L^p(\mathbb{R})$, $p>1$. To this end, we construct random samples from the overcomplete basis
\begin{equation}\label{eq:PDEsample}
u^{\textrm rand}(x)=N\sum_j e^{-a_j(x-b_j)^2}\sum_k\frac{A_{jk}}{k^3}\sin\frac{k\pi x}{L}
\end{equation}
where all parameters $a_j,\ b_j, A_{jk}$ are appropriately sampled from uniform distributions, $L$ is the length of the truncated spatial domain, and the coefficient $N$ ensures a unit norm in the space $L^1(\mathbb{R})$. In this way, we try to verify the compatibility of our operators with functions sampled from the function space $C^{3}(\mathbb{R})\cap L^p(\mathbb{R})$, $p>1$ and with equivalent masses in the sense of $L^1(\mathbb{R})$. We ensure smoothness of the samples by the decay of the Fourier coefficients $A_{j,k}/k^3$ while also regularizing by the fact that we only take the sum over $k$ to be finite. This smoothness is used for computational tractability in the evaluation of differential operators in the loss. We find $k=10$ to be sufficient in the numerics.

The second modification is the operator hypothesis. In general, Lax pairs are linear differential operators (in $x$) with coefficients dependent on $x,\ u,$ and derivatives of $u.$ A fairly wide class of operators is given by
$$
	\tilde{L}=\xi_1u+\xi_2\partial_x+\xi_3\partial_x^2,\quad \tilde{P}=\sum_j \sum_k \sum_m \zeta_{ j k m} u^{j-1} \left(\partial_x^{k-1} u\right) \partial_x^{m-1}
$$
where the indices in the sum all start at one. It may seem like the hypothesis on $\tilde{L}$ is overly restrictive. However, we must keep in mind that we seek to reproduce an evolution equation involving $\partial_t\tilde{L}$. Should any higher powers of $u$ enter into the hypothesis of $\tilde{L},$ then the likelihood of finding an \textit{explicit} equation of the form $u_t=F(u,u_x,u_{xx},...)$ decreases substantially.  In this sense, we seek to preserve the
semi-linear (in time) nature of the PDE of interest. To compute the spatial derivatives in the operator hypothesis, we use the Fast Fourier transform. That is, we use the formula
$$\partial^j_xu=\mathcal{F}^{-1}\left\{(ik)^j\mathcal{F}\{u\}\right\}$$
to compute the derivatives spectrally, where $k$ denotes the grid-dependent wavenumbers.

In our training, we find that only using $20$ samples is sufficient in our cross-validation. After training on these samples, we find that the loss, on average and without sparsification, is on the order of $10^{-11}$ when evaluated on unseen samples from $\Omega.$ We show, in Figure~\ref{fig:KdVResult}, a visual comparison between $\frac{\partial \tilde{L}}{\partial u} Q \frac{\delta H}{\delta u}$ and $[\tilde{L}, \tilde{P}]$ for four unseen sample functions. Note that despite the unit norms of the samples in $L^1(\mathbb{R})$, the generalized Poisson bracket and matrix commutators have arbitrary norms, which account for the variance of the scales across these images.
\begin{figure}[htbp]
\centering
\begin{minipage}{0.48\textwidth}
  \centering
  \includegraphics[width=\textwidth]{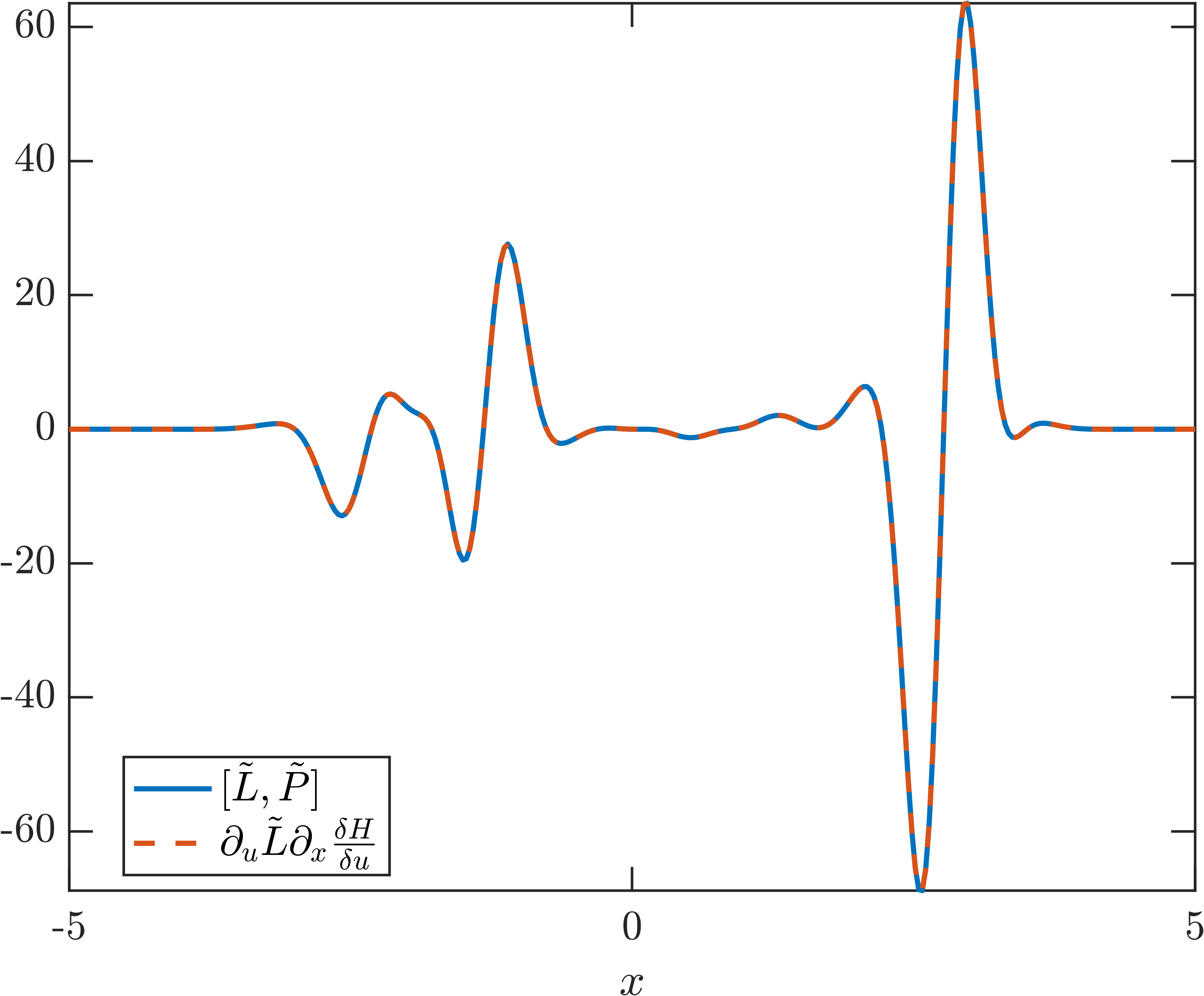}
\end{minipage}
\hfill
\begin{minipage}{0.48\textwidth}
  \centering
  \includegraphics[width=\textwidth]{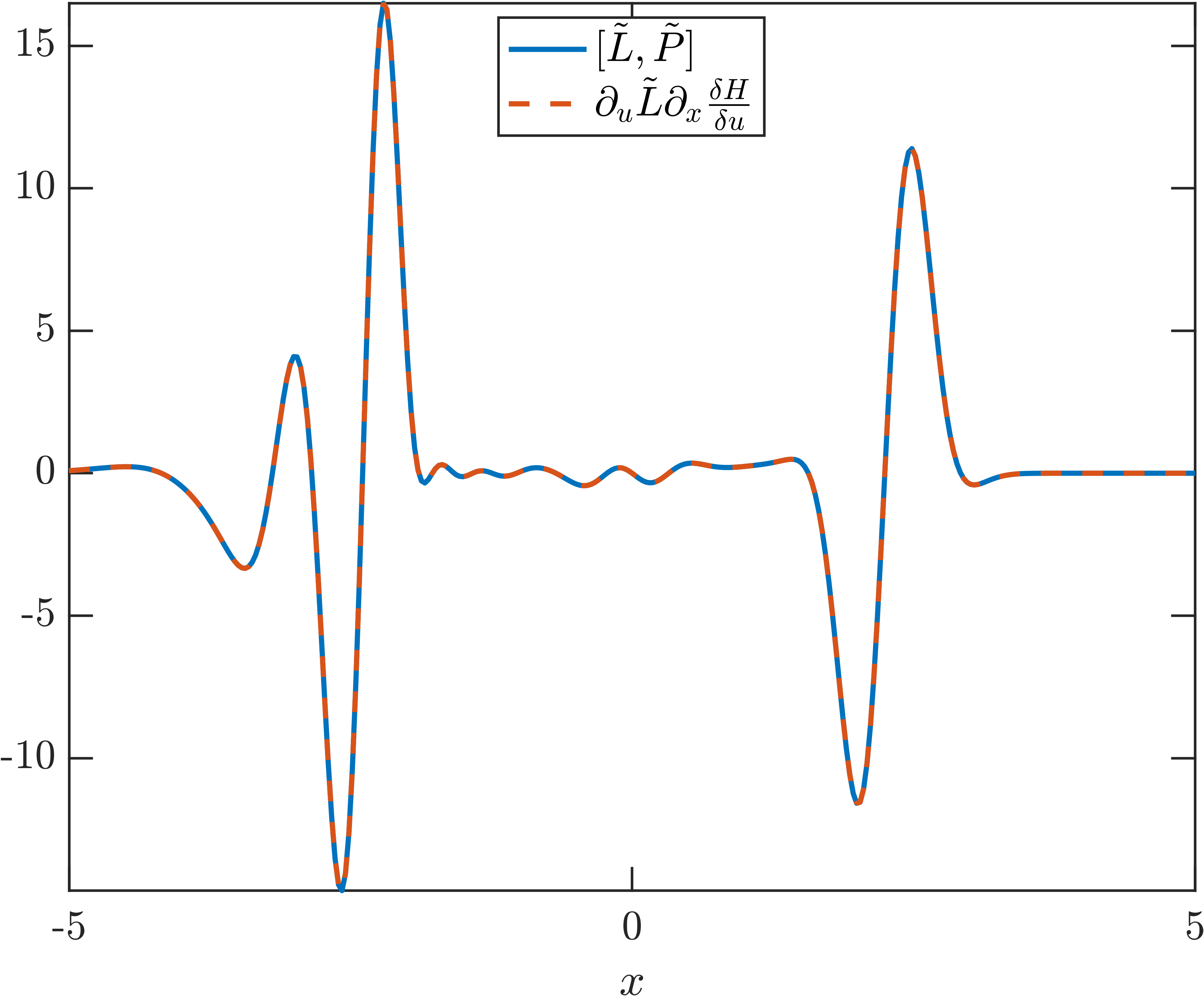}
\end{minipage}
\vspace{1em}

\begin{minipage}{0.48\textwidth}
  \centering
  \includegraphics[width=\textwidth]{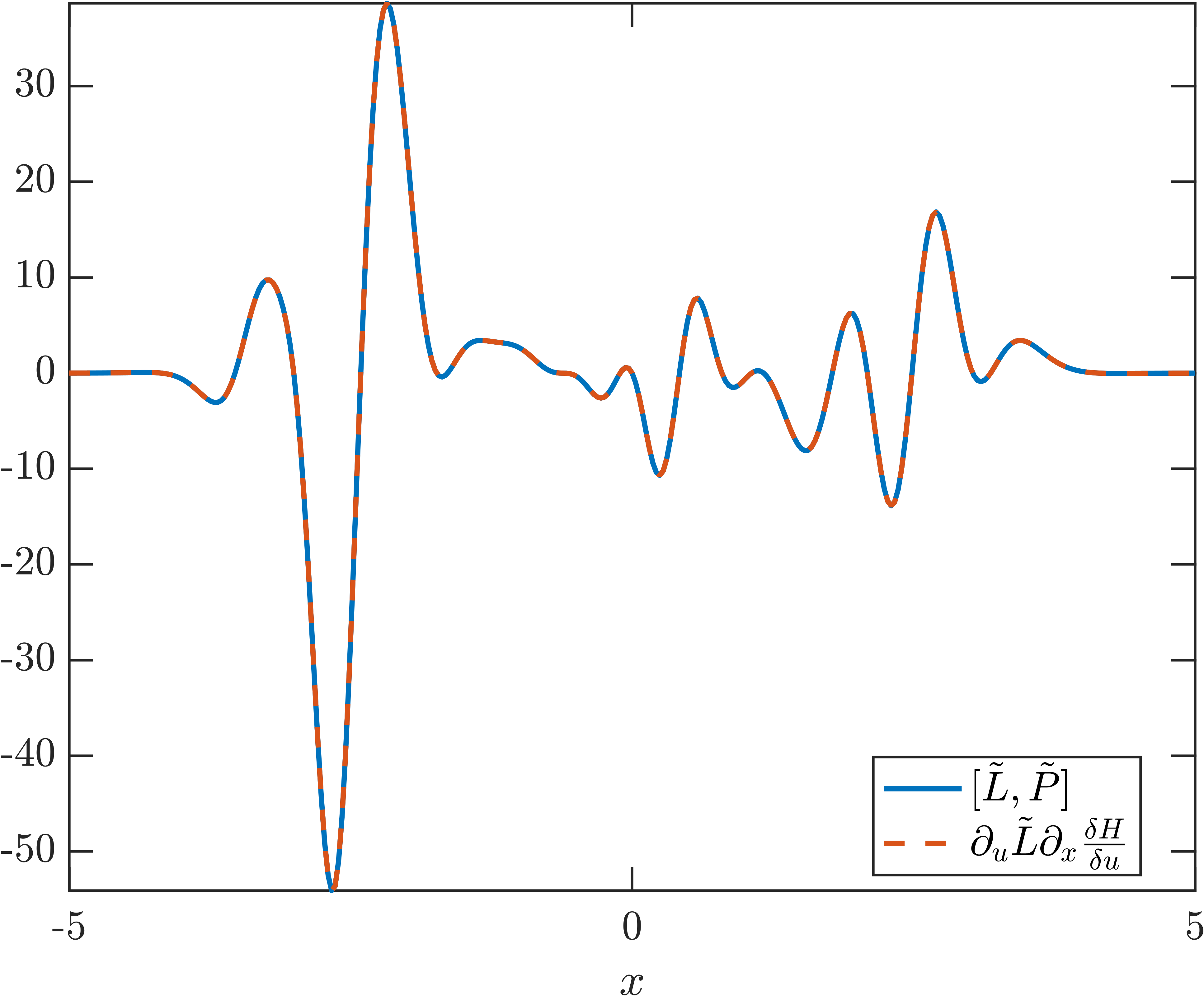}
\end{minipage}
\hfill
\begin{minipage}{0.48\textwidth}
  \centering
  \includegraphics[width=\textwidth]{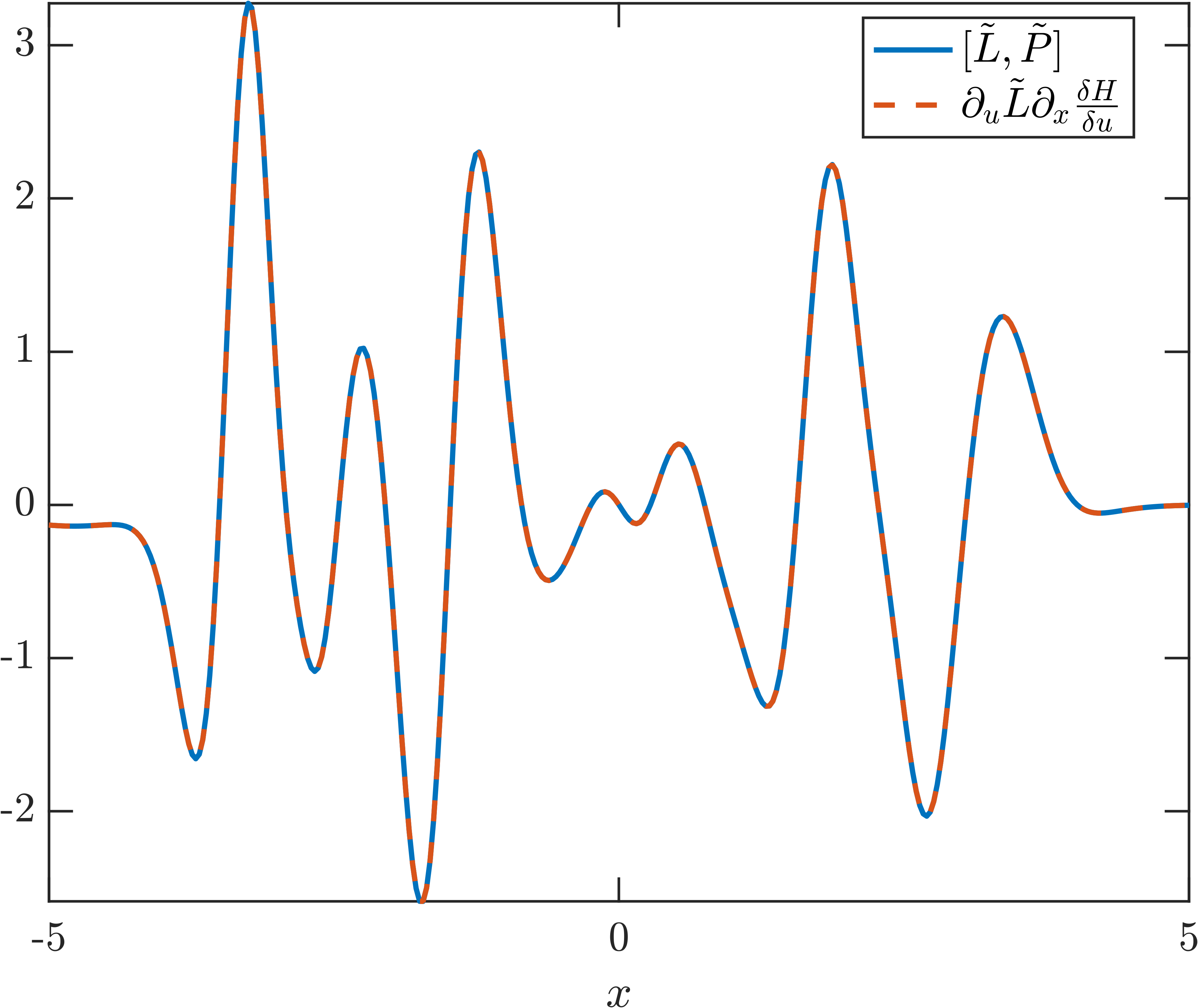}
\end{minipage}

\caption{A numerical result of solving Problem~\eqref{eq:KdVProb} without sparsification from the work of~\cite{adriazola2025computer}. Visualized here is a cross-validation study displaying the generalized Poisson brackets and commutators evaluated at the optimal point $\eta^*$ and on four samples from the function space $\Omega$ that were unseen during training. For all four cases, the loss is on the order of $10^{-11}$. Reproduced from~\cite{adriazola2025computer}. CC BY 4.0.}
\label{fig:KdVResult}
\end{figure}

To further demonstrate the validity of our approach, we show that our framework is sensitive enough to detect the known integrability of the KdV equation, similarly to how this was done for the HH system. Consider the non-integrable perturbation $h_1=\frac{1}{2}\left(\partial_x^2 u\right)^2$. We solve Problem~\eqref{eq:KdVProb}, again without sparsification, for Hamiltonian densities $h+\varepsilon h_1$, where $h$ is defined in Equation~\eqref{eq:KdVHam} and $\varepsilon\in[-.01,.01]$. Figure~\ref{fig:KdVPerturb} shows that the loss has a nearly smooth dependence on the parameter $\varepsilon$ with a minimum at the expected integrable point $\varepsilon=0$. Again, the integrability point is privileged, as the loss is several orders of magnitude smaller than the nearby values of $\varepsilon.$

\begin{figure}[htbp]
  \centering
  \includegraphics[width=0.8\textwidth]{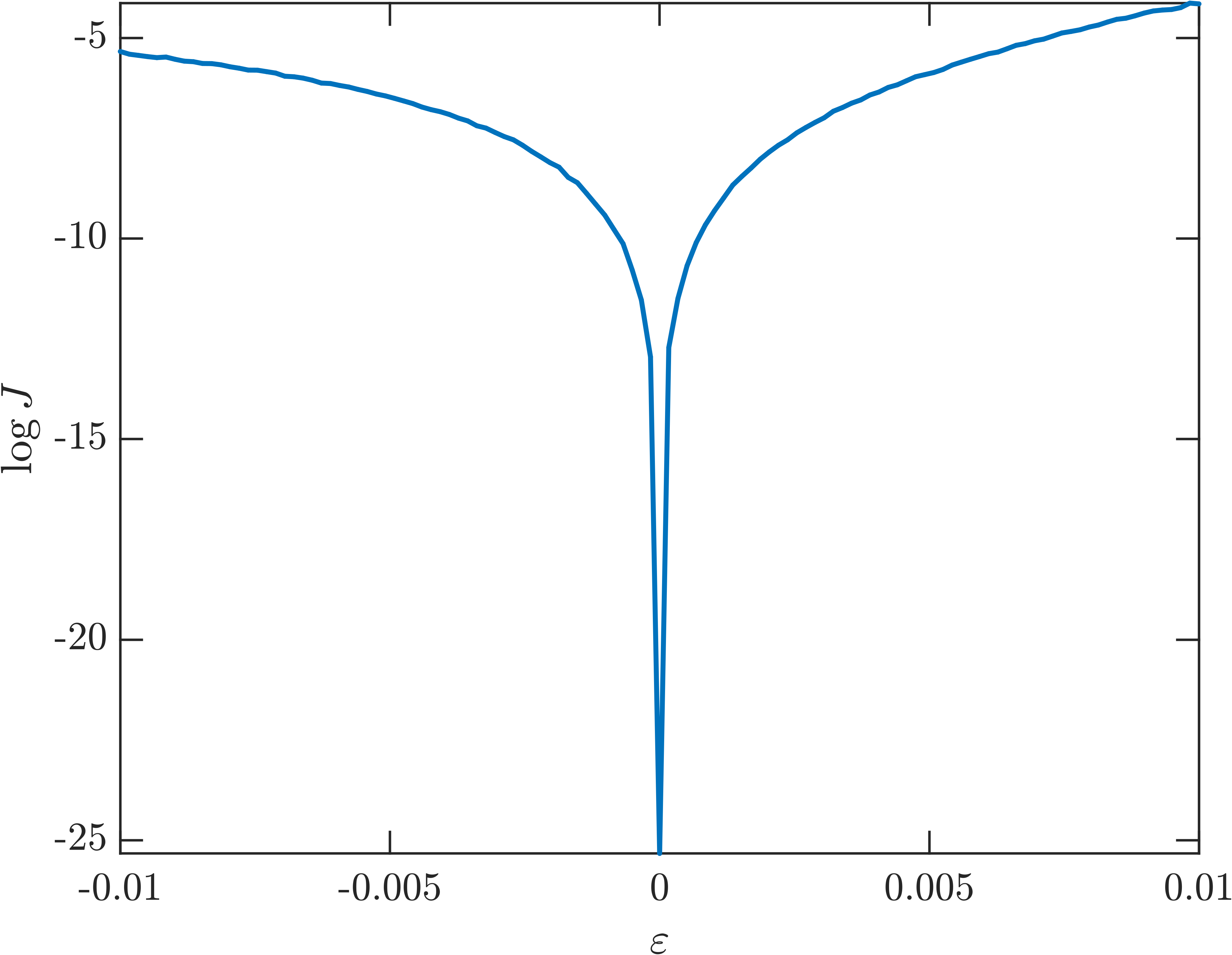}
  \caption{A perturbation study the perturbed Hamiltonian density $h_1 = \frac{\varepsilon}{2}(\partial_x^2 u)^2$. Shown here is the numerical solution of Problem~\eqref{eq:KdVProb}, without sparsification, using the full density $h + \varepsilon h_1$. We observe a near-smooth dependence on $\varepsilon$ with a clearly discernible ``special'' point associated with the detection of integrability at $\varepsilon = 0$. Reproduced from~\cite{adriazola2025computer}. CC BY 4.0.}
  \label{fig:KdVPerturb}
\end{figure}

We now discuss the interpretability of building Lax pairs from solving Problem~\eqref{eq:KdVProb}. Without sparsification, it is not surprising that all 39 coefficients in our operators are activated. Therefore, even with computer algebra systems such as Mathematica, we have no chance to interpret the PDE that the Lax pair is producing.

Once again, we use the same sparsification from the finite-degree-of-freedom setting to aid us in our interpretation of our discovered Lax pairs. Alongside the expected Lax pair, which we tend to rediscover only when constraining $L$ to be self-adjoint,
SILO discovers an entirely new family of Lax pairs, communicated by the following theorem.

\begin{theorem}[Existence of a first order KdV Lax pair]\label{thm:strongthm}
For every $u\in C^1([0,T];C^3(\mathbb{R}))$ and every $\alpha,\beta,\gamma,\delta,\epsilon,\kappa\in\mathbb{R}$ with $\alpha\neq0$, there exists a parameter $v\in\mathbb{R}$ such that the pair of operators
    $$\begin{aligned} 
	&L=\alpha u+\beta\partial_x,\\
	&P=\gamma u+\delta u^2+\epsilon u_{xx}+\kappa\partial_x,
\end{aligned}
$$
satisfying Lax's equation $\partial_tL=[L,P],$ understood as acting on the function space $C^3(\mathbb{R}),$ reproduces the KdV equation
$$
u_t=\frac{2 \beta  \delta}{\alpha }  u u_x+\frac{\beta  \epsilon}{\alpha }u_{\text{xxx}}
$$
in the co-travelling reference frame $x\to x-vt$.
\end{theorem}

A natural question raised by our discussion of Lax pairs is whether one should also consider ``fake'' Lax pairs, i.e. formulations that do not lead to a nontrivial IST. This is indeed a subtle point: in certain cases, such as the Euler top, it is possible to construct a Lax representation whose associated scattering data are trivial, and thus no meaningful integrable structure is obtained. The new Lax pair that we report above is indeed a fake Lax pair, which can be verified by hand when manually constructing the associated Jost functions and their dynamics. In this context, it is especially
relevant for future work to consider the integration of penalizations that steer us away from the space of fake Lax pairs during training.

\section{Conclusions and outlook}
\label{sec7}

In this review, we have surveyed the emerging interplay between nonlinear wave dynamics and ML, emphasizing how modern data-driven tools can be strengthened by embedding analytical structure. Section~\ref{sect:Pillars} outlined the main pillars of SciML that form the majority of the foundation for the methods used throughout this review. Section~\ref{sec3} reviewed advances in PINNs and their variants, highlighting both the potential and limitations of incorporating PDE constraints directly into learning architectures. Section~\ref{sec4} and section 5 turned to reduced order modeling, with a particular focus on SINDy~\cite{brunton2016discovering} and related sparse regression approaches, as well as autoencoder-based methods, which provide interpretable surrogates for complex dynamics. Section~\ref{sec5} examined methods for learning structural properties such as conservation laws, Hamiltonian and metriplectic structure, and related physical invariants. Section 7 presented a wave-centered perspective on the wide purview of applications of the Koopman operator theory. Finally, section~\ref{sec6} discussed the discovery of integrability, exploring recent attempts to learn Lax pairs, integrability, and conserved-quantity hierarchies.

There are several future directions in nonlinear waves to consider, including the study of another important scenario in nonlinear waves, namely that of soliton gases. These thermodynamic ensembles of interacting solitons offer a natural bridge between integrable PDEs and statistical mechanics~\cite{El2016,El2017}. ML methods may accelerate inference of effective kinetic equations, provide reduced surrogates for ensemble dynamics, and extend the scope of statistical soliton theory to non-integrable perturbations. By leveraging neural operators and generative models, it may become possible to simulate large soliton ensembles more efficiently, uncover new statistical closures, and explore thermodynamic regimes that remain inaccessible to current analytical techniques.

Another major thread of intense recent exploration concerns the dynamics, interactions (between them, as well as with solitary waves), higher-dimensional, as well as discrete realizations of dispersive shock waves (DSWs). Here, there is a central theoretical development that still merits extensive study, namely the so-called Whitham modulation theory~\cite{whitham1999linear}. Despite crossing the half-century mark of numerous hallmark developments associated with this theory, the field remains extremely active to this day, in part due to the development of novel theoretical tools (such as the so-called DSW fitting~\cite{el2005resolution}), and also due to the addition of numerous experimental platforms where such structures can be identified; see, e.g. the review of~\cite{el2016dispersive}. Nevertheless, in many cases, the equations of modulation theory remain extremely difficult to identify (especially so in non-integrable models, where, e.g. analytical waveforms for periodic waves may not be available), or in other cases can be derived, yet are extremely difficult to analyse and understand the features/structural characteristics thereof. Indeed, beyond the simpler playground of 1D continuum models, e.g. even in 1D discrete settings, or 2D continuum ones, it is fair to say that a deeper development and understanding of the theory and its implications still pose extensive and formidable challenges. It is hoped that, e.g. the techniques discussed herein and their particular relevance in especially inverse, but also direct problem solutions may be of significant assistance in this vein of research.

ROM within the nonlinear waves context is another key direction to consider. The recently proposed SHRED-ROM framework~\cite{tomasetto2025shredrom}, based on recurrent decoder networks with compressive training, provides a sensor-driven route to compact surrogates for dispersive systems and merits careful benchmarking across integrable, near-integrable, and turbulent regimes, as well as extensions to higher-dimensional wave equations. At the same time, hyperbolic and transport-dominated problems remain notoriously challenging for projection onto linear subspaces because the singular values of snapshot ensembles decay slowly-reflecting the slow Kolmogorov $n$-width decay of these solution manifolds~\cite{Peherstorfer2022breaking}. Recent strategies to overcome this include co-moving or transport-aware representations such as shifted POD~\cite{Reiss2018sPOD} and its robust variants~\cite{Krah2025}, registration and optimal-transport alignment of features before reduction~\cite{Blickhan2024LOT}, calibration methods---e.g. in hyperbolic problems characterized with multiple travelling discontinuities---that allow one to transform the original solution manifold into a lower dimensional one~\cite{Nonino2024ALECalibration}.

A further avenue involves climbing the conserved-quantity hierarchy. Although recent methods can recover conserved quantities such as mass, momentum, and energy, future algorithms should aim to automatically discover higher-order invariants, thereby reconstructing larger portions of the integrable tower. Such developments would also allow for the systematic study of how integrability breaks under perturbation. Beyond fundamental insights, these advances could provide powerful diagnostic tools for distinguishing chaotic from near-integrable dynamics in high-dimensional systems. Such developments could also play a central role towards a deeper understanding of the action-angle formulations of integrable and near-integrable systems. An important open direction there concerns especially how non-integrable perturbations may form short- (or progressively longer-) networks of connections between the actions, a perspective that has been recently occasionally argued in the literature~\cite{flach1,flach2}, but has yet to be quantified more broadly.

Symbolic learning frameworks such as AI-Descartes~\cite{Cornelio2023AIDescartes} and AI-Hilbert~\cite{CoryWright2024AIHilbert} show a strong potential to complement sparse regression approaches. By combining data with background knowledge, these systems have already demonstrated the ability to rediscover canonical scientific laws. A key ingredient in their design is the integration of algebraic geometry into the learning process. Hilbert's nullstellensatz provides a theoretical foundation for certifying the validity of candidate relations, while Gr\"obner basis methods enable systematic elimination of variables and simplification of polynomial structures. This marriage of symbolic reasoning and computational algebra allows the algorithms not only to fit data but also to ensure that the recovered expressions satisfy fundamental consistency conditions. For nonlinear dispersive PDEs, such capabilities could be transformative, opening pathways to the automated discovery of algebraic invariants, the construction of statistical closures that respect underlying constraints, and the reduction of kinetic models into interpretable symbolic forms. In the broader landscape of nonlinear waves and ML, the promise of these approaches lies in their ability to bridge rigorous mathematical theory with flexible data-driven pipelines, ensuring that discovered models remain both physically grounded and algebraically consistent.

Meanwhile, operator learning continues to provide a broad frontier. FNOs and related architectures~\cite{Kovachki2021Survey,Li2021FNO} have demonstrated significant success in fluid mechanics and wave propagation. Extending operator learning to dispersive and integrable systems may open new possibilities for efficient simulation and real-time control. Closely related generative models such as variational autoencoders and Boltzmann generators~\cite{Kingma2014,Noe2019} could help explore invariant measures, approximate soliton gases, and rare event statistics in nonlinear wave ensembles.

Briefly expanding upon generative models, we note that methods based on variational autoencoders and GANs are opening new frontiers in synthesizing realistic wave field statistics in turbulent or stochastic regimes, while reinforcement learning is emerging as a tool for optimal wave control in nonlinear PDEs~\cite{10886215}. Some of these recent advances in generative modeling have opened new avenues in fundamental wave physics, particularly for canonical systems such as water waves and BECs. For example, physics-informed diffusion models have shown promise in reconstructing high-fidelity fluid dynamics fields---including shallow-water and Navier--Stokes regimes---by enforcing equation constraints during sampling~\cite{zhou2024pifusion,wang2025fundiff}. Neural ODE frameworks extended with wave-equation-inspired architectures (e.g. ``neural wave equation'') provide continuous-time evolution learning suitable for BEC dynamics and dispersive hydrodynamics~\cite{chen2018neural,majumdar2025neuralwave}. Recent advances in physics-based flow matching provide a powerful framework for generating wavefield ensembles that explicitly respect the governing PDEs, making them highly suitable for surrogate modeling, uncertainty quantification, and accelerated simulation in wave phenomena~\cite{baldan2025pbfm}. Finally, stochastic interpolation via diffusion in function spaces, such as FunDiff, enables the generation of continuous wavefield realizations that obey conservation laws, which is critical for high-fidelity modeling in quantum and classical nonlinear wave systems~\cite{wang2025fundiff}.

A significant point of contact with realistic applications is, naturally, the ability to generalize the considerations presented herein to higher dimensional settings, as many of the case examples presented involve one spatial dimension. Interestingly, in the nonlinear wave realm---perhaps, arguably, per the prevalence of models in one spatial dimension within integrability and related techniques---, there have been surprisingly limited forays in higher-dimensional settings. A particularly instructive example, including due to its systematic comparison to traditional numerical techniques such as the finite element method (FEM), consists of the work of~\cite{schonlieb} which---in addition to numerous linear and nonlinear one-dimensional examples--- also considers 2d semilinear Schr{\"o}dinger examples. The comparison of FEM with PINNs reveals that for such ``standard PDE settings'', it is unclear that PINNs offer a competitive edge in comparison to standard methods; in fact, one might be inclined to conclude the opposite inasfar as ``traditional timestepping'' is concerned. To this date, studies of higher-dimensional nonlinear wave model settings addressed with ML methods remain few and far between. Indeed, another recent comparison study that emerged between the submission and resubmission of this manuscript also focused on the efficiency of a variety of 1d KdV, NLS or Klein--Gordon variants~\cite{haight2025solitonprofilesclassicalnumerical}. However, that being said, one should recognize that there exist settings where deep learning methods may present very noteworthy advantages in comparison to traditional methods, especially due to their potential to address the curse of dimensionality and due to their meshfree form. A particularly notable example stems from the realm of financial mathematics and the work of~\cite{SIRIGNANO20181339}, where, in order to study systems of (numerous, i.e. up to $d=200$) options, the authors developed a so-called Deep Galerkin meshfree method that leads to an accurate solution of the high-dimensional free boundary PDEs involved. This feature would have not been possible with standard (e.g. finite-element) methods. These authors have systematized their findings and presented a platform (including also numerical codes) for addressing such problems in the very recent book~\cite{SpiliopoulosSowersSirignano2025MathFoundationsDL}. This is, in our view, a ground-breaking development in this direction. Yet, at the moment, it is not immediately evident to us what implications it may enable in the realm of Nonlinear Wave studies. This is clearly an exciting avenue for further study.

Here, we also briefly touch upon a number of works that have sought to extend ML-based considerations to 2+1-dimensional settings. In the early study of~\cite{yongchen}, it was shown that PINNs could be used to higher-dimensional settings such as the Kadomtsev--Petviashvili equation leading to ``acceptably'' small approximation errors. On the other hand, the work of~\cite{Liu:24} used necklace patterns of integer-, half-integer and even fractional topological charge in 2+1-dimensional NLS models. Its aim was to illustrate that PINNs can effectively emulate the solution of such PDEs, potentially also enabling the prediction of the nonlinear dynamics of such physically relevant patterns. As is often expected in the realm of PINNs, they have worked in such higher-dimensional settings quite well in terms of solving inverse (coefficient identification) problems, even when data from multi-component settings as well as multi-dimensional equations are used, as shown in the work of~\cite{WANG2024114509}. However, in some cases where proximal models have been used such as the Zakharov--Kuznetsov and the regularized long-wave equation model in the work of~\cite{Nakamula_2025}, then there may be misidentification issues due to the structural similarities between the models. In such cases, initial condition variations, as well as other conservation-law-inspired amendments of the method may improve equation identification. Moreover, PINN-type methodologies have also been used to detect phase transitions within 2+1-dimensional Boussinesq equations, between periodic solutions in the form of lump chains and transformed waves (including different types of solitary waves). It will be particularly interesting to examine the potential application of other classes of techniques, including Neural Operator ones~\cite{Li2021FNO}, in such problems.

We hope to have illustrated herein that ML approaches for nonlinear wave equations have been particularly valuable on synthetic benchmarks. Nevertheless, their application to laboratory and field measurements introduces additional and highly nontrivial challenges. For instance, it is natural to expect that measurement noise can significantly degrade reconstruction accuracy and model discovery unless explicitly accounted for through potential regularization, noise-aware training, or physics-based constraints~\cite{brunton2016discovering,raissi2017physicsinformeddeeplearningi}. It is also important to recall that sparse and irregular sampling---which is common in experimental settings---further limits performance in tasks such as model (or, e.g. conservation law or integrability) identification, particularly for dispersive systems where fine spatial and temporal scales play a critical role. Some promising perspectives along this vein arise through recent operator-learning frameworks which partially alleviate this issue by learning resolution-independent mappings between function spaces~\cite{Li2021FNO}. However, it should be noted that generalization properties remain contingent upon adequate coverage of the input distribution. More broadly, extrapolation to unseen parameter regimes, boundary conditions, or dynamical behaviours remains a particularly challenging direction for future study. As has been discussed also recently, e.g. in~\cite{singha2025learninglawsconstraintprojectedneural}, purely data-driven models may produce solutions that fit observations while violating physical invariants or conservation laws. It is for that reason that this very recent work proposes a constraint-projected learning [a representative example of which was also presented in the work of a subset of the present authors in~\cite{zhu2022neural}]. Physics-informed and hybrid approaches can indeed improve physical fidelity and robustness, but they introduce additional modeling and optimization tradeoffs, particularly in balancing data fidelity against enforced constraints~\cite{raissi2019physics}. These considerations highlight the importance of careful validation on real data and the continued need for development of noise-robust, physically consistent ML methodologies for realistic nonlinear wave applications.

Taken together, these directions suggest that the integration of the rich mathematical structure of nonlinear wave systems with ML is still in its infancy, with much more ahead than behind. Approaches that respect conservation laws, integrability, and operator structure will remain central in guiding data-driven models so that they reflect (and shed further light on), rather than obscure, the underlying physics. Indeed, by embedding these principles, we not only safeguard interpretability and fidelity but also open new avenues for discovery, expanding the class of models and phenomena that can be meaningfully explored. The convergence of classical analysis and modern ML thus points toward a future in which our theoretical understanding and computational capabilities advance hand in hand. We hope that this review will encourage researchers from applied mathematics, nonlinear physics, and SciML to join in shaping this emerging landscape, using the complex, rich and ever-intriguing platform of nonlinear waves as their exploration ground of choice.

\section*{Acknowledgments}
This material is based upon work supported by the U.S. National Science Foundation under the award PHY-2316622 (JA), PHY-2110030, PHY-2408988 and DMS-2204702 (P G K) and DMS-2502900 and DMS-2540370 (WZ) and by the Air Force Office of Scientific Research (AFOSR) under Grant No. FA9550-25-1-0079 (WZ). Finally, this work was also supported by a grant from the Simons Foundation [SFI-MPS-SFM-00011048, P G K].

\section*{References}
\bibliography{bibliography3}
\bibliographystyle{siam}

\end{document}